\documentclass[superscriptaddress,twocolumn,showpacs,preprintnumbers,%
  amsmath,amssymb,aps]{revtex4-2}

\usepackage{subfig}
\usepackage{wrapfig}
\usepackage{caption}
\usepackage{color}
\usepackage{graphicx} 
\usepackage{amsfonts}
\usepackage{slashed}
\usepackage{bbm}
\usepackage[breaklinks,colorlinks=true]{hyperref}
\usepackage[nameinlink,capitalize]{cleveref}
\usepackage[utf8]{inputenc}
\usepackage{amscd}
\usepackage{latexsym}
\usepackage{textcomp}
\usepackage{bm}
\usepackage{commath}			
\usepackage{dsfont}			
\usepackage{extarrows}
\graphicspath{{figures/}}

\def\Eq#1{\Cref{#1}}
\def\Eqs#1{\Cref{#1}}

\def\eq#1{\Cref{#1}}

\def\Fig#1{\Cref{#1}}

\def\sec#1{\Cref{#1}}

\def\0#1#2{\frac{#1}{#2}}

\newcommand{\imag}{\text{i}}
\newcommand{\skipthis}[1]{}
\renewcommand{\d}{{\text{d}}}

\newcommand{\LQCD}{\Lambda_{\text{\tiny QCD}}}

\newcommand{\muq}{\mu_{\text{q}}}
\newcommand{\mud}{\mu_{\text{d}}}
\newcommand{\mub}{\mu_{\text{b}}}
\newcommand{\mubo}{\bar{\mu}_{\text{b}}}
\newcommand{\mudo}{\bar{\mu}_{\text{d}}}
\newcommand{\muqo}{\bar{\mu}_{\text{q}}}
\newcommand{\MN}{M_{\text{N}}}

\newcommand{\Md}{M_{\text{d}}}
\newcommand{\Mq}{M_{\text{q}}}
\newcommand{\Mb}{M_{\text{b}}}
\newcommand{\Mqb}{M_{\text{q}/\text{b}}}
\newcommand{\energyq}{\varepsilon_{\text{q}}}
\newcommand{\energyb}{\varepsilon_{\text{b}}}

\newcommand{\MeV}{\;\text{MeV}}
\newcommand{\GeV}{\;\text{GeV}}
\newcommand{\fm}{\;\text{fm}}

\newcommand{\Nc}{N_{\mathrm{c}}}

\newcommand{\tr}{\text{tr}}
\newcommand{\Tr}{{\text{Tr}}}

\newcommand{\Phifund}{\Phi_{\text{\tiny{fund}}}}
\newcommand{\rhos}{\rho_{\rm \phi}}
\newcommand{\rhod}{\rho_{\rm d}}
\newcommand{\Zq}{Z_{q}}
\newcommand{\Zphi}{Z_{\phi}}
\newcommand{\Zd}{Z_{d}}
\newcommand{\ZV}{Z_{V}}
\newcommand{\Zo}{Z_{\omega}}
\newcommand{\Zb}{Z_{b}}
\newcommand{\lambdas}{\lambda_{q\phi}}
\newcommand{\lambdad}{\lambda_{qd}}
\newcommand{\lambdasd}{\lambda_{q \phi/ d}}
\newcommand{\lambdaV}{\lambda_{qV}}
\newcommand{\lambdao}{\lambda_{q\omega}}
\newcommand{\lambdab}{\lambda_{qb}}
\newcommand{\lambdabd}{\lambda_{qdb}}
\newcommand{\lambdabo}{\lambda_{b\omega}}

\newcommand{\hs}{h_{q\phi}}
\newcommand{\hd}{h_{qd}}
\newcommand{\hV}{h_{qV}}
\newcommand{\ho}{h_{q\omega}}
\newcommand{\hsd}{h_{q \phi/d}}
\newcommand{\hsb}{h_{q/b \phi}}
\newcommand{\hb}{h_{b\phi}}
\newcommand{\hbo}{h_{\omega \rm b}}
\newcommand{\hdo}{h_{\omega \rm d}}
\newcommand{\hqo}{h_{\omega \rm q}}
\newcommand{\hqdb}{h_{qdb}}
\newcommand{\hqb}{h_{qb}}
\newcommand{\dynA}{D}
\newcommand{\dynB}{D}
\newcommand{\bolda}{\boldsymbol{a}}
\newcommand{\boldabar}{\bar{\boldsymbol{a}}}
\newcommand{\feyn}[1]{
  \setbox0=\hbox{\ensuremath{#1}}
  \hbox to\wd0{\hbox to0pt{\hbox to\wd0{\hss/\hss}\hss}\box0}}

\newcommand{\be}{\begin{align}}
\newcommand{\ee}{\end{align}}
\newcommand{\bea}{\begin{eqnarray}}
\newcommand{\eea}{\end{eqnarray}}

\renewcommand\d{\delta}

\def\di{\displaystyle}

\def\bg{\begin{eqnarray}\begin{array}{rcl}\displaystyle}
\def\eg{\end{array} &\di    &\di   \end{eqnarray}}

\def\Tr{{\rm Tr}}
\def\0#1#2{\frac{#1}{#2}}

\def\R{{{\rm l}\!{\rm R}}}

\begin{document}
\title{Emergent Hadrons and Diquarks}

\author{Kenji Fukushima}
\affiliation{Department of Physics, The University of Tokyo, 
7-3-1 Hongo, Bunkyo-ku, Tokyo 113-0033, Japan}
\author{Jan M. Pawlowski}
\affiliation{Institut f\"{u}r Theoretische Physik,
             Universit\"{a}t Heidelberg,
             Philosophenweg 16, 69120 Heidelberg, Germany}
\affiliation{ExtreMe Matter Institute EMMI, GSI, Planckstr. 1, 64291
  Darmstadt, Germany}
\author{Nils Strodthoff}
\affiliation{Fraunhofer Heinrich Hertz Institute,
Einsteinufer 37,
10587 Berlin, Germany}

\begin{abstract}
  We discuss the emergence of a low-energy effective theory with quarks, mesons, diquarks and baryons at vanishing and finite baryon density from first principle QCD\@. The present work also includes an overview on diquarks at vanishing and finite density, and elucidates the physics of transitional changes from baryonic matter to quark matter including diquarks. This set-up is discussed within the functional renormalisation group approach with dynamical hadronisation. In this framework it is detailed how mesons, diquarks, and baryons emerge dynamically from the renormalisation flow of the QCD effective action. Moreover, the fundamental degrees of freedom of QCD, quarks and gluons, decouple from the dynamics of QCD below the respective mass gaps. The resulting global picture unifies the different low energy effective theories used for low and high densities within QCD, and allows for a determination of the respective low energy constants directly from QCD\@.   
\end{abstract}

\maketitle

\tableofcontents
\vfill
\eject 

\section{Introduction}

There are a number of speculative scenarios for the large density phases of strongly interacting matter with quarks and gluons, see e.g.~\cite{Rajagopal:2000wf, Alford:2007xm, Fukushima:2010bq, Fukushima:2011jc, Fukushima:2013rx, BraunMunzinger:2011ze, Pawlowski:2014aha, Buballa:2014tba, Fischer:2018sdj, Fu:2019hdw, Dupuis:2020fhh} for reviews. In QCD, only one of these many scenarios is singled out. However, it is still a tough problem to reduce uncertainties in theoretical treatments at finite baryon density.  

In lattice-QCD, Monte-Carlo simulations break down for a baryon chemical potential comparable or larger than the temperature scale, $\mub/T\gtrsim 1$, see~\cite{Muroya:2003qs, Ejiri:2012vy, Aarts:2014fsa, Berger:2019odf, Alexandru:2020wrj} for reviews on attempts to evade the sign problem.  

By now, functional first principles approaches to QCD, such as the functional renormalisation group (fRG) approach or Dyson-Schwinger equations (DSEs), provide quantitative predictions for QCD correlation functions in the vacuum, at finite temperature, as well as moderate densities with baryon chemical potential $\mub$ restricted by $\mub/T\lesssim 4$, see~\cite{Fischer:2018sdj, Fu:2019hdw, Braun:2019aow, Isserstedt:2019pgx, Gao:2020qsj, Gao:2020fbl, Fu:2021oaw}. 

Moreover, the fRG-approach allows to describe naturally the emergence of low energy effective theories (LEFTs) from first principle QCD: in this diagrammatic approach, \textit{off-shell} quantum, thermal and density fluctuations are taken into account in terms of one-loop diagrams with full, momentum-dependent vertices and propagators. At lower momentum scales, gapped degrees of freedom successively decouple from the \textit{off-shell} dynamics of the theory. Finally, at very low momentum scales, only the lowest lying degrees of freedom survive. For example, in vacuum QCD these are the pions. This picture of \textit{emergent} low energy degrees of freedom and the \textit{decoupling} of fundamentals ones, and in particular the gluons, already explains the success of chiral perturbation theory, or chiral effective theory ($\chi$EFT). 

These processes and the dynamical emergence of the respective relevant degrees of freedom from QCD are naturally formulated with the functional renormalisation group approach: this approach allows the introduction of composite fields as dynamical degrees of freedom in the fundamental theory at hand. This is a very efficient reformulation of the fundamental theory in terms of dynamically relevant degrees of freedom. For example, if the latter completely dominate the dynamics of the theory, the fundamental degrees of freedom simply decouple. In turn, if either this dominance is not complete, or the composite fields are simply present as asymptotic states, the fundamental degrees of freedom still play a r$\hat{\textrm{o}}$le. This approach (rebosonisation or dynamical condensation) has been put forward in \cite{Gies:2001nw, Gies:2002hq}, and has been further developed in \cite{Pawlowski:2005xe, Floerchinger:2009uf}. In QCD it is called \textit{dynamical hadronisation}, for applications see \cite{Gies:2002hq, Braun:2014ata, Mitter:2014wpa, Rennecke:2015eba, Cyrol:2017ewj, Fu:2019hdw}. Its current form with further conceptual developments has been discussed in \cite{Fu:2019hdw}, for a recent review see \cite{Dupuis:2020fhh}.  

Importantly, for smaller RG-scales the dynamics or RG-flows of full QCD resembles closely the RG-flow of low energy effective theories (LEFTs) with emergent hadronic and di-quark degrees of freedom, for a detailed discussion see \cite{Fu:2019hdw, Dupuis:2020fhh}. Such LEFTs have been studied intensively in the past decades, for QCD-related fRG reviews see \cite{Berges:2000ew, Pawlowski:2005xe, Gies:2006wv, Schaefer:2006sr, Rosten:2010vm, Braun:2011pp, Pawlowski:2014aha, Dupuis:2020fhh}, a rather complete list of works can be found in \cite{Dupuis:2020fhh}. 

However, while not hampered by a sign problem, at larger densities the reliability of the expansion order used so far in these functional works is successively lost. This is in one-to-one correspondence to an increasing numbers of potentially resonant hadronic interaction channels of multi-quark interactions, that have to be taken into account with dynamical hadronisation. While this is in principle possible systematically, it is still technically very challenging. Indeed, the present work will provide some of the necessary further developments of the expansion schemes. 

Consequently, for large densities with $\mub/T\gtrsim 4$, this leaves us with the urgent task to device phenomenological studies within LEFTs, that can be systematically embedded into QCD\@. With the emergence of LEFTs from functional QCD via dynamical hadronisation, this systematics is provided. It is left to detail the relation of LEFTs to QCD, and to successively improve respective LEFTs via their connection to first principle QCD for $\mub/T\lesssim 4$ at present, but hopefully soon also larger densities. This leads to a very appealing situation with exciting possibilities: while LEFTs at large density can be benchmarked and improved systematically with their emergence (at lower densities) within first principle QCD for $\mub/T\lesssim 4$, phenomenological studies and improvements of LEFTs can guide the respective first principles studies. This combined analysis should allow to push the current  $\mub/T\lesssim 4$ border further and further, finally resulting in a first principle access to QCD at all densities. 

Note in this context, that LEFT-studies had played a similar guiding role in regimes with $T\gg \mub$, and many of the findings where later confirmed by lattice QCD or functional QCD studies. Naturally we expect that in the same way as for the high-$T$ studies, low energy effective field theory studies guide us to a correct intuition of the dense state of QCD matter. 

Therefore, we proceed by discussing respective LEFTs used so far at large densities. We have already mentioned the emergence of $\chi$EFT within first principle QCD\@. Accordingly, a theoretical approach  based on the chiral effective theory, see e.g.~\cite{Hebeler:2013nza}, should be a valid framework as long as the system is not far from cold nuclear matter, that is up to twice of normal nuclear density and temperature around the pion mass. Such a theoretical formulation in terms of hadronic degrees of freedom should be an alternative of the first principles approach based on the systematic low-energy expansion of QCD, but should lose its foundation at high density or high temperature where partonic degrees of freedom would become dominant.  See also \cite{Drews:2013hha, Drews:2014wba} for a similar calculation using the renormalisation group equation based on \cite{Floerchinger:2012xd}, see also \cite{Fukushima:2014lfa} for fluctuation estimate within the same model. More recently, the nuclear equation of state (EoS) has been investigated within the fRG, \cite{Otto:2019zjy, Otto:2020hoz}, and in a combination of  $\chi$EFT and fRG in QCD, see \cite{Leonhardt:2019fua}.

At moderate densities we may then investigate the properties of matter entirely out of quarks using the chiral quark models including fluctuations, for fRG work in this direction we refer to the reviews \cite{Berges:2000ew, Pawlowski:2005xe, Gies:2006wv, Schaefer:2006sr, Rosten:2010vm, Braun:2011pp, Pawlowski:2014aha, Dupuis:2020fhh}, at asymptotically high  
densities much work has been done within (resummed) perturbative QCD calculations~\cite{Andersen:2002jz, Fraga:2004gz, Kurkela:2009gj, Fujimoto:2020tjc}, for respective fRG-work in QCD including deviations due to non-perturbative condensation effects see \cite{Leonhardt:2019fua}.

It is a big puzzle even today how these two descriptions of dense matter in terms of baryons and quarks could be (possibly
smoothly~\cite{Kojo:2014rca, Fujimoto:2019sxg}) connected in an intermediate density region.  In other words, there is no clear picture of quark deconfinement induced by large density effects, while deconfinement is a well-studied phenomenon along the temperature axis in lattice QCD and effective models. If we have a quark-model picture in mind, deconfinement simply means that
bound states of quarks melt without phase transition~\cite{Wang:2010iu} or the mobility of quarks gets increased by percolation~\cite{Andronic:2009gj}, see also \cite{Fukushima:2020cmk} for modernisation with quantum percolation.  As we will argue later, such a picture should need more refinement taking account of differences between meson and baryon interactions.

A key ingredient to shed light on a modern picture of high-density deconfinement is the diquark correlation (see \cite{Anselmino:1992vg} for a comprehensive review on diquark physics). Diquarks form a condensate above the threshold, having an impact on the phase diagram.  For example, see~\cite{Hatsuda:2006ps, Abuki:2010jq} for a breach of the chiral phase boundary due to the diquark coupling. There are already some attempts to build an effective model in terms
of diquarks as well as mesons and quarks (i.e., Quark-Meson-Diquark models).  In conventional model set-ups as in \cite{Bentz:2003qz} for example, it is common to accommodate diquarks as explicit (and point-like) degrees of freedom and generate baryons as bound states of quarks and diquarks~\cite{Ebert:1994mf, AbuRaddad:2002pw}.  See also attempts~\cite{Blaschke:2014zsa, Dubinin:2016wvt} in which diquark dissociation or the Mott transition has been taken into account, see \cite{Strodthoff:2011tz, Strodthoff:2013cua} for an application of the Quark-Meson-Diquark model for two-colour QCD, see \cite{Cichutek:2020bli} for fRG studies of an SO(6)-symmetric diquark model.  Obviously, however, it is always possible to represent the baryon interpolation field as a combination of the diquark and the quark fields, see e.g.~\cite{Nagata:2007di, Chen:2012vs} for its chiral properties. 
We emphasise that this does not entail the interpretation of diquarks as bound objects or asymptotic states. Clearly, they are not due to their colour. Instead, diquarks are dynamically generated (off-shell) excitations, and hence we must take into account these  microscopic structures of diquarks.   

We will formulate these processes of the formation and the dissociation of diquarks by means of the fRG equations. In this work we will give a theoretical foundation with correct physical degrees of freedom and physical boundary conditions satisfied.  Here, we will limit ourselves to a rather simplified framework of the Lagrangian similar to the Quark-Meson-Diquark model.  As we will explicitly see later, this simple framework is already nontrivial enough and needs technical developments beyond the simpler two-colour QCD case~\cite{Strodthoff:2011tz, Strodthoff:2013cua, Khan:2015etz, Khan:2015puu}.  

In \sec{sec:overview} we provide an overview on diquark constituents in hadrons and in dense quark matter. In \sec{sec:ContDual} we discuss the continuous quark-hadron duality. In \sec{sec:Emmesons} we iniate our analysis of emergent hadrons and diquarks with that on mesons and diquarks.  In \sec{sec:Emmesons} we iniate our analysis of emergent hadrons and diquarks with that on mesons and diquarks. It is shown, how resonant four-quark interaction channels in QCD can be described in terms of dynamical low energy degrees of freedom with dynamical hadronisation. In \sec{sec:Emhadrons} this analysis is extended to baryons. This includes a discussion of the different formation processes of baryons via diquark-quark scattering and three-quark scattering. This analysis is continued in \sec{sec:QuarkHadron}, specifically concentrating on quark-hadron mixing.

\section{Overview: Diquarks at Zero and Finite Baryon Density}\label{sec:overview}

Here we briefly review basic properties of diquarks in the
  contexts of hadronic structures and the ground state constituents in
  dense quark matter.

\subsection{Classification of diquarks}

The idea of the diquark can be traced back to a classic paper by
Gell-Mann~\cite{GellMann:1964nj} and the word ``diquark'' was used
already in a paper in 1966~\cite{Ida:1966ev} in the same way as we do
today.  It is a long-standing problem how to define
diquarks, or precisely speaking, how to quantify diquark correlations
in a baryon or in baryonic (nuclear) matter.  In fact, it
is a widely accepted knowledge to introduce a ``constituent diquark''
to solve a three-body bound state problem (i.e., the Faddeev equation)
of baryons~\cite{Ishii:1993np,*Ishii:1993rt,*Ishii:1995bu}.  See also
\cite{Nagata:2005qb} for the two nucleon problem.  The diquark
in this formulation is, however, not necessarily a physical object and
there are a number of theoretical attempts to seek for a strong and
localised correlation of diquarks in hadrons.

One of the most well-known examples that suggest the diquark
correlation is physical is the inverted mass ordering in the scalar
nonet channel~\cite{Jaffe:1976ig,*Jaffe:1976ih} and this idea is also
supported by the instanton-induced interactions~\cite{Hooft:2008we}
(see \cite{Wakayama:2014gpa} for a recent lattice study).  There is
also a proposal that some of the exotic $X$, $Y$, $Z$ mesons might have 
a significant overlap with the diquark--anti-diquark state on top of the meson molecule
state (see \cite{Chen:2013wva} for an overview), though the coupled channel analysis 
with interactions from lattice QCD has clarified that $Z_c(3900)$ results from threshold 
enhancement~\cite{Ikeda:2016zwx}.  Another example
that indicates the diquark correlation is the $\Delta I=1/2$ rule in
the non-leptonic weak decay, that is, the decay amplitude in the
$\Delta I=1/2$ channel is by an order of magnitude enhanced than that
in the $\Delta I=3/2$ channel.  One possible interpretation for the
$\Delta I=1/2$ rule is provided by the localised triplet-diquark in
the hyperon wave-function that couples to the $\Delta I=1/2$
channel~\cite{Stech:1987fa,Stech:1990fx,Dosch:1988hu,Neubert:1990gh,Neubert:1991zd}.

As diquarks are coloured
objects, they do not exist as asymptotic states, and hence are not
manifest observables in the vacuum. This leaves us with the physics
question of whether diquarks exist as collective modes in some
particular environments.  In what follows below let us make a brief
summary of a diquark classification in different channels.
\vspace{0.2em}

\textit{Colour ---}
Because quarks belong to the colour triplet, a system with two quarks
can be decomposed to a colour anti-triplet and a sextet:
(${\bf 3}_c\otimes{\bf 3}_c=\bar{\bf 3}_c\oplus{\bf 6}_c$). One-gluon
exchange potentials suggest an attractive (and repulsive) force in the
triple $\bar{\bf 3}_c$ (and sextet ${\bf 6}_c$, respectively) channel, so
that we can discard colour ${\bf 6}_c$ pairing in the first
approximation.  In addition, it is only the $\bar{\bf 3}_c$ which can be
coupled to another ${\bf 3}_c$ quark to form a colourless baryon.
\vspace{0.2em}

\textit{Flavour ---}
For $N_f=2$ two quarks can form
${\bf 2}_f\otimes{\bf 2}_f={\bf 1}_f\oplus{\bf 3}_f$ and for $N_f=3$ the
irreducible decomposition reads as
${\bf 3}_f\otimes{\bf 3}_f=\bar{\bf 3}_f\oplus{\bf 6}_f$.  For a given colour
and flavour representation, the symmetry properties of the spin-orbit
part is fixed by the Pauli principle.  For instance, for a colour $\bar{\bf 3}_c$ and
two-flavour ${\bf 1}_f$ or three-flavour $\bar{\bf 3}_f$, both colour and flavour indices are
anti-symmetric under exchanging two quarks, and so the spin-orbit part
of the diquark wave-function must be anti-symmetric to satisfy the
Fermi statistics. Likewise a colour $\bar{\bf 3}_c$ and the symmetric two-flavour
${\bf 3}_f$ or three-flavour ${\bf 6}_f$ requires a spin-orbit part that is symmetric with respect to the exchange 
of two quarks.
\vspace{0.2em}

\textit{Spin ---} The colour and flavour analysis above entails that the
spin-orbit part should be anti-symmetric in the colour anti-symmetric and flavour anti-symmetric channel, and the allowed
quantum numbers are $0^+$ (for $L=0$, $S=0$), $1^-$ (for $L=1$,
$S=0$), etc.  If the colour is anti-symmetric and the flavour is symmetric, on the
other hand, the spin-orbit part should be also symmetric.
Then, $1^+$ (for $L=0$, $S=1$), $0^-$ (for $L=1$, $S=1$), etc.\ should
appear.  \vspace{0.2em}

It is chromo-magnetic interaction that favours
positive-parity (i.e., $0^+$ and $1^+$) diquarks rather than
negative-parity (i.e., $0^-$ and $1^-$) ones.  Moreover, it is
straightforward to confirm that the colour-spin and flavour-spin
interaction would stabilise the $0^+$ diquark more, which is concisely dictated by the Breit interaction in QCD proportional to the inner product of two spins, $(\boldsymbol{s}_1\cdot \boldsymbol{s}_2)$.  For the spin singlet state $(\boldsymbol{s}_1\cdot \boldsymbol{s}_2)$ takes a value of $-3/4$, while it is $+1/4$ for the spin triplet state, leading to an energy difference between $0^+$ and $1^+$.  Thus, the $0^+$ scalar diquark called ``good diquark'' is the most favoured,
and the $1^+$ axial-vector diquark called ``bad diquark'' is the second
favoured.  It is important to distinguish the good and the bad
diquarks for phenomenological applications; see, for example,
\cite{Selem:2006nd} for high-spin hadron spectra nicely
parametrised with the good and the bad diquarks. 

Since diquarks are not gauge invariant, it is indispensable to choose
a certain gauge to observe diquarks in the lattice-QCD simulation.
(This is the biggest difference from diquarks in two-colour QCD, where
diquarks are colourless baryons~\cite{Kogut:2000ek,Strodthoff:2011tz}.)
In Ref.~\cite{Hess:1998sd} the diquark mass has been estimated in
lattice-QCD (with $m_\pi\simeq 350\MeV$) to be
$m_{\text{good}}=694(22)\MeV$ from the exponential decay in the
diquark-diquark correlation, and also an expected relation of the mass
difference between $S=1$ and $S=0$ channels, i.e.\
$m_{\text{bad}}-m_{\text{good}}\approx \frac{1}{2}(m_\Delta-m_{\text{N}})$,
was investigated.
Here, $m_{\text{good}}$ and $m_{\text{bad}}$ represent the masses of the
good diquark in the scalar channel and the bad diquark in the
axial-vector channel. Also the maximum entropy method has been applied
to calculate the diquark spectral functions~\cite{Wetzorke:2000ez} and
a prominent peak stands at the diquark mass
$m_{\text{good}}=0.60(2)\GeV$.  Interestingly, the spatial size
of the diquark correlation in a hadron has been also evaluated from the charge or
baryon density distribution measured on the
lattice~\cite{Leinweber:1993nr,Alexandrou:2006cq}.  The characteristic
size of the diquark correlation which is the strongest in the $0^+$
channel is, however, found to be comparable to the hadron size
$\sim 1\fm$.  Therefore, it is definitely true that the diquark
``correlation'' exists inside of baryons (or even more manifested in
systems of one heavy quark and two light quarks~\cite{Copley:1979wj}),
but diquarks do not necessarily appear as point-like particles but
they should look like spatially extended objects. It is important to
keep in mind that diquarks could be extended objects and this is why
we need to solve a dynamical problem to treat diquarks as emergent
degrees of freedom.

\subsection{Diquarks in colour superconductivity}
The fact that diquarks are not point-like is evident in the
consideration of colour superconductivity that is a ground state of QCD
matter at asymptotically high
density~\cite{Rapp:1997zu,Alford:1997zt}, for reviews see
\cite{Fukushima:2010bq,Rajagopal:2000wf,Alford:2007xm}.  Among
various pairing patterns~\cite{Iida:2003cc,Fukushima:2004zq}, the
two-flavour colour superconducting (2SC) phase and the colour-flavour
locked (CFL) phase, which are both characterised by condensation of
the good diquarks, have special importance on the phase diagram of
high-density matter~\cite{Fukushima:2005fh}.  Also for one-flavour
colour superconductivity, the flavour sector is symmetric and so only
the bad diquark is possible.  Then, the rotational symmetry is broken
by condensation of axial-vector diquark, leading to a spin-one colour
superconductivity~\cite{Schmitt:2003xq}.

The 2SC phase should be a ground state if the $s$-quark mass is
sufficiently large.  This state, however, does not break global
${\rm U(1)_V}$ symmetry and so it is not a superfluid. Moreover one can
prove that chiral symmetry is not broken in the 2SC phase;
anti-symmetric combination of two ${\bf 2}_f$ forms a flavour singlet.
Therefore, if there is an interface between the hadronic phase and the
2SC phase, there should be a chiral phase transition.  There is also a
possibility of a coexisting state of the 2SC gap and the chiral
condensate that breaks chiral symmetry.  Besides, a mixture of
spin-one and tensorial diquark condensates may be possible, and then it is difficult
to distinguish baryonic matter (in which
$\langle pp\rangle\sim \langle uuudud\rangle$ and
$\langle nn\rangle \sim \langle ddudud\rangle$ pairings exist)
from the 2SC phase (with $\langle ud\rangle$ in the colour-triplet
channel and $\langle uu\rangle$ and $\langle dd\rangle$ in the
colour-sextet channel) from the symmetry arguments not only for $^1 S_0$ 
but also $^3 P_2$ states, see \cite{Fujimoto:2019sxg}.

The CFL phase is also indistinguishable from baryonic matter with
strangeness superfluid components.  Unlike the 2SC phase the CFL phase
itself spontaneously breaks chiral symmetry and ${\rm U(1)_V}$ symmetry
associated with superfluidity~\cite{Alford:1998mk}.  So, all of
${\rm U(3)_L\times U(3)_R}$ symmetry in flavour space is broken by the
condensates.  It has been an established knowledge that the chiral
symmetry breaking is partially restored at normal nuclear density
about $\sim 30\%$ (see Ref.~\cite{Hayano:2008vn} and references
therein), but there is so far no convincing evidence for full chiral
restoration~\cite{Fiorilla:2012bc}.  In fact, contrary to our naive
intuition of symmetry restoration at higher density, once the CFL
phase occurs, chiral symmetry must be always broken
even at high baryon densities.
Interestingly, physical degrees of freedom may have one-to-one
correspondence between baryonic and quark
matter~\cite{Schafer:1998ef,Alford:1999pa,Hatsuda:2006ps}; for
example, mesons in the hadronic phase are composed from a quark and an
anti-quark, while in the CFL phase mesons are mostly tetra-quark
objects of a diquark and an anti-diquark~\cite{Fukushima:2004bj} just
like the hypothesised nonet scalar mesons.  The counterparts of baryons in the
hadronic phase are quarks in the CFL phase.  Recently, the continuity
in terms of superfluid vortices has also been studied;  some arguments
disfavour the continuity~\cite{Cipriani:2012hr,Cherman:2018jir}, and
others support the continuity~\cite{Alford:2018mqj,Hirono:2018fjr}.

For both cases of the 2SC phase and the CFL phase, the important point
is that a scenario with no first-order phase transition between
baryonic matter and quark matter can be a logical possibility.  In
other words, baryonic and quark matter is indistinguishable,
\textit{in principle}, from the symmetry (and also from the topology,
see \cite{Hirono:2018fjr}).  On the formal level, such a
hypothesis of the Quark-Hadron Continuity is based on the fact that
there is no gauge invariant order parameter
to define colour superconductivity.  In lattice-QCD simulation or
in any experimental measurements what we can observe is an expectation
value of gauge-invariant operators.  Then, the 2SC and the CFL phases
could be characterised by condensation of four-quark and six-quark
operators, but these quantities would be non-vanishing already in
baryonic matter with a superfluid component.

  \section{Continuous Duality Between Baryonic And Quark Matter with Diquarks}\label{sec:ContDual}

We know that at zero temperature the first onset of baryon density
appears with a first-order phase transition from vacuum to
normal and symmetric nuclear matter as the baryon chemical potential $\mub$
increases.  It happens at $\mub=\MN-\epsilon_B\simeq 923\MeV$ for
symmetric nuclear matter, where $\MN\simeq 939\MeV$ is the nucleon
mass and $\epsilon_B\simeq 16\MeV$ is the binding energy.
Right on the phase transition, a mixed state realises and various spatial
structures may occur~\cite{Tatsumi:2011tt}.  The question is where we
can expect other degrees of freedom of quarks and diquarks.

\subsection{From a bag model picture to a modern view}

If there are any new degrees of freedom in matter coupled to the baryon
density, it would emerge when the baryon chemical potential exceeds
the corresponding mass threshold.  If density-induced deconfinement is
a smooth phenomenon just as observed along the temperature axis,
as argued in the previous section, confinement would be gradually lost
in baryonic matter.  Then, the quark onset should be located at three
times the (in-medium) constituent quark mass (if the interaction effect is 
renormalised in the effective mass).  In the same way, as
soon as $\mub$ gets greater than three halves times the (in-medium)
constituent diquark mass, the system should accommodate the diquark
degrees of freedom.  We should note that the diquarks immediately form
a condensate because they are bosons, leading to a colour
superconducting state.  So, the onset of diquarks degrees of freedom
is essentially the onset of colour superconductivity.

The old-fashioned picture of deconfinement is as follows.  The pressure
of quark matter, $P_{\text{deconfined}}$, and the pressure of baryonic
matter, $P_{\text{confined}}$, are compared and the one with
larger pressure is identified as the favoured state.  In this bag model
picture physical degrees of freedom are switched sharply at the
density with $P_{\text{deconfined}}=P_{\text{confined}}$.  In this
picture the double-counting problem of relevant physical degrees of
freedom is avoided too strongly, and any coexistence of baryons,
quarks, and diquarks is excluded by construction.

In a dynamical picture which allows for smooth dissociation of bound
states, however, coexistence is not excluded, and it would be a
natural anticipation to consider successive onsets.  For the ensuing
discussion let us introduce some notations here; the constituent quark
mass is denoted by $\Mq$, the constituent diquark mass by $\Md$, the
baryon binding energy by $\epsilon_N$, the diquark mass difference by
$\epsilon_ d$, so that $\MN=3\Mq-\epsilon_N$ and
$\Md=2\Mq-\epsilon_d$, where $\epsilon_N>0$ and $\epsilon_d<0$
supposedly.  Then, we have an ordering pattern as $\MN < 3\Mq <
\frac{3}{2}\Md$.  We postulate $\epsilon_d<0$ because the diquark
shows up not as a bound state in configuration space but as a
correlation in momentum space.

If the diquark correlation is strong enough, there is a possibility of 
diquarks being localised bound states with $\epsilon_d>0$.  In this case
one might think that bound diquarks appear earlier than quarks, but
because of the gap, the quark energy dispersion relation is modified.
Then, there is no longer a sharp Fermi surface of quarks even at zero
temperature, and the onset of quarks is blurred.  We should consider
that the onset of quarks is thus determined by the onset of diquarks in
this case.  Besides, diquarks in QCD are coloured objects and strongly
interacting.  Therefore, it is not clear whether there is any sharp
onset of such interacting diquarks, which also underlies the idea of the
Quark-Hadron Continuity.

\subsection{A further push -- continuous duality}

The large-$\Nc$ (where $\Nc$ is the number of colours) approximation
for QCD matter provides us with interesting pictures that are more or
less consistent with the above. There may be no particular onset
behaviour of quarks and diquarks but these degrees of freedom may be
gradually developing as the baryon density grows up.  In other words,
in the large-$\Nc$ world, interacting nuclear matter and quark matter
share the same properties.  This is the state-of-the-art understanding
of dense QCD matter as Quarkyonic Matter recognised in
\cite{McLerran:2007qj}.  The baryons interact as strongly as
$\sim\mathcal{O}(\Nc)$ enhanced by a combinatorial factor due to quark 
exchanges, while mesons become free particles when $\Nc$ is infinitely
large.  Thus, interestingly, baryon interactions lead to the same
$\Nc$ counting as quark matter and so quark degrees of freedom should
be already there through baryon interactions.  The duality between
nuclear matter and quark matter is seemingly a
radical interpretation of the large-$\Nc$ observation, but we
emphasise that such a quarkyonic picture is very natural.  The baryon
interaction is mediated by mesons which are objects of a quark and an
anti-quark.  So, by this interaction, quarks can hop from one baryon to
the other, and this is nothing but a mobility characterisation of
deconfinement at large densities. The traditional picture in nuclear
physics is composed of baryons with meson exchanges, which may well be
regarded as partial deconfinement by quark exchanges, see soft 
deconfinement arguments in \cite{Fukushima:2020cmk}.

To implement such continuous duality in the field-theoretical
calculation, we should deal with hadrons, diquarks, and quarks in a
way without the double-counting problem, and also we should ensure the
correct boundary condition that degrees of freedom at ultraviolet (UV)
scale should be quarks and gluons only, which are the subjects we are
discussing in the following sections.

\section{Emergent mesons and diquarks}
\label{sec:Emmesons}

In the following two sections, \sec{sec:Emmesons} and \sec{sec:Emhadrons},
we discuss how hadrons emerge as dynamical degrees of freedom in the
low energy regime of QCD\@. One illuminating setting for emergent degrees
of freedom is the functional renormalisation group, fRG, approach, where the
theory is solved in an iterative procedure by reducing a flowing
energy or momentum scale from the initial perturbative UV energy
scales down to non-perturbative infrared (IR) energy scales. Within such a
procedure dynamical low energy degrees of freedom emerge as resonant
interaction channels. Hadrons, as well as diquarks can be identified
by the quantum numbers of the interaction channels under
investigation. The introduction of effective degrees of freedom for
the respective channel interactions turns out to be a very efficient
book-keeping device~\cite{Gies:2001nw,Gies:2002hq,Pawlowski:2005xe,Floerchinger:2009uf}
for general condensation phenomena. This procedure is called dynamical
hadronisation. As
indicated above, it accounts for the dynamical adjustment of the
physical degrees of freedom of the theory at the flowing energy scale $k$.

\subsection{QCD flows}

\begin{figure}
	\includegraphics[width=0.8\columnwidth]{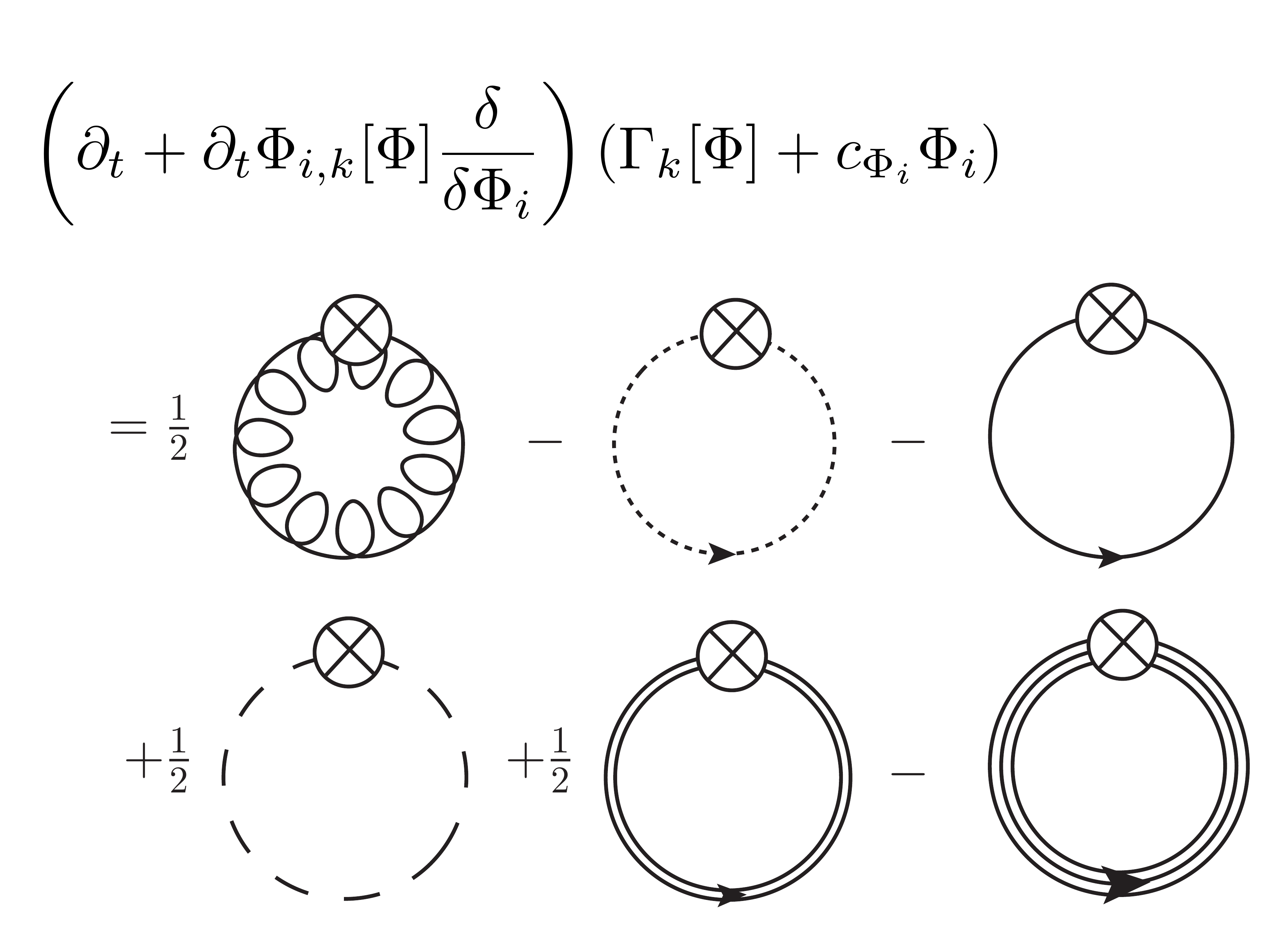}
	\caption{Flow equation for the effective action of QCD\@. The lines
		stand for the full propagators of gluons, ghost, quarks, and
		hadronic degrees of freedoms, the crossed circle stands for the
		cutoff insertion, see \Eq{eq:flow}. The loops
		encode the contributions of fluctuations of gluons, ghost, quarks,
		and hadrons introduced by dynamical hadronisation. \hspace*{\fill}}
	\label{fig:fRG-QCD_rebosonised}
\end{figure}

By now, the fRG approach to QCD is a well-established formalism for
computing correlation functions by integrating momentum fluctuations
out from a UV scale, $\Lambda \gg \LQCD$, down toward an IR scale.
See \cite{Pawlowski:2014aha, Mitter:2014wpa, Braun:2014ata,
  Rennecke:2015eba, Cyrol:2016tym, Cyrol:2017ewj, Cyrol:2017qkl,
  Corell:2018yil, Fu:2019hdw} for recent quantitative works in
Yang-Mills theory and QCD\@. The master flow equation for the effective
action of QCD including dynamical hadronisation is given by \cite{Fu:2019hdw}
\begin{align}\nonumber 
  &\left( \partial_t+\partial_t\Phi_{i,k}[\Phi]\0{\delta}{\delta \Phi_i}\right) \left( 
\Gamma_k[\Phi] + c_{\Phi_i}\Phi_i\right)  \\[2ex]\nonumber 
 = &\,\012 \Tr\,
  G_A[\Phi]\partial_t R_A -\Tr\, G_c[\Phi]\partial_t R_c-\Tr\,
  G_q[\Phi]\partial_t R_q \\[2ex]&\, +\012 \Tr\,
  G_\phi[\Phi]\partial_t R_\phi+\012 \Tr\,
  G_d[\Phi]\partial_t R_d-\Tr\, G_b[\Phi]\partial_t R_b\,, 
  \label{eq:flow} 
\end{align} 
where the effective action contains linear terms in $\Phi_i$. These terms generically occur in the presence of explicit chiral symmetry breaking. In their absence the effective action is that in 
the chiral limit, for a details discussion see  \cite{Fu:2019hdw}. Accordingly,  the combination $\Gamma_k[\Phi] + c_{\Phi_i}\Phi_i$ does not contain terms linear in the $\Phi_i$, and the flow of the theory in the presence of explicit chiral symmetry breaking is that of the chirally symmetric one in the chiral limit. The flow \eq{eq:flow} is schematically depicted in \Fig{fig:fRG-QCD_rebosonised},  see
\cite{Pawlowski:2014aha, Fu:2019hdw, Dupuis:2020fhh} for an overview. 

The $t$-derivative on the right hand side where $\partial_t=k\partial_k$
is taken at fixed field
$\Phi$ which encodes all fields, $\Phi=(A\,,\,c\,,\,\bar
c\,,\,q\,,\,\bar q\,,\,\sigma\,,\,\vec \pi\,,\,d\,,\,d^*\,,\, \bar b\,,
b,\ldots)$. The field derivative term on the left hand side relates to
dynamical hadronisation, and its occurrence is detailed in
\sec{sec:dynhadmesdi}. The standard flow equation for QCD is
obtained, when dropping the dynamical rebosonisation terms that come
with the introduction of effective low energy fields. Then, QCD is
solely described in terms of gluons, ghosts and quarks. 
Then, on the left hand side of \Eq{eq:flow} only the
$\partial_t \Gamma_k$ term is present, and only the loops of the
fundamental degrees of freedom survive on the right hand side: the
third line in \Eq{eq:flow} disappears. In
\Fig{fig:fRG-QCD_rebosonised} this simply amounts to dropping the
third line.

Flow equations for correlation functions are then obtained by taking
derivatives with respect to the fields. \Eq{fig:fRG-QCD_rebosonised}
is a one-loop exact differential equation, where each loop encodes the
fluctuation effect of gluons, ghosts, quarks, and also that of
collective degrees of freedom with hadronic quantum numbers.
The propagators in the loops are taken in general
background fields $\Phi$, and the loops are peaked at momenta $p$
about a sliding momentum scale $p^2 \approx k^2$ due to the cutoff
insertion $\partial_t R_\Phi(p)$. The second line encodes the
off-shell fluctuations of the fundamental degrees of freedom, gluons,
ghosts and quarks. The third line encodes that of hadronic degrees of
freedom, in the present case the lightest scalar and pseudoscalar
mesons, i.e., $\phi=(\sigma,\vec \pi)$, and the lightest baryons, i.e., the
nucleons $b$. Finally we also include collective modes of the diquarks
$d$ in the present case.

We emphasise that the hadronic and collective mode contributions in
\Fig{fig:fRG-QCD_rebosonised} do not signal an effective model
approach but rather stand for an efficient book-keeping of resonant
channels of multi-quark interactions. This procedure, called dynamical
hadronisation or rebosonisation~\cite{Gies:2001nw,Gies:2002hq,Pawlowski:2005xe,Floerchinger:2009uf},
has been extended further in the case of mesons in
\cite{Mitter:2014wpa,Braun:2014ata}
(see a review~\cite{Boettcher:2012cm} and references therein for
application to molecules and trimers).
In the present work we generalise the approach to diquark and baryon
channels in four-quark and six-quark interactions, respectively.

The local composite operators in this work, i.e.,
$\Phi_i =\sigma,\, \vec\pi,\, d,\, b$ carry the quantum numbers of
mesons, diquarks, and baryons, respectively. They are introduced via
one-meson, one-diquark, and one-baryon exchanges, taking care of
specific momentum channels (with the coupling
$\lambda_{\Phi_i}$) of four-quark and six-quark scattering
vertices. Due to the single-particle exchange relation amplitudes in
these channels read $|\lambdasd|\propto \hsd^2/m_{\phi/d}^2$ for
mesons and diquarks, and $|\lambda_{b}| \propto \hqdb^2/m_{b}^2$ for
baryons at vanishing exchange momentum. This is detailed in
\sec{sec:effectiveDOFS}. Here $m_{\phi/d/b}$ is the mass gap of
the meson/diquark/baryon dispersion, and $\hsd$ or $\hqdb$ is
the Yukawa coupling between a $\bar q q$ (meson), a $q q$ (diquark),
or a $qd$ (baryon) and the respective composite fields.  Hence,
decreasing multi-quark scattering amplitudes in the UV region
simply implies that corresponding states acquire mass-type dispersions
with increasing masses (at fixed Yukawa coupling) at large momentum
scales. Note in this context that also chiral channels can acquire a
mass gap in the above sense.

We expect that within the present approach we smoothly flow from high
energy QCD with quarks and gluons toward low energy QCD with dynamical
quarks and hadrons.  The low energy degrees of freedom get potentially
dynamical at about the chiral symmetry breaking scale,
$k_\chi\approx 0.5-0.4\GeV$.  Evidently, for $k < k_\chi$, only the
lowest-lying mesons of $(\sigma,\vec\pi)$ can contribute to the
loops.  Hence, we expect semi-quantitative results already with an
approximation with dynamical gluons, quarks, and $(\sigma,\vec\pi)$ at
least at vanishing chemical potential.  In turn, asymptotic states
have to include higher meson multiplets and baryons.  Nevertheless,
diquarks may play a role in the baryon formation and should be taken
into account especially at finite density, though they are not
asymptotic states as not being colourless.

\subsection{The effective action at large momentum scales}\label{sec:UVflows}

At a large flowing scale $k\gg\LQCD$ the effective action $\Gamma_k$
is well-described by the classical (finite) action of QCD as
\begin{align}
  \Gamma_{k\gg \LQCD} \simeq &\int_x\;\biggl\{
   \frac{1}{4}F^a_{\mu \nu} F^a_{\mu \nu} 
   + \bar{q} \Bigl( \slashed{D} + m_q \Bigr) q \notag\\ 
  & \qquad\quad +
  \bar c^a \partial_\mu D^{ab}_\mu c^b
  + \frac{1}{2\xi} (\partial_\mu A_\mu^a)^2 \biggr\}\,,
  \label{eq:Sqcd}
\end{align} 
with covariant derivative $D_\mu=\partial_\mu -igA_\mu^a t^a$, and
$\int_x = \int d^4 x$. Here we have normalised the fields and
couplings such that wave-function renormalisations and coupling
renormalisations can be taken to be unity at the given scale. In the
present approach the UV effective action also features a gluon mass
term due to the cutoff term that leads to modified Slavnov-Taylor
identities. This mass term has to vanish for vanishing cutoff scale,
and we suppress it in the following qualitative discussion (for more
details see \cite{Cyrol:2016tym} and references therein). The quark
mass $m_q$ is the running current quark mass in the
present fRG scheme.

\begin{figure}
 \includegraphics[width=0.7\columnwidth]{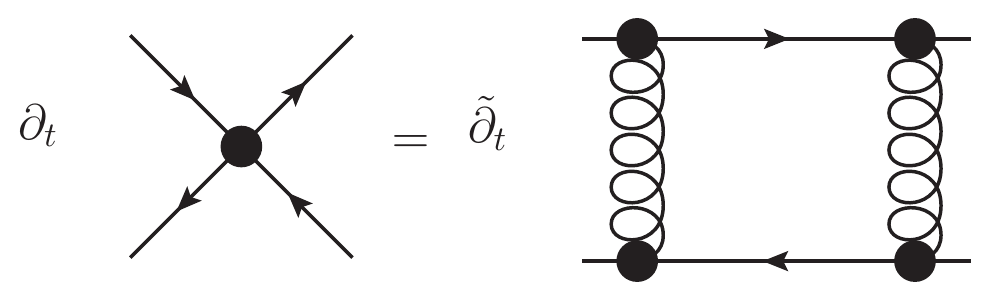}
 \caption{Generation of four-quark couplings in the first RG-step via 
 the gluonic box diagram (permutations, signs, and diagram
   multiplicities are not shown). The derivative
   $\tilde \partial_t$ only hits the explicit $t$-dependence in the
   regulators of the involved propagators.  \hspace*{\fill}}
 \label{fig:rebos_step1}
\end{figure}

Inserting the UV-effective action~\eq{eq:Sqcd} into the right hand
side of \Eq{eq:flow} generates all-order scattering terms within one
RG-step. The flow of the corresponding vertices,
\begin{align}
\Gamma^{(n)}_{\Phi_1\cdots\Phi_n,k}(p_1,\ldots,p_{n-1}) = 
\int_{p_n}\0{\delta^n\Gamma_k[\Phi]}{\delta\Phi_1(p_1)
    \cdots\delta\Phi_n(p_n)}\,,
\label{eq:GaN}\end{align}
can be obtained by taking the derivatives of \Eq{eq:flow} with respect
to the fields, $\Phi_1,\dots,\Phi_n$. The integral $\int_{p_n}= \int d^4
p_n/(2 \pi)^4$ on the right hand side of \Eq{eq:GaN} simply eliminates
$(2 \pi)^4 \delta(p_1+\cdots +p_n)$ that signals momentum
conservation, and $p_n=-(p_1 +\cdots +p_{n-1})$ is understood on the
left hand side. Now we use this structure for the four-quark
interactions. A flow step towards lower cutoff scales, $k\to k-\Delta
k$, immediately generates momentum-dependent four-quark interaction terms as
depicted in \Fig{fig:rebos_step1}. The diagram on the right hand side
of \Fig{fig:rebos_step1} is the only contribution to the four-quark
flow if we insert the UV-effective action \eq{eq:Sqcd} on the right
hand side of \Eq{eq:flow}. The four-quark term on the left hand side
of \Fig{fig:rebos_step1} includes all possible four-quark tensor
structures that are not forbidden by the symmetry properties of the
flow as well as the symmetry properties of the initial effective
action \eq{eq:Sqcd}. Hence, these terms have to be added to the
effective action \eq{eq:Sqcd}. Concentrating only on purely fermionic
terms this reads very schematically
\begin{align}\nonumber 
  \Gamma_{k} \simeq &\,\Gamma_{k\gg \LQCD} +\Gamma_{\bar qq \bar q
    q,k}+O(1/k^3) \,,\\[2ex] 
\Gamma_{\bar q q \bar q q,k} [\bar q, q]
  =&\,\014 \int_{p_1,p_2,p_3} \bar q_{p_1} q_{p_2} \bar q_{p_3} \bar
  q_{p_4} \Gamma^{(4)}_{\bar q q\bar q q,k}(p_1,\ldots,p_3)\,.
\label{eq:4fermi}\end{align}
In \Eq{eq:4fermi} the internal spinor and group indices are suppressed
and momentum conservation $p_4 = -(p_1+p_2+p_3)$ is implied. The
vertex function $\Gamma^{(4)}_{\bar q q\bar q q}(p_1,\ldots,p_3)$ carries
all tensor structures allowed by symmetries; for $N_f=2$ the complete
Fierz basis includes 10 tensor structures (at vanishing
momentum). Here we used a common basis also used in
\cite{Mitter:2014wpa}.  For more details see
Appendix~\ref{app:4f}.  Note that at large cutoff scales the four-quark
terms are irrelevant.  For dimensional reasons they run with $1/k^2$. 
Moreover,
$\Gamma^{(4)}_{\bar q q\bar q q,k}(p_1,\ldots,p_3)\propto \alpha_{s,k}^2$
with $\alpha_{s,k}=g_k^2/(4 \pi)$, which follows straightforwardly
from the flow in \Fig{fig:rebos_step1}. In the same way higher fermionic
interaction terms are generated as well as general quark-gluon
interaction terms. At large momentum scales all these terms are even
more suppressed by powers of the cutoff scale and the running strong
coupling $\alpha_{s,k}$ due to perturbative power-counting.

\begin{figure}
 \includegraphics[width=\columnwidth]{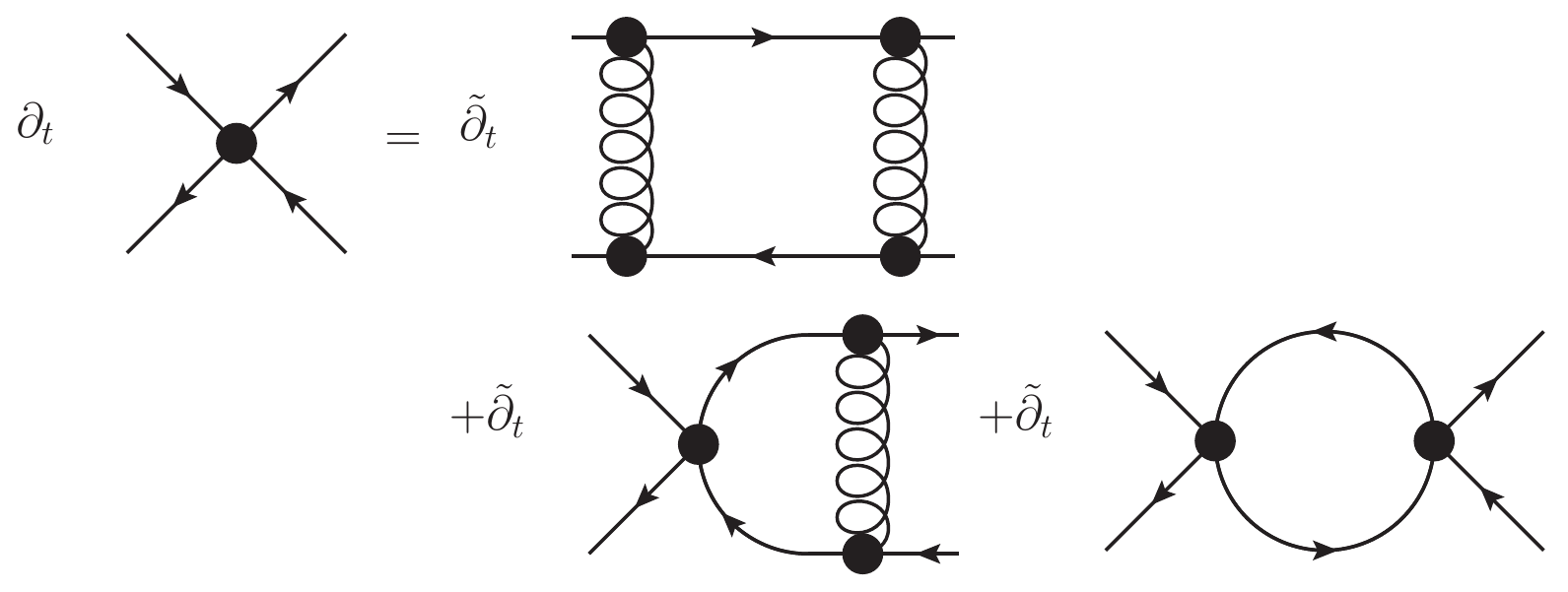}
 \caption{Schematic representation of the flow equation for the four-quark
 coupling involving just QCD degrees of freedom (permutations, signs, and diagram
   multiplicities are not shown). The derivative
   $\tilde \partial_t$ only hits the explicit $t$-dependence in the
   regulators of the involved propagators. \hspace*{\fill}}
 \label{fig:rebos_step2}
\end{figure}

For now let us concentrate on the four-quark terms, and drop the other
perturbatively irrelevant terms for the current qualitative
argument. Furthermore we go to the chiral limit. The four-quark
couplings generate further diagrams in their flow equation, as well as
in the flows of other couplings. In the approximation indicated in
\Eq{eq:4fermi} the full flow of
$\Gamma^{(4)}_{\bar q q\bar q q,k}(p_1,\ldots,p_3)$ is now given by
terms as shown in \Fig{fig:rebos_step2}. Again the
perturbative ordering is extracted straightforwardly: the last two
diagrams with two and one four-quark coupling are proportional to
$\alpha_{s,k}^4$ and $\alpha_{s,k}^3$, and hence are suppressed by
additional powers of the coupling in comparison to the first quark-gluon
diagram.

\subsection{Resonant interactions and low energy degrees of
  freedom}\label{sec:effectiveDOFS}

Here we discuss dynamically emergent mesons and emergent diquarks.

\subsubsection{Emergent scalar-pseudoscalar mesons}

\begin{figure}
	\includegraphics[width=\columnwidth]{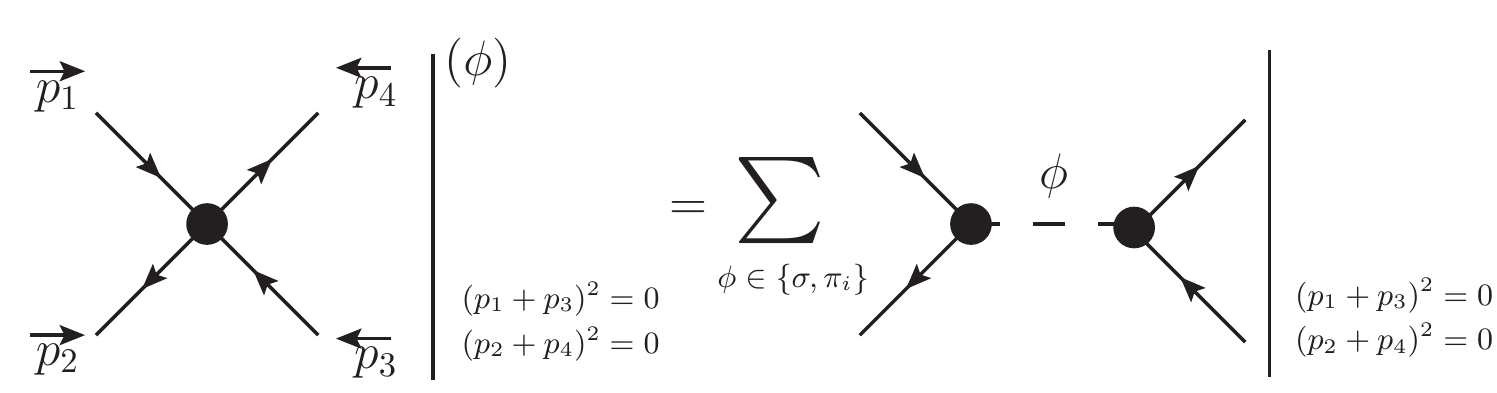}
	\caption{Rewriting the scalar-pseudoscalar momentum-dependent
		four-quark interaction in the
		diquark-free-channel as an effective meson
		exchange. \hspace*{\fill}}
	\label{fig:rebos_definition1}
\end{figure}

When the cutoff scale approaches the hadronisation scale
$k_{\text{\tiny{had}}}\approx 0.5\GeV$, the four-quark interaction
gets resonant at least for the scalar-pseudoscalar tensor structure
with the pion ($\vec \pi$) and sigma ($\sigma$) quantum numbers. 
Within a Fierz-complete basis given in Appendix~\ref{app:4f} the
scalar-pseudoscalar tensor structure 
can be read off from
the Lagrangian which is for $N_f=2$ given by
\begin{align}\label{eq:meson}
  \mathcal{L}_{(\bar q q)^2}^{(\phi)}
  &=-\frac{\lambdas}{2}\Zq^2
    \left[(\bar q q)^2-(\bar q \gamma^5
    \vec \tau q)^2\right] \nonumber\\[1ex]
  &=-\frac{\lambda_s}{2}\Zq^2\, \bar q^{\boldabar_1} q^{\bolda_1}
    \bar q^{\boldabar_2} q^{\bolda_2} \,
    T_{\rm s}^{\boldabar_1 \bolda_1 \boldabar_2 \bolda_2} \,,
\end{align}
where bold indices, $\bolda_i$ and $\boldabar_i$, run over Dirac,
colour and flavour indices. For the embedding of 
$T_{\rm s}^{\boldabar_1 \bolda_1 \boldabar_2 \bolda_2}$ in the
complete Fierz basis used here see also \Eq{eq:sigmapiT}.  Now we
restrict ourselves to a subspace of full momentum space that is
parametrised by only two momentum variables. A distinguished choice
is a channel with vanishing momentum of the diquark,
where the sum of both (anti)quark
momenta vanishes, see \Fig{fig:rebos_definition1}.
This momentum condition is given by
$(p_1+p_3)^2=(p_2+p_4)^2=0$ or $p_1=q=-p_3$ and
$p_2=-q+p=-p_4$, where $p$ is the exchange momentum. This can be
interpreted as projecting the scalar-pseudoscalar tensor structure on
collective $ q \bar q$ --mesonic-- degrees of freedom.
This suggests to rewrite the full scalar-pseudoscalar four-quark
interaction
for vanishing momentum in the diquark channel
into $\Gamma_{\bar q q\bar q q,k}^{(4),(\phi)}$ in terms of an exchange of
effective $SU(2)$-flavour field $\phi=(\sigma,\vec \pi)$ with the
$\sigma$ and $\vec \pi$ quantum numbers.
\Fig{fig:rebos_definition1} indicates,
\begin{align} \nonumber
  & \Gamma^{(4),(\phi)}_{\bar q q\bar q q,k}(q,-q+p,-q) \\[2ex]
  & = \sum_{\phi=\sigma,\vec \pi} \Gamma^{(3)}_{ q\bar q
    \phi,k}(q,p-q) \,G_{\phi,k}(p)\, \Gamma^{(3)}_{ q\bar q
    \phi,k}(-q,q-p)\,,
\label{eq:pion4}\end{align}
with  $G_{\phi,k}(p)$  being  the  propagator,  $G_{\sigma,k}(p)$  and
$G_{\vec \pi,k}(p)$ of  the effective $\sigma$ and  $\vec \pi$ fields,
respectively.

\Eq{eq:pion4} introduces a complete
bosonisation of the respective four-quark interaction channel in terms
of a one-meson exchange at all cutoff scale $k$. In the present
approach also a partial bosonisation of this channel is possible. Such
a partial bosonisation can be implemented by enforcing
\Eq{eq:pion4} only at one cutoff scale $k\neq 0$, or adding some
specific function of $p,q$ to \Eq{eq:pion4}. In the present work
we will only consider complete bosonisations for the sake of
simplicity.  Partial bosonisation may have advantages in the presence
of competing order effects such as colour superconductivity.

Even though we have introduced $\Gamma^{(3)}_{ q\bar q \phi,k}$ and
$G_{\phi,k}$ as mere parametrisations of the four-quark interaction
$\Gamma^{(4),(\pi)}_{\bar q q\bar q q,k}$ for
vanishing diquark momentum,
this set-up can be elevated to a consistent set-up in terms of effective meson fields
in full QCD with the identity
\begin{align}\label{eq:effQCD}
  \Gamma_{\text{\tiny{QCD}},k}[A,c,\bar c, q,\bar q]
  = \Gamma_{\text{\tiny{QCD}},k}^{(\phi)}[A,c,\bar c, q,\bar q,\phi_{\text{\tiny{EoM}}}]\,.
\end{align}
\Eq{eq:effQCD} entails that the full effective action of QCD with the
fundamental fields is given as the full effective action of QCD,
$\Gamma_{\text{\tiny{QCD}},k}^{(\phi)}[A,c,\bar c, q,\bar q,\phi]$,
evaluated on the solution, $\phi_{\text{\tiny EoM}}$, of the equations
of motion of the effective mesonic fields $\phi$.
\Eq{eq:effQCD} holds both for complete and partial
bosonisations. Formally, the introduction of the local composite
fields, $\sigma,\vec \pi,d, b$, can be done by introducing
corresponding source terms to the generating functionals in the sense
of density functional theory (for respective fRG work see \cite{Schwenk:2004hm, Kemler:2013yka, Liang:2017whg, Yokota:2018wue, Yokota:2020sfd}) or two-particle point Irreducible (2PPI) and 3PPI effective action (see \cite{Pawlowski:2005xe} for more details). Related fRG for $n$-particle irreducible (nPI) actions can be found in e.g.\ \cite{Wetterich:2002ky, Pawlowski:2005xe, Dupuis:2005ij, Blaizot:2010zx, Carrington:2012ea, Dupuis:2013vda, Carrington:2014lba,  Rentrop:2015tia, Carrington:2019fwp, Alexander:2019cgw,  Alexander:2019quf, Blaizot:2021ikl}. Note that in the standard dynamical hadronisation approach the sources are coupled to the full operators, instead of their connected or 1PI part in the nPI and nPPI approaches, for a discussion see \cite{Pawlowski:2005xe}.

In summary, in this framework the Bethe-Salpether-type equation
\eq{eq:pion4} holds on the equations of motion
$\phi_{\text{\tiny EoM}}$ of $\Gamma_k^{(\phi)}$ up to higher order
terms. Within the fRG approach it is possible to avoid higher order
terms beyond one loop in the dynamical hadronisation framework
explained in the next section. Within this framework we are led to the
one-loop exact flow equation \eq{eq:flow}. The only signals of the
dynamical hadronisation are the additional one-loop terms in the
effective fields as well as the $\partial_t\Phi_{i,k}$ term on the
left hand side, that carries the scale-dependent change of the
dynamical degrees of freedom.

Then, $\Gamma^{(4),(\phi)}_{\bar q q\bar q q,k}$ on the left hand side
of \Eq{eq:pion4} is understood as the fourth fermionic derivative of
$\Gamma_{\text{\tiny{QCD}},k}$, while the vertices on the right hand
side are that of
$\Gamma_{\text{\tiny{QCD}},k}^{(\phi)}$. Reparametrisations in terms
of effective degrees of freedom as indicated in \Eq{eq:pion4} and
\eq{eq:effQCD} with mesonic, diquark, and baryonic degrees of freedom,
do exist, with the problem not being its formal existence but rather its
practical consistent construction. The latter task concerns in
particular potential double-counting issues often present in LEFTs.
In the present form this is related to the
missing higher order terms in \Eq{eq:pion4} if derived from
$\Gamma_{\text{\tiny{QCD}},k}^{(\phi)}$ as well as the consistent
definition of the residual four-quark interaction terms. Within the
flow equation approach both tasks are solved systematically by the
dynamical hadronisation procedure explained in detail in
\sec{sec:dynhadmesdi}.

The non-trivial momentum dependence in these channels is then carried
by the Yukawa coupling $\Gamma^{(3)}_{ q\bar q \phi}$ of
quark--anti-quark to sigma and pion, and the sigma and pion
propagator.
\Eq{eq:pion4} with vanishing momentum in the diquark
channel allows us to determine
the Yukawa couplings $\Gamma^{(3)}_{ q\bar q \phi}(p,p-q)$ and the
propagators $G_{\phi,k}(p)$ up to positive momentum-dependent functions
$e^{2 \dynB_\phi(p)}>0$, that can be absorbed in propagators and vertices, 
\begin{align}
  \begin{split}
    \Gamma^{(3)}_{ q\bar q \phi}(p,p-q) \;\;\to\;\;
  &  e^{-\dynB_\phi(p)} \Gamma^{(3)}_{ q\bar q \phi}(p,p-q)\,, \\[2ex]
    G_\phi(p) \;\;\to\;\; & e^{2 \dynB_\phi(p)}G_\phi(p)\,,
  \end{split}
  \label{eq:rescale}
\end{align}
leaving \Eq{eq:pion4} invariant.  Such a rescaling can be used
either to arrange for a classical dispersion of the effective fields at
all scales, or to reduce the momentum- and cutoff-scale dependence of
the Yukawa vertices. Note that the exchange diagrams also give contributions
away from the
diquark-free-channel momentum configurations and we define the
$\sigma$-$\vec \pi$ part of the four-quark vertex,
$\Gamma^{(4),(\phi)}_{\bar q q\bar q q,k}$, as
\begin{align} \nonumber
  & \hspace{-0.2em}\Gamma^{(4),(\phi)}_{\bar q q\bar q q,k}(q,-q+p,-q')  \\[2ex]
  & \hspace{-0.2em}= \sum_{\phi=\sigma,\vec \pi} \Gamma^{(3)}_{ q\bar q \phi,k}(q,p-q) \,G_{\phi,k}(p)\,
  \Gamma^{(3)}_{ q\bar q \phi,k}(-q',q'-p)\,,
\label{eq:pion4full}
\end{align}
where $\Gamma^{(3)}_{ q\bar q \phi,k}$ and $G_{\phi,k}$ are now
fixed by \Eq{eq:pion4} together with a particular choice in
\Eq{eq:rescale}. Note that the difference from \Eq{eq:pion4} is the occurance of $q'$ in \eq{eq:pion4full}. In the current set-up higher scattering terms of the hadronised scalar-pseudoscalar interaction is represented in
terms of an effective potential $U_k(\phi)$. The flow of the latter is technically far more easily computed as the
corresponding purely fermionic vertex $\Gamma^{(2n),(\phi)}_{\text{\tiny{QCD}},\bar q q\cdots \bar q
  q}$. This is one of the reasons for the technical efficiency of the
dynamical hadronisation procedure described in the next
section~\ref{sec:dynhadmesdi}. It also carries the direct physical
interpretation of the (multi-)exchange of light mesonic degrees of
freedom, that allows for a direct connection to standard low energy
effective theories such as Chiral Perturbation Theory, and Quark-Meson
models. With \Eq{eq:pion4full} the residual four-quark vertex,
$\Gamma^{(4),\text{{res}}_\phi}_{\bar q q\bar q q,k}$, is defined with
\begin{figure}
  \includegraphics[width=.9\columnwidth]{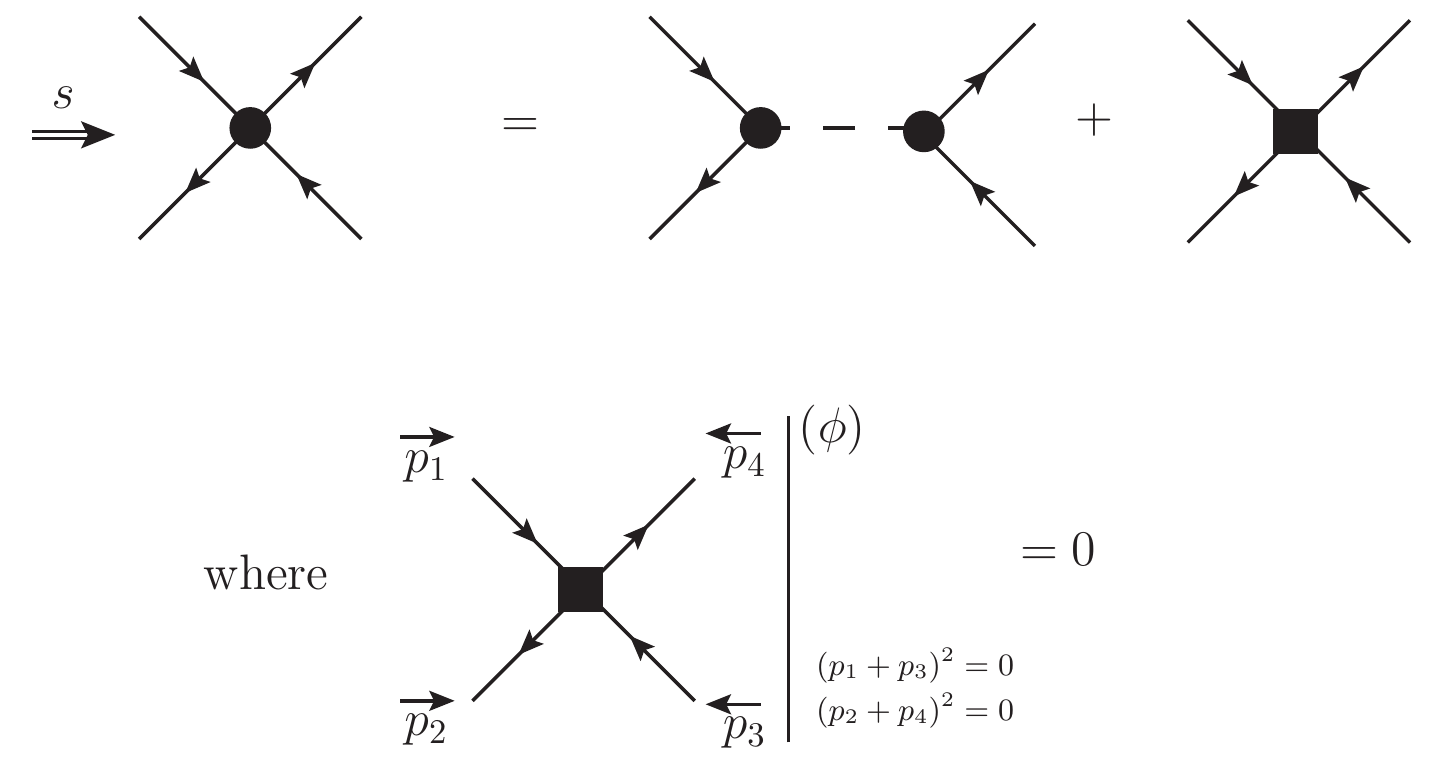}
  \caption{Rewriting the full four-quark interaction in the
    diquark-free-channel
    as an effective scalar-pseudoscalar meson exchange and a residual
    four-quark interaction (square), $\Gamma_{\bar q q\bar q
      q}^{(4),\text{{res}}_\phi}$ [see \Eq{eq:4fermi-ps}]. For better
    accessibility we also indicate the direction of the $s$-channel.  \hspace*{\fill}}
 \label{fig:rebos_definition2}
\end{figure}
\begin{align}\nonumber
 &\, \Gamma_{\bar q q\bar q q,k}(q,-q+p,-q')  \\[2ex] 
=& \, \Gamma^{(4),{\text{{res}}}_\phi}_{\bar q q\bar q
    q,k}(q,-q+p,-q') +\Gamma^{(4),(\phi)}_{\bar q q\bar q
    q,k}(q,-q+p,-q')
  \label{eq:4fermi-ps} \,, 
\end{align}
see also \Fig{fig:rebos_definition2}. From the derivation it is
evident that the scalar-pseudoscalar projection of the residual vertex
vanishes on the channel with vanishing diquark momentum,
\begin{align}\label{eq:nsps0}
  \left.  \Gamma^{(4),{\text{{res}}}_\phi}_{\bar q q\bar q
      q,k}(q,-q+p,-q)\right|^{(\phi)} \equiv 0\,.
\end{align}
\Eq{eq:4fermi-ps} and \eq{eq:nsps0} imply in particular that
the scalar-pseudoscalar part of the residual four-quark interaction
vanishes at vanishing momentum, $p_1=p_2=p_3=p_4 =0$. Hence the
one-meson exchange captures the leading four-quark term in a
derivative expansion that supposedly works at low energies. Note also
that $\Gamma^{(4),\text{{res}}_\phi}_{\bar q q\bar q q,k}$ still has a
scalar-pseudoscalar part. Firstly, the scalar-pseudoscalar $t$-channel
is still present.  Secondly, the full four-quark vertex carries more
momentum dependence as is captured by the $t,s,u$-channels. Still, the
channel where the diquark momentum vanishes is the dominant momentum structure, which suggests the
above reparametrisation. The identity \eq{eq:4fermi-ps} with
\Eqs{eq:pion4} and \eq{eq:nsps0} is depicted in
\Fig{fig:rebos_definition2}. We emphasise that the above formulation
avoids any double-counting problem present in effective field theories
and the effective mesonic field is merely a book-keeping device for
potentially resonant channels in scattering vertices.

\Eq{eq:4fermi-ps} defines a reparametrisation of the
effective action; on the level of momentum-independent approximations
for vertices and classical dispersions for all fields involved (also
the effective ones) this leads us to an effective action that also
involves the mesonic fields. This reparametrisation can be performed
at all scales $k$, thus removing the scalar and pseudoscalar diquark
channel in the four-quark interaction. This $k$-dependent procedure
can be systematically implemented, (see \cite{Gies:2001nw,
  Pawlowski:2005xe, Floerchinger:2009uf}), and is called dynamical
hadronisation in the context of QCD\@. It has been developed further
in \cite{Mitter:2014wpa, Braun:2014ata}. Here we extend it
explicitly to diquarks and baryons in Secs.~\ref{sec:dynhadmesdi} and
\ref{sec:dynhadb}.

Carrying out this procedure for the $\sigma$-$\vec \pi$ channel leads
to an effective Lagrangian with constituent quarks and light mesons as
effective degrees of freedom. For the sake of simplicity we consider only
momentum-independent couplings and classical dispersions. For $N_f=2$ 
this leads us to
\begin{align}\nonumber 
\Gamma^{\rm (qm)}_k & = \int_x \biggl\{ \Zq\, \bar{q}
   (  \slashed{D} -\muq\gamma^0 )q+ \frac{\Zphi}{2}\, 
   \partial_\mu \phi_i \partial^\mu \phi_i 
    \\[2ex]\nonumber 
  & \quad - \frac{\lambdas}{2}\,\Zq^2 
  \Bigl[ (\bar{q}q)^2+(\bar{q} i\gamma^5\vec\tau q)^2 \Bigr]\\[2ex]
  & \quad + \frac{\hs}{2}\, \Zq\Zphi^{1/2} \, \bar{q} ( \sigma
  + i \gamma^5 \vec\tau\cdot\vec\pi ) q+ U_k(\phi) \biggr\}\,.
\label{eq:qm}
\end{align}
The first line comprises the kinetic terms of the quarks and the
effective meson fields, where $\Zq, \Zphi$ are the wave-function
renormalisations of quarks and mesons, respectively. The Dirac operator $\slashed{D}(A)$ indicates 
a potentially non-trivial background gauge field at finite temperature. In the current
approximation they only carry a scale-dependence which is not
explicitly shown in \Eq{eq:qm} for notational simplicity. 
This also allows us to identify $\muq$ with the
RG-invariant quark chemical potential. The second line is the
scalar-pseudoscalar four-quark interaction that is generated from the
QCD flow in the chiral limit. We have also pulled out the appropriate
$\Zq^2$-factor such that the four-quark coupling $\lambdas$ is
RG-invariant but cutoff scale-dependent. The last line carries the
one-meson exchange interaction with the Yukawa coupling $\hs$. 
As for the four-quark coupling it is RG-invariant. Finally
we allow for a mesonic effective potential, which describes
multi-meson self-interactions. In a purely fermionic language this
relates to higher-order fermionic scatterings: $\phi^{(2n)}$ interaction terms 
relate to fermionic scatterings up to $(\bar q q)^{2n}$. Its precise form will
be discussed later. Here we simply note that it includes mass terms
for the mesons as
\begin{align}\label{eq:Ukmass}
  U_k({\phi^2}/{2}) = \012 \Zphi m_\phi^2 \phi^2 +O\left((\phi^2)^2
  \right)\,. 
\end{align}
In the regime with chiral symmetry breaking the expectation value of
the meson field is
$\phi_{\text{\tiny{EoM}}}^2=\sigma_{\text{\tiny{EoM}}}^2\neq 0$. Then
we have different masses, wave-function renormalisations, etc., for
$\sigma$ and $\vec \pi$.

The form~\eq{eq:qm} can be reduced with the help of dynamical
hadronisation. As described above, it allows us to rewrite certain
momentum channels partially or completely in terms of effective
mesonic interactions.  Residual four-quark interactions in other
channels have to be kept in principle.

At this point we would like to stress the necessity of including a
wave-function renormalisation factor $\Zphi$ corresponding to the
composite mesonic operator to distribute the original four-quark
coupling properly onto mass and Yukawa coupling. For the
approximation~\eq{eq:qm}, \Eq{eq:pion4} reduces to
\begin{align}
  \lambdas(p^2)=\frac{1}{2}\frac{\hs^2}{p^2+m_\phi^2}\,.  
  \label{eq:ident}
\end{align}
\Eq{eq:ident} is one of the basic equations in the present set-up and
we discuss it here in some detail: Let us also remark that in the full
theory both sides will be field-dependent. For example, if we only
consider a constant mesonic background $\phi$, we arrive at
\begin{align}
  \lambdas(\phi,p^2)=\frac{1}{2}\sum_{\phi_i=\sigma,\vec \pi}\frac{\left(
      \hs(\phi)+h_{q\phi\phi_i}^{(1)}(\phi)\phi_i\right)^2}{p^2+
    m_\phi^2(\phi)}\,,
  \label{eq:identfull}
\end{align} 
and $h_{q\phi\phi_i}^{(1)}=\partial_{\phi_i}\hs$. The field-dependence of the 
Yukawa couplings encodes multi-meson--$q\bar q$ scattering 
processes in the scalar-pseudoscalar channel. \Eq{eq:identfull}
is relevant for an evaluation on the equations of motion of the
mesons, $\phi_{\text{\tiny{EoM}}}=(\sigma_{\text{\tiny{EoM}}},\vec 0)$
with a vanishing $\vec\pi$ expectation value and a non-vanishing $\sigma$
one. Then, the derivative term vanishes for the $\vec \pi$-directions
and gives contributions in the $\sigma$-direction. Note also that the
constituent quark mass is $\Mq=\frac{1}{2}\hs\, \sigma$, and hence is proportional to
$\Gamma^{(3)}_{q\bar q\vec \pi}(\phi=(\sigma,\vec 0))$ while
$\Gamma^{(3)}_{q\bar q \sigma}(\phi=(\sigma,\vec 0))$ is
proportional to $ \hs(\sigma)+h_{q\phi\sigma}^{(1)}(\sigma)\,\sigma$. Moreover,
seemingly the constituent quark mass $\Mq=\frac{1}{2}\hs \sigma$
vanishes for $\sigma\to 0$. However, it has been shown in
\cite{Pawlowski:2014zaa} that in the full theory the Yukawa coupling
$\hs(\sigma)$ has a non-trivial $\sigma$-dependence leading to
$\Mq(\sigma)\geq \Mq(\sigma_{\text{\tiny{EoM}}})$ for all $\sigma$.

Moreover, seemingly \Eq{eq:ident} does not depend on $\Zphi$; however, the
flow of the coefficient of the mass term, $\Zphi m_\phi^2$, is that of
$\Gamma_{\phi\phi,k}^{(2)}(p=0)$. Then the flow of the mass $m_\phi^2$ is
obtained from
\begin{align}
  \partial_t m^2_\phi = \partial_t\Gamma_{\phi\phi,k}^{(2)}(p=0)
  +\eta_\phi m^2_\phi\,,\quad
  \eta_\phi=-\0{\partial_t \Zphi}{\Zphi}\,.
\end{align}
If we work within the approximation $\Zphi=1$, the term proportional
to $\eta_\phi$ disappears. However, the diagrammatic flow
$\partial_t\Gamma_{\phi\phi,k}^{(2)}$ is that of $\Zphi m_\phi^2$, and
still carries the running of $\Zphi$. Only by subtracting the
$\eta_\phi$-term do we project on the RG-invariant flow of $m_\phi$,
and we are led to an interesting observation: by definition $m_{\phi}$
is the Euclidean curvature mass (for a detailed discussion of the pole,
screening, and curvature masses, see \cite{Helmboldt:2014iya}). There it
has been shown that indeed the curvature and pole masses of the pion
are very close in fully-momentum dependent approximations in
\cite{Helmboldt:2014iya}, with the discrepancy of curvature and pole masses
being on the percent level. This is related to the fact that the pion
pole is the first pole in the complex plane and the momentum
dependence of the pion wave-function renormalisation is weak in the Euclidean
domain.

In the present approximation with scale-dependent, but
momentum-independent, wave-function renormalisation $Z_{\phi}$, the
pole mass of the pion equals the curvature mass on the level of the
effective action. Moreover, due to the weak momentum dependence its
value is already close (percent level) to the one computed within
fully momentum-dependent approximations, see
\cite{Helmboldt:2014iya}. For the reduced approximation with
$\eta_\phi=0$ the latter property does not hold. While pole and
curvature masses still agree on the level of the effective action,
their values are off by roughly 30\%.

\subsubsection{Emergent diquarks}
We now consider an equivalent procedure for diquarks. It is evident
from the discussion so far, that the effective exchange fields do not
need to be colourless as they only represent exchange channels in a
given four-quark scattering vertex. Hence, the introduction of these
composite fields does not imply their existence as asymptotic
states. Following our discussions on diquarks we are particularly
interested in four-quark interactions corresponding to a scalar
diquark channel. For $N_f=2$ this is a colour triplet and flavour
singlet channel. A simple form is given by the Lagrangian,
\begin{align}\label{eq:diquark}
  \mathcal{L}_{(\bar q q)^2}^{(d)}=-\frac{1}{2}\Zq^2\lambdad ( q^T \epsilon^a \tau^2 C
  \gamma^5 q) ( \bar q \epsilon^a \tau^2 C \gamma^5 \bar q^T) \,,
\end{align}
where $C=\gamma^2\gamma^0$ denotes the charge conjugation matrix,
$\{\epsilon^a\}_{bc}= \epsilon^{abc}$ and $\tau^2$ the second
Pauli matrix in flavour space. Note that the related tensor structure
$T^{(1)}_{\rm d}$ mixes with the scalar-pseudoscalar one, $T_{\rm s}$,
derived from \Eq{eq:meson} in the current Fierz basis in
Appendix~\ref{app:4f}. While this is not a conceptual problem, and it
may be even advisable to keep \Eq{eq:diquark} for explicit
computations, some of the following equations and discussions are more
complicated in the presence of a mixing. Orthogonality can be achieved
by one of the following procedures: we either build-up an orthogonal
Fierz basis with the first two basis elements,
$T_{\rm s}$ and $T^{(1)}_{\rm d}$. A further possibility for enforcing
orthogonality of the scalar-pseudoscalar tensor structure and the
diquark one is to stick to the Fierz basis in Appendix~\ref{app:4f}
and project $T^{(1)}_{\rm d}$ on the complement of $T_{\rm s}$
such that
\begin{figure}
	\includegraphics[width=.9\columnwidth]{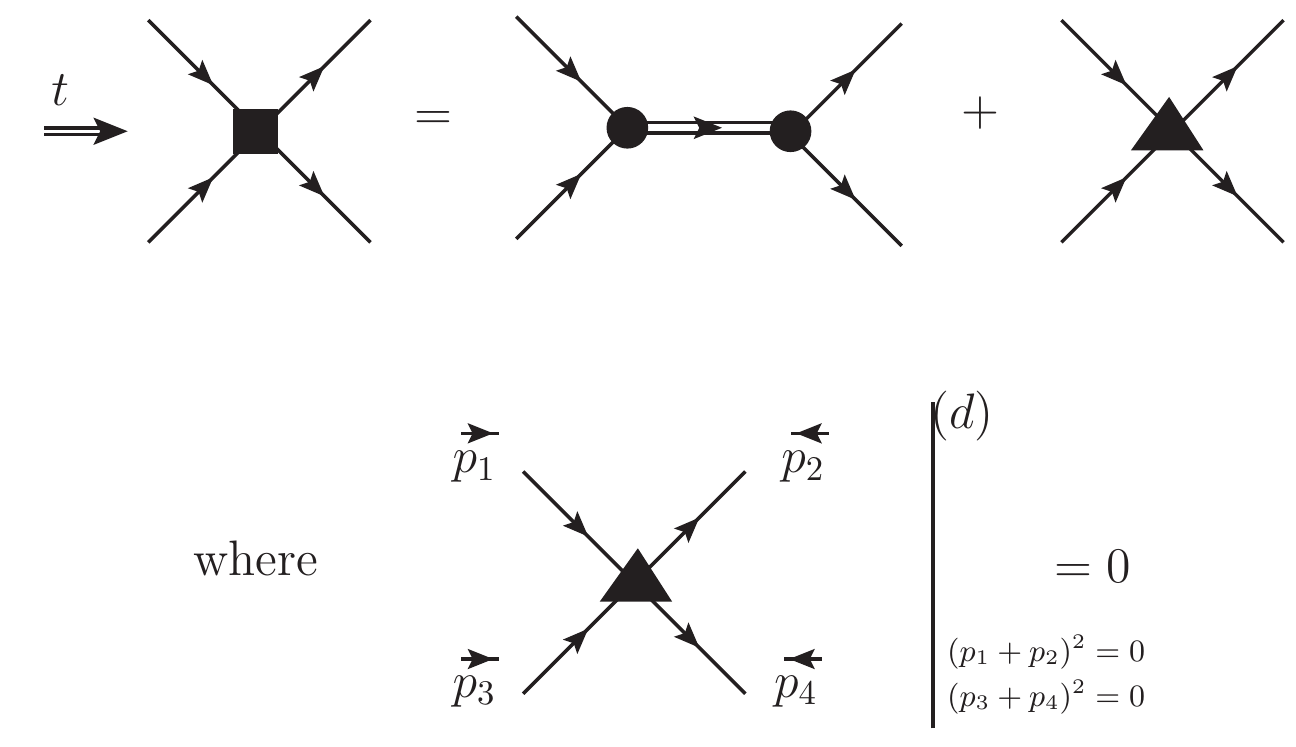}
	\caption{Rewriting the residual momentum-dependent four-quark
		interaction in the $s=0$ channel in \Eq{eq:4fermi-ps} (square) as
		an effective diquark exchange in the diquark tensor structure
		\eq{eq:diquark} and a further residual four-quark interaction
		(triangle). For better accessibility we also indicates the
		direction of the $t$-channel. \hspace*{\fill}}
	\label{fig:rebos_definition_diquark}
\end{figure}
%
\begin{align}
T_{\rm d}^{(2)}= (1- P_{\rm s})
  T^{(1)}_{\rm d} (1- P_{\rm s}) \,,
 \label{eq:dproject} 
\end{align}
where $P_{\rm s} $ is the projection on the scalar-pseudoscalar
subspace with $P_{\rm s} T_{\rm s} = T_{\rm s} P_{\rm s}= T_{\rm s}$
and $T_{\rm s} (1 - P_{\rm s})=0$. The corresponding Lagrangian is now
simply
\begin{align}\label{eq:diquark2}
  \mathcal{L}_{(\bar q q)^2}^{(d)}=-\frac{\lambdad}{2}\Zq^2 \,
  \bar q^{\boldabar_1} q^{\bolda_1} \bar q^{\boldabar_2} q^{\bolda_2}
  \,T_{\rm d}^{(2), \boldabar_1 \bolda_1 \boldabar_2 \bolda_2}\,.
\end{align}
For the rest of our formal analysis both procedures are equivalent,
and we shall refer to both possibilities as $T_{\rm d}$. Hence we
proceed analogously to \Eq{eq:pion4} and subtract an effective scalar
diquark exchange e.g.\ in a channel where $s=0$ from the residual
four-quark vertex defined in \Eq{eq:4fermi-ps}. This channel is given by
$p_2 = - p_1 = q$ and $p_4 = - p_3 = p-q$. The scalar-pseudoscalar
meson exchange does not contribute as the two tensor structures are
orthogonal in the chosen Fierz-complete basis. The
one-diquark--exchange equation reads
\begin{align}\nonumber
 & \Gamma^{(4),{\text{res}}_{\phi}}_{\bar q q\bar q q,k}(-q',q,q'-p)=
 \Gamma^{(4),{\text{{res}}}_{\phi d}}_{\bar q q\bar q
    q,k}(-q',q,q'-p) \\[2ex]
 &\quad +\Gamma^{(3)}_{q q d,k}(q,p-q,p) G_{d,k}(p) 
\Gamma^{(3)}_{\bar q \bar q d^*,k}(q',p-q',p) \,,
  \label{eq:4fermi-d}
\end{align}
with 
\begin{align}\label{eq:nd0}
  \left.  \Gamma^{(4),\text{{res}}_{\phi d}}_{\bar q q\bar q
      q,k}(-q,q,q-p)\right|^{(d)} \equiv 0\,.
\end{align}
For the pictorial representation, see
\Fig{fig:rebos_definition_diquark}.  As for the meson exchange the
above relation can be embedded in an effective action
$\Gamma_{\text{\tiny QCD},k}^{(\phi d)}$
which agrees with $\Gamma_{\text{\tiny QCD},k}$
on the equations of motion for $\phi$ and $d$,  
\begin{align}\label{eq:effQCDphid}
  \Gamma_{\text{\tiny{QCD}},k}[A,c,\bar c, q,\bar q]=
  \Gamma_{\text{\tiny{QCD}},k}^{(\phi d)}[A,c,\bar c, q,\bar
  q,\phi_{\text{\tiny{EoM}}},d_{\text{\tiny{EoM}}}]\,.
\end{align}
Then, \Eq{eq:4fermi-d} includes further higher order terms, as already
discussed below \Eq{eq:4fermi}. This will be fully resolved within dynamical
hadronisation in \sec{sec:dynhadmesdi}. Note also
that the orthogonal projection with $T_{\rm d}$ and $T_{\rm s}$ is already
paid off: if the two tensor structures would have an overlap the two
bosonisations as described here would not commute, and the results
would depend on the order. In turn, with the present orthogonal tensor
structures the two bosonisations commute and the order is irrelevant. 

The corresponding terms of the effective action leading to \eq{eq:qm},
\eq{eq:4fermi-d}, and \eq{eq:nd0} read for $T_{\rm d}^{(1)}$,
\begin{align}
\Gamma^{\rm (qd)}_k = &\,\int d^4x \biggl\{ \Zd
  (D_\mu-\mud\delta_{\mu 0}) d^a
  (D^\mu+\mud\delta^{\mu 0}) d^{a \ast} \notag\\[2ex]
 &\hspace{.8cm}-\frac{\lambdad}{2}\, \Zq^2\, \left( q^T \epsilon^a \tau^2 C \gamma^5 q\right)
\left(\bar q \epsilon^a \tau^2 C \gamma^5 \bar q^T\right)\notag\\[2ex]
 &\hspace{-1.1cm}+ \frac{\hd}{2\sqrt{2}}\, \Zq \Zd^{\frac12} \left[ d^{a\ast}\! \left(q^T \!\epsilon^a \tau^2
 C \gamma^5 q\right) \!+ \! d^a\!\left(\bar q\epsilon^a\tau^2 C \gamma^5 \bar q^T\right)\!
 \right] \!\biggr\},
\label{eq:qd}
\end{align}
where $d^a$ denotes the complex scalar diquark field with the diquark
chemical potential given by $\mud=2\muq$. As for the quark-meson
action we have pulled out the appropriate powers of the wave-function
renormalisations $\Zq,\Zd$ for having RG-invariant running couplings
$\lambdad, \hd$. The action defined in \Eq{eq:qd} does not contain an
explicit mass term for the diquark.  Instead, in \Eq{eq:qm}, we
include an effective potential of the form,
\begin{align}
 U_k = U^{\rm (md)}_k(\rhos,\rhod)\,,
\end{align}
as a function of two invariants, 
\begin{align}\label{eq:rhos}
\rhos:= \frac{1}{2} \sum_i\phi_i^2\,,\qquad 
\rhod:= \sum_a d^{a \ast}d^a\,.
\end{align} 
Linear terms in $\rhos$ and $\rhod$ in the potential
give rise to mass terms $m_\phi^2 \rhos$ and $m_d^2 \rhod$ in
the symmetric phase for mesons and diquarks, respectively.  Higher
order terms are interpreted as inter-meson, inter-diquark, and
meson-diquark interactions arising from mixed terms in the
potential. The corresponding action for $T_{\rm d}^{(2)}$ is just
obtained by the projection in \Eq{eq:dproject}. Again we emphasise
that the difference is not the projection but rather the different
Fierz bases.

\subsubsection{Emergent vector mesons}
In the same line as above we can introduce an effective field $\phi_\mu$ for
vector mesons
\begin{align}\label{eq:Vmu}
  \phi_\mu= \omega_\mu+V_\mu\,,\quad V_\mu = \rho_\mu^a \tau^a+a_{1\mu}^a
  \tau_5^a\,,
\end{align} 
where $\tfrac{1}{2}( \tau^a \pm \tau_5^a)$ are the generators of $SU(2)_{R/L}$
respectively. The vector field $\phi_\mu$ has the entries
$(\phi_\mu^i)= (\omega_\mu,\vec \rho_\mu,\vec {a_1}_\mu )$. In the
present context the field $V_\mu =(\rho_\mu,a_{1\mu})$ is added for the sake of
completeness and for later applications beyond the scope of the
present paper. The four-quark tensor structure corresponding to the $\omega_\mu$
is derived straightforwardly from the Fierz basis in Appendix~\ref{app:4f}: it
is given by the symmetric combination of $ \mathcal{L}^{(V-A)}_{(\bar
  q q)^2}+ \mathcal{L}^{(V+A)}_{(\bar q q)^2}$ in
\Eq{eq:fourfermi_sym}, and we dropped the axial-vector tensor structure 
$ \mathcal{L}^{(V-A)}_{(\bar
  q q)^2}-\mathcal{L}^{(V+A)}_{(\bar q q)^2}$. In turn, the $\rho,
a_1$-tensor structures are more complicated in terms of the basis
\eq{eq:fourfermi_sym}, and we refrain from showing them explicitly.

The different components of the vector field $\phi_\mu$ are relevant
for an appropriate description of the onset of nuclear matter via
$\omega$, as well as the access to $\rho$-physics via $V_\mu$. The
latter has been treated within dynamical hadronisation in
\cite{Rennecke:2015eba}. The quark-vector meson part of the action
reads in analogy to \Eq{eq:qm}, \eq{eq:qd},
\begin{align}\nonumber 
 & \Gamma^{\rm (qV)}_k = \int_x \Biggl\{ \frac{\Zo}{4}\, (\partial_\mu
  \omega_\nu -\partial_\nu \omega_\mu)^2+\frac{\ZV}{4}\, (\partial_\mu
  V^i_\nu -\partial_\nu V^i_\mu)^2 \\[2ex]\nonumber 
&\quad - \frac{\lambdao}{2}\,\Zq^2
 \left(\bar{q}\gamma^\mu q\right)^2-
  \frac{\lambdaV}{4}\,\Zq^2 \Bigl[ \left(\bar{q}\gamma^\mu \vec\tau q\right)^2+
  (\bar{q} \gamma^\mu \gamma^5 \vec \tau  q)^2\Bigr]\\[2ex]\nonumber 
&\quad + \frac{\ho}{2}\,\
  \Zq\Zo^{1/2} \, \bar{q}\gamma^\mu \omega^\mu q \\[2ex]
  & \quad + \frac{\hV}{2}\,\ \Zq\ZV^{1/2} \, \bar{q} \left( \gamma^\mu
  \vec\tau\cdot\vec\rho^\mu + \gamma^\mu\gamma^5 \vec \tau\cdot \vec
  a_1^{\mu} \right) q \Biggr\}\,,
\label{eq:qV}
\end{align}
where the potential $U_k$ in \Eq{eq:qm} now also depends on the vector
mesons $\phi_\mu$, i.e., $U_k=U_k(\phi,\phi_\mu, d)$. Most notably, in
the regime with vector meson dominance (VMD) we have
\begin{align}
 \014  (\partial_\mu V^i_\nu -\partial_\nu
V^i_\mu)^2+U_k(\phi,V_\mu,d) = \012 \tr \,F_{\mu\nu}(V_\mu)^2+\cdots
\end{align}
with the non-Abelian field strength $F_{\mu\nu}(V_\mu)$. In the
present dynamical hadronisation approach to first principle QCD,
vector meson dominance is not assumed. It is or is not the outcome of
the dynamical computation, for a first work in this direction, see
\cite{Rennecke:2015eba}.

\subsection{Dynamical hadronisation for mesons and
  diquarks}
\label{sec:dynhadmesdi}

Dynamical hadronisation has been invented under the name of
rebosonisation for bilinear composite operators in
\cite{Gies:2001nw}. In \cite{Pawlowski:2005xe} it has been
extended to general composite operators, for further work see
\cite{Floerchinger:2009uf}. In the present work we build on the
formulation \cite{Braun:2014ata, Mitter:2014wpa,
  Cyrol:2017ewj,Fu:2019hdw} for QCD that also takes into account the
full momentum dependence of vertices.

\subsubsection{Dynamical hadronisation for scalar-pseudoscalar
  mesons}\label{sec:dynhadspsmes}
Let us first describe the essential idea behind dynamical
hadronisation at the standard QCD example of the light
scalar-pseudoscalar mesons $\phi=(\sigma,\vec\pi)$ along the
discussion in \cite{Braun:2014ata,Mitter:2014wpa, Cyrol:2017ewj,
  Fu:2019hdw}: first we note that the fields $\sigma,\vec\pi $ have
the quantum numbers of $\bar{q}q\,,\, \bar{q}i \gamma_5 \vec\tau q$
states. The simplest identification would be
$\phi \propto \dynA_{\bar q q,k} \, (\bar q q\,,\, \bar{q} i \gamma_5
\vec\tau q)$. In the last section we have introduced $\phi$ as an
effective field for a one-meson exchange that describes the related
resonant scalar-pseudoscalar
channel with vanishing diquark momentum in the full
four-quark scattering vertex $\Gamma^{(4)}_{\bar q q\bar q q}$, see
\Eq{eq:pion4full}. It is clear from the discussion there, that the
prefactor $\dynA_{\bar q q,k}$ is related to the interaction strength
of the related four-quark interaction. Moreover, a rescaling as
indicated in \Eq{eq:rescale} also changes the normalisation of the
fields with $\phi\to {\dynB}_\phi \phi$. This leads to the notion of a
$k$-dependent change of the composite field $\phi_k$, which we
parametrise as
\begin{align}
 \partial_t \begin{pmatrix} \sigma_k \\ \vec{\pi}_k \end{pmatrix}\!(r)
  = \dot\dynA_{\bar q q}(r) \0{\Zq}{\Zphi^{\frac12}}\!\begin{pmatrix} \bar{q}q \\
  \bar{q}i\gamma^5 \vec{\tau}q \end{pmatrix}\!(r) + \dot\dynB_\phi(r)\!
  \begin{pmatrix} \sigma_k \\ \vec{\pi}_k \end{pmatrix}\!(r) \,,
\label{eq:scaldynhad}
\end{align}
where $(\bar{q}q)(r) = \int_l \bar q(l) q(r-l)$, and
$(\bar{q}i\gamma^5 \vec{\tau}q)(r)$ is defined similarly. The field
$(\Zq/\Zphi^{1/2}) (\bar q q\,,\, \bar{q}i\gamma^5 \vec{\tau}q)$ has
the same RG-running as the mesonic field, and hence the coefficients
of the dynamical hadronisation, $\dot\dynA_{\bar q q},\dot \dynB_\phi$, are
RG-invariant. \Eq{eq:scaldynhad} is the reduction of the fully general
dynamical hadronisation with local operators in the $t=0$ channel with
$\dot \dynA_{\bar q q}(l,r-l)$ to the plain $u$-channel discussed in
\cite{Mitter:2014wpa}. The general case will be considered
elsewhere. The coefficients $\dot{\dynA}_{\bar q q} = \partial_t \dynA_{\bar q q}$ and
$\dot{\dynB}_\phi= \partial_t \dynB_\phi$ may have a matrix structure in the chiral
broken phase where the degeneracy with respect to $\sigma_k$ and
$\vec{\pi}_k$ is lost. As the flow equation for the effective action
was derived for $k$-independent fields, the left hand side of the flow
equation now changes to
\begin{align}\label{eq:rebos} 
  \partial_t \Gamma_{\text{\tiny{QCD}},k}\to \partial_t
  \Gamma_{\text{\tiny{QCD}},k}^{(\phi)}[\Phi] +\int_x \partial_t \Phi_{i,k}[\Phi]
  \0{\delta\Gamma_{\text{\tiny{QCD}},k}^{(\phi)}[\Phi] }{\delta\Phi_{i}}\,,
\end{align}
where the $t$-derivatives are always taken at fixed fields. This
already explains the dynamical hadronisation term on the left hand
side of \Eq{eq:flow}. We also emphasise that the argument of the
effective action, the mean field $\Phi$, is $t$-independent. 

Let us discuss the related modification of the flows at the example of
the full mesonic two-point function,
\begin{align}\label{eq:G2phi} 
\Gamma^{(2)}_{\phi,k}(p)=\Zphi(p)\left( p^2 +m^2_\phi\right)\,,
\end{align}
with a momentum-dependent wave-function renormalisation $\Zphi(p)$
and a momentum-independent, but cutoff scale-dependent mass
$m_\phi^2$.  Note that the parametrisation in \Eq{eq:G2phi} is not
unique, we may even absorb the mass $m_\phi^2$ completely in the wave
function renormalisation: $\Zphi\to \Zphi\, p^2/(p^2
+m_\phi^2)$. The rescaling property of $\dynB_{\phi,k}$ becomes evident from the
form of the flow equation for the mesonic two-point function at vanishing fields, 
\begin{align}\nonumber 
  & \0{1}{\Gamma^{(2)}_{\phi,k}(p)}\left( \partial_t +2 \dot \dynB_\phi(p)\right)
  \Gamma^{(2)}_{\phi,k}(p) \\[2ex]&= \left( 2 \dot \dynB_{\phi}(p)
    -\eta_\phi(p)\right)+\0{\partial_t m_\phi^2}{p^2+m_\phi^2} =
  \0{1}{\Gamma^{(2)}_{\phi,k}(p)} {\rm Flow}^{(2)}_{\phi}(p)\,,
\label{eq:dotG2phi}
\end{align}
with the anomalous dimension 
\begin{align} \label{eq:etaphi} 
\eta_\phi(p)=-\0{\partial_t \Zphi(p)}{\Zphi(p)}\,.
\end{align}
%
\begin{figure}
 \includegraphics[width=.98\columnwidth]{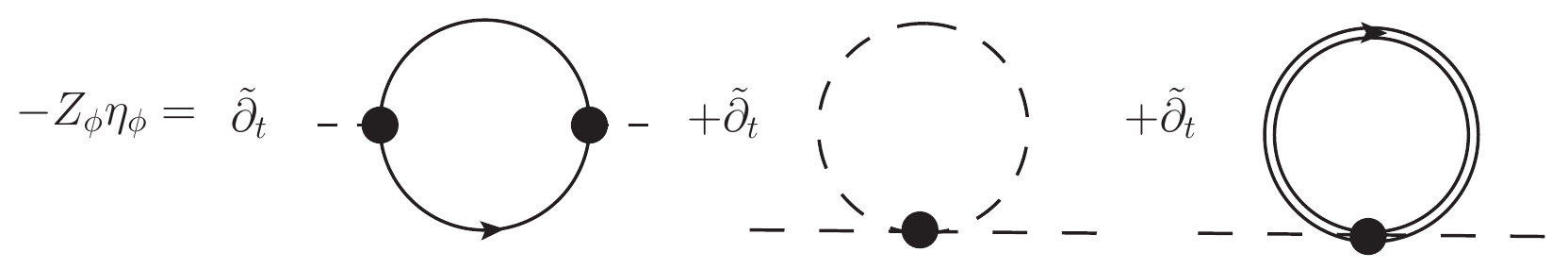}
 \caption{Schematic representation of the flow equation for the
   bosonic anomalous dimension in the symmetric phase (permutations,
   signs, and diagram multiplicities are not shown). \hspace*{\fill}}
 \label{fig:etaphi_rebos}
\end{figure}
The left hand side of \Eq{eq:dotG2phi} is obtained straightforwardly
from taking the second $\phi$-derivative of \Eq{eq:rebos} at $\Phi=0$,
while the right hand side stands for the diagrams contributing to the
flow of the mesonic two-point function. Evidently, the mesonic
two-point function is proportional to a factor $\exp(2 \dynB_{\phi,k}(p))$ as
introduced in \Eq{eq:rescale}. The contributing terms include
contributions from the quark loop proportional to $\hs$, see
\Fig{fig:etaphi_rebos}, and hence ${\Zphi}_k\neq 0$ even with
$\Zphi=0$ at the initial UV scale. In turn, $-2\dot\dynB_\phi$
accounts for the rescaling shift of the anomalous dimension. If we
choose $\dot\dynB_\phi$ such that it compensates for the $\partial_t
m_\phi^2$ term and the diagrammatic contribution,
\begin{align} \label{eq:eta0} 
2 \dot\dynB_\phi(p) = -\0{\partial_t
    m_\phi^2}{p^2+m_\phi^2}  +\0{1}{\Gamma^{(2)}_{\phi,k}(p)} {\rm
    Flow}^{(2)}_{\phi}(p)\,, 
\end{align} 
the anomalous dimension vanishes identically for all cutoff scales,
$\eta_\phi(p)\equiv 0$. This is in line with the rescaling symmetry in
\Eq{eq:rescale} which can be used to shift the dependence on the
exchange momentum between the Yukawa vertex and the meson
propagator. 

For the sake of simplicity, the discussion above was done in the case
of vanishing backgrounds in the symmetric phase. If expanding the
flows about non-vanishing backgrounds, i.e.,  in the regime with chiral
symmetry breaking, the equations receive further straightforward
contributions.

For the determination of $\dot{D}_{\bar{q}q}$, see later
  discussions in \sec{sec:dynhadcomplete}.

\begin{figure*}
 \includegraphics[width=0.9\textwidth]{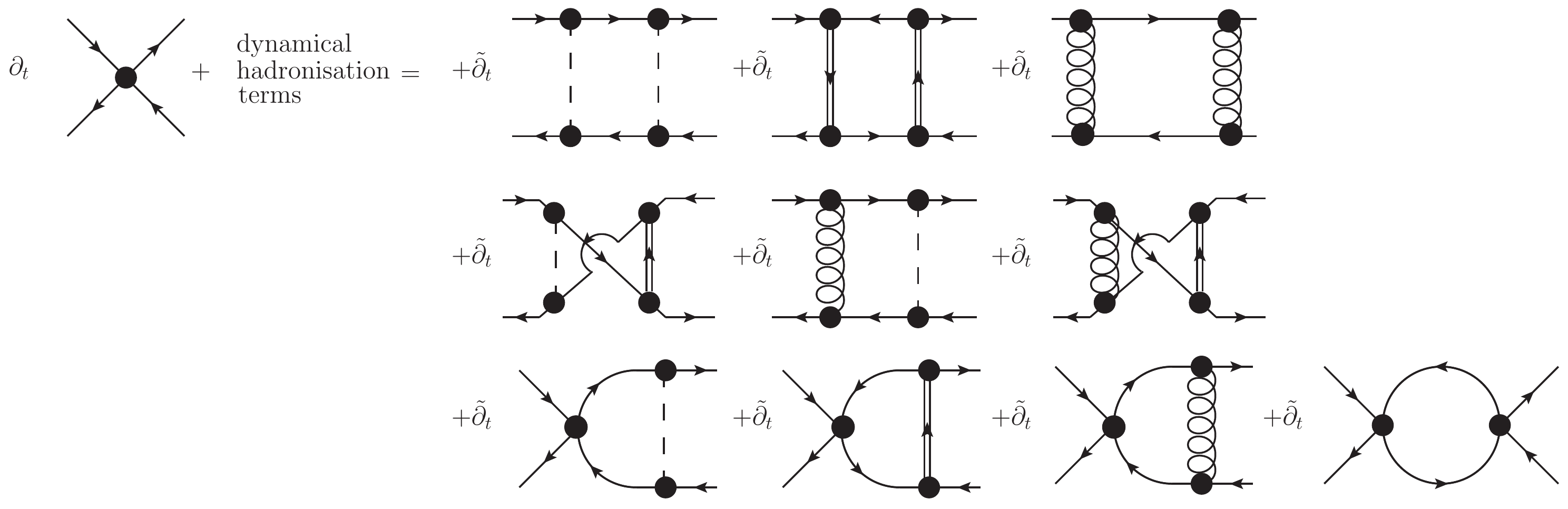}
 \caption{Schematic representation of the flow equation for the
   four-quark coupling with dynamical hadronisation of the 
   scalar-pseudoscalar and the scalar diquark channel (permutations,
   signs, and diagram multiplicities are not shown). \hspace*{\fill}}
 \label{fig:lambdaphi_rebos}
\end{figure*}

\subsubsection{Dynamical hadronisation for diquarks}\label{sec:dynhadd}

In summary, the above introduction of composite meson fields allows
for partial and full bosonisations of the respective four-quark
interaction channels with a specific choice of the dispersion of the
composite fields. It is the constraint equations for
$\dot\dynA_{\bar q q},\dot\dynB_\phi$ which settles the related
choice. Before we discuss these constraints, we extend the above
argument to the diquark sector. In analogy to \Eq{eq:scaldynhad} we write,
\begin{align}
  \partial_t \begin{pmatrix} d^a_k \\ d^{a\ast}_k \end{pmatrix} =
  \dot\dynA_{qq} \0{\Zq}{\Zd^{\frac12}}\sqrt{2}\begin{pmatrix} q^T
    \epsilon^a\tau^2 C\gamma^5 q \\
    \bar q \epsilon^a\tau^2 C\gamma^5 \bar q^T \end{pmatrix} +
  \dot{\dynB}_d
  \begin{pmatrix} d^a_k \\ d^{a\ast}_k \end{pmatrix} \,.
\label{eq:ddynhad}
\end{align}
\Eq{eq:scaldynhad} and \eq{eq:ddynhad} encode the
mixing effect of the four-quark interaction and the Yukawa interaction
during the fRG-flow. They can be used to impose that the respective
residual channels of the four-quark interaction vanish, see
\Eq{eq:nsps0} and \eq{eq:nd0}. This is complete bosonisation of
the respective channels, and it entails in particular that the
respective four-quark couplings vanish for all scales $k$ at vanishing
momenta, $\lambdas=\lambdad=0$. Due to the consistent book keeping
during the flow any double counting problem is avoided.

\subsubsection{Dynamical hadronisation for vector
  mesons}\label{sec:dynhadvecmes}

The dynamical hadronisation flows that comes with the
quark-vector meson action \eq{eq:qV} read
\begin{align}
  \partial_t \omega_{k}^\mu(r)
  = \dot\dynA_{\bar q\gamma q}(r) \0{\Zq}{\Zo^{\frac12}}(\bar{q}
    \gamma^\mu q) (r) + \dot\dynB_\omega(r)\,\omega_{k}^\mu(r) \,,  
\label{eq:oynhad}
\end{align}
and
\begin{align}
  &\partial_t \begin{pmatrix}\vec \rho_{ k}^\mu\\
    \vec a_{1 k}^\mu\end{pmatrix}(r)\nonumber\\[2ex]
  &= \dot\dynA_{\bar q\gamma\tau q}(r)
  \0{\Zq}{\ZV^{\frac12}}\begin{pmatrix} \bar{q}
    \gamma^\mu\vec \tau q \\
    \bar{q}\gamma^\mu\gamma^5\vec{\tau}q \end{pmatrix}(r) +
  \dot\dynB_V(r) \, \begin{pmatrix}\vec \rho_{k}^\mu\\\vec a_{1
      k}^\mu\end{pmatrix}(r) \,.
\label{eq:Vdynhad}
\end{align}
%

\subsubsection{Complete dynamical hadronisation}\label{sec:dynhadcomplete}

Let us now elaborate on the case of a complete dynamical hadronisation,
concentrating on the scalar-pseudoscalar and the vector channel first, i.e.,
on the $\lambdasd=0$ constraints. The flow
equations for $\lambdasd$ is obtained by taking for quark--anti-quark
derivatives of the flow \eq{eq:flow}. These four derivatives also hit
the rebosonisation term on the right hand side of \Eq{eq:rebos}, and
generate $\dot\dynA_{qq} \hs$ and $\dot \dynB_d\hd$ terms,
respectively. Then the four-quark flows can be derived from
\Eq{eq:flow} including the dynamical hadronisation terms similarly to
the flow of the mesonic two-point function \eq{eq:dotG2phi}. For the
sake of simplicity we restrict ourselves to the vanishing momentum
part and vanishing backgrounds $\Phi=0$ in the symmetric phase, to wit
\begin{align} \nonumber
  \partial_t \lambdas-2 \eta_q\,\lambdas - 2\hs\dot{\dynA}_{\bar q q}
  =&\,\0{1}{\Zq^2} {\rm
    Flow}^{(4)}_{\phi,qq \bar q \bar q}\,,\\[2ex]
  \partial_t \lambdad-2\eta_q\,\lambdad - 2\hd\dot{\dynA}_{qq}
  =&\,\0{1}{\Zq^2} {\rm Flow}^{(4)}_{\text{d},q q \bar q \bar q}\,,
\label{eq:dotLambda}\end{align}
where ${\rm Flow}^{(4)}_{\phi/\rm d}$ stands for the
diagrammatic contributions coming from the right hand side of
\Eq{eq:flow}, see \Fig{fig:lambdaphi_rebos}. 

For large cutoff scales the four-quark couplings vanish, and in
particular we have $\lim_{k\to \infty}\lambdasd\to 0$. If we adjust
the dynamical hadronisation functions $\dot\dynA_{\phi/d}$ such, that
the flows $\left( \partial_t -2 \eta_q\right)\lambdasd\equiv 0$, these couplings vanish
identically also for vanishing cutoff scale $k=0$. This is obtained by
imposing
\begin{align}
  \dot{\dynA}_{\bar q q} =-\0{1}{\Zq^2}\frac{{\rm Flow}^{(4)}_{\phi,qq \bar q \bar q}}{2\hs} \,,\qquad
  \dot{\dynA}_{qq} = -\0{1}{\Zq^2}\frac{{\rm Flow}^{(4)}_{\text{d},qq \bar q \bar q}}{2\hd}\,,
\label{eq:rebosphid}
\end{align}
where the right hand sides are RG-invariant by construction. This leads us to  
\begin{align}\label{eq:lambdasd0}
\lambdas\equiv 0\,,\qquad \qquad \lambdad\equiv 0\,.
\end{align} 
In conclusion we have shown how to eliminate momentum channels of
specific tensor structure of the four-quark interaction with the
choice of the dynamical hadronisation functions $\dot
\dynA_{\bar qq/qq}$. Furthermore, the scaling functions
$\dot\dynB_{\phi/d}$ can be used to re parametrise the meson and diquark 
propagators, or, more generally, the meson and diquark fields in a
momentum-dependent way. This may help to optimise the physics content
of expansion schemes where only part of the full momentum dependence
is taken into account.

Subject to an appropriate diagonalisation of the Fierz basis the
vector meson channel of the four-quark interaction can be eliminated
with the condition,
\begin{align}
  \dot{\dynA}_{\bar q\gamma q} =-\0{1}{\Zq^2}\frac{{\rm
      Flow}^{(4)}_{\omega,q q\bar q \bar q }}{2\ho}\,,\quad \dot{\dynA}_{\bar q\gamma\tau q} =-
  \0{1}{\Zq^2}\frac{{\rm Flow}^{(4)}_{\text{V},q q\bar q \bar q }}{2\hV}\,, 
\label{eq:rebosV}
\end{align}
where we have restricted ourselves to $\Phi=0$ in the symmetric phase
as for the mesons and diquarks. For later purposes we already mention,
that for $\omega$ in the broken phase, apart from terms
proportional to $\Gamma^{(5)}_{\omega q q\bar q \bar q}$, the
denominator in \Eq{eq:rebosV} is changed to $\ho
+h^{(1)}_{q \omega\omega_0} \,\omega_0$ similarly to the
scalar-pseudoscalar meson case.

\Eq{eq:rebosV} together with
\eq{eq:rebosphid} leads to complete dynamical hadronisation of the
scalar-pseudoscalar and vector meson channels as well as the scalar
diquark channel of the four-quark interaction. The effective potential
$U_k(\phi,\phi_\mu,d)$ then carries multi-meson-diquark scatterings
in these channels. The link to the effective action of QCD in terms of
the fundamental degrees of freedom $\Phifund=(A,c,\bar c,q,\bar q)$ is
now given by
\begin{align}\label{eq:effQCDphiVd}
  \Gamma_{\text{\tiny{QCD}},k}[\Phifund]=\Gamma_{\text{\tiny{QCD}},k}^{(\phi\phi_\mu
    d)}[\Phifund,\phi_{\text{\tiny{EoM}}},\phi_{\mu,\text{\tiny{EoM}}},d_{\text{\tiny{EoM}}}]\,.
\end{align}
In the present context we are most interested in the $\omega_0$-part
of the vector meson action. It couples to higher correlation functions
of the quark, and in particular to the baryon channel of the six-quark
interaction. There, it provides a density field for the baryon density
in addition to the chemical potential. A non-vanishing expectation
value of $\omega_0$ hence shifts the onset of the baryon density
and influences the liquid-gas transition. We shall come back to this
point at the end of the next section on emergent baryons.

For later reference we present at this point in more detail the four-quark interaction
channels relevant in this work, namely the four-quark couplings
corresponding to the scalar-pseudoscalar channel $\lambdas$, the
scalar diquark channel $\lambdad$ and the vector $\lambdao$, evaluated
on the solution presented in Ref.~\cite{Mitter:2014wpa}. In terms of the
Fierz-complete basis from App.~\ref{app:4f} these are given by
\begin{align}
  \lambdas&=\lambda_{(S-P)_+}+\lambda_{(S+P)_-}\,,\nonumber\\[2ex]
  \lambdad&=\023 \lambdas-\023 \lambda_{(S-P)_-}+
  \023 \lambda_{(S+P)_-}+\023 \lambda_{(V-A)}\,,\nonumber\\[2ex]
  &\quad-2 \lambda_{(S+P)_-^{\rm adj}}-2\lambda_{(V-A)^{\rm adj}}\,,\nonumber\\[2ex]
  \lambdao&=\lambda_{(V-A)}+\lambda_{(V+A)}\,.
\label{eq:4fcouplingsexplicit}
\end{align}
It turns out that the scalar diquark channel is attractive, see
\cite{Mitter:2014wpa,Cyrol:2017ewj}, consistent with weak-coupling
arguments~\cite{Barrois:1977xd,Bailin:1983bm,Iida:2000ha}.

\section{Emergent Baryons}\label{sec:Emhadrons}

\begin{figure}
 \includegraphics[width=0.8\columnwidth]{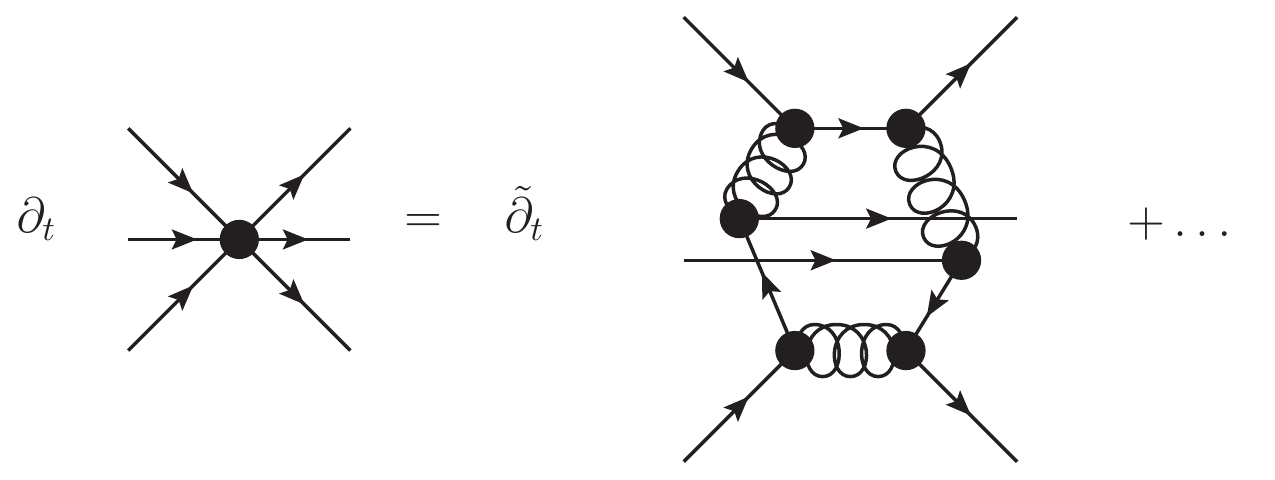}
 \caption{UV-flow of six-quark interactions via quark-gluon ring diagrams. \hspace*{\fill}}
 \label{fig:6qflow}
\end{figure}

Having discussed the four-quark interactions along with its
reformulation in terms of effective mesonic/diquark degrees of freedom
in details, we now turn to six-quark interactions. For this discussion
we start from the QCD framework, where the scalar-pseudoscalar
channel, and the scalar diquark channel have been treated within
dynamical hadronisation. Here we consider the baryon channel of the
six-quark interaction.

\subsection{Qualitative features of baryon flows}

At large scales, analogously to the generation
of the four-quark interaction, the six-quark interaction is generated and sustained by the
quark-gluon ring diagrams, see \Fig{fig:6qflow}. This diagram runs
with $\alpha_{s,k}^3$. All other diagrams in terms of quarks and
gluons are suppressed with additional powers of $\alpha_{s,k}$, and
hence will be neglected in the current UV analysis.

\begin{figure*}
 \includegraphics[width=.8\textwidth]{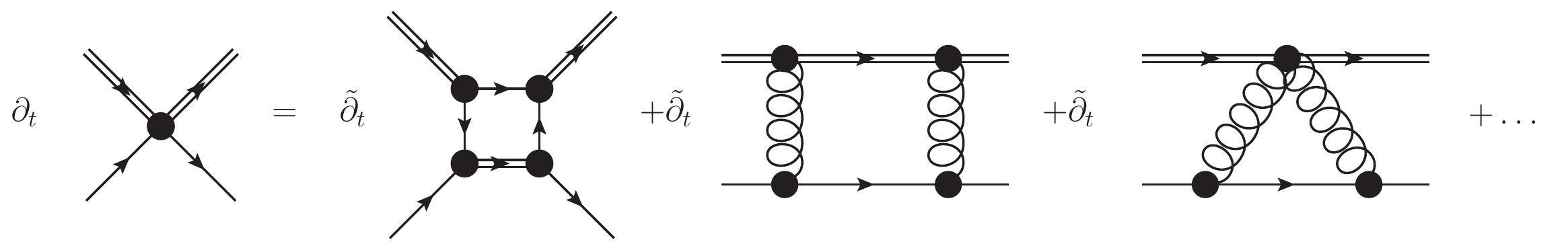}
 \caption{UV-flow of two-quark--two-diquark interactions via
   quark-diquark box, quark-diquark-gluon box, and quark-gluon
   triangle diagrams. \hspace*{\fill}}
 \label{fig:2q2dqflow}
\end{figure*}

Due to the introduction of the effective diquark field $d \propto qq$
the UV-flow also generates diquark-quark box diagrams that have the
same effective quark content as the quark-gluon ring, see
\Fig{fig:2q2dqflow}. A comparison of their relative strength in terms
of the running coupling is only possible after the equations of motion
for $\phi, d$ are invoked. Without this the hadron correlation
functions have the scaling symmetry \eq{eq:rescale} or, in the
dynamical hadronisation context, the choice of $\dot\dynB_{\phi/d}$ in
\Eq{eq:scaldynhad} and \eq{eq:ddynhad}. Only combinations of hadron
correlation functions that are symmetric under these rescalings may
relate directly to diagrams in QCD without dynamical hadronisation.

\begin{figure}[b]
 \includegraphics[width=0.7\columnwidth]{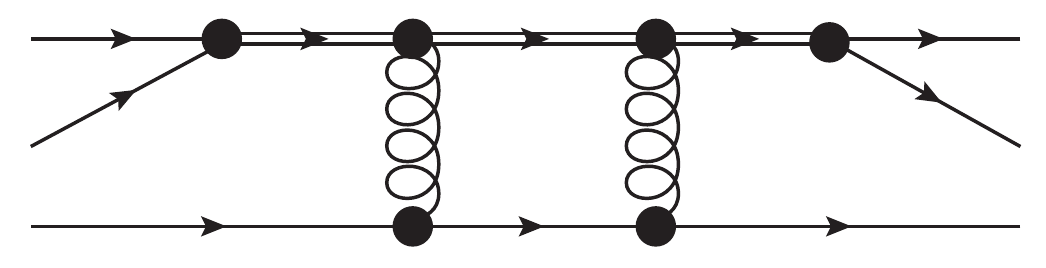}
 \caption{Relevance counting for two-quark--two-diquark interactions
   compared to six-quark interactions for the example of the
   quark-diquark-gluon box diagram.  \hspace*{\fill}}
 \label{fig:6qto2q2dq}
\end{figure}

If the equations of motion, $\phi_{\text{\tiny{EoM}}},
d_{\text{\tiny{EoM}}}$ are invoked, external meson and diquark lines
receive at leading order a further scattering in $\bar q q$ and $qq$,
respectively, via the Yukawa interaction, see, e.g., \Fig{fig:6qto2q2dq}
for the quark-diquark-gluon box. The upper line in \Fig{fig:6qto2q2dq} now
describes the scattering of two quarks into a diquark, the subsequent
emission of two gluons and then the rescattering of the diquark into
two quarks. We also observe that the lower part of the quark-gluon box
and the triangle diagram are the same. The upper line of the
combination of these diagrams follows from two gluon field derivatives
of a diquark propagator, i.e.,
\begin{align}\nonumber 
\left. \0{\delta^2 }{\delta A^2} \right|_{\hd} \hd G_{\rm
  d}\hd  = &\,- \hd G_{\rm
  d} \Gamma_{dd^* AA}^{(4)} G_{\rm
  d} \hd \\[2ex]
&\hspace{-0.5cm}+\hd G_{\rm
  d}\Gamma_{dd^* A}^{(3)} G_{\rm
  d} \Gamma_{dd^* A}^{(3)} G_{\rm
  d} \hd \,.
\label{eq:dexchangeA2}
\end{align}
Now we utilise that the tree-level diagram of the quark-diquark
(re-)scattering $\hd G_{\rm d}\hd$ scales as $\alpha_{s,k}^2$ in the UV as
it relates to the four-quark vertex that is driven by the gluonic box
diagram, see \Fig{fig:rebos_step1} and
\Fig{fig:rebos_definition_diquark}. In summary this leads to an
$\alpha_{s,k}^4$-scaling for both the triangle diagram
and the quark-diquark-gluon box.

The $\alpha_{s,k}$ scaling of the quark-diquark box follows similarly:
attaching the diquark-quark scattering arising from the equations of
motion the diagram is constructed solely from
\begin{align}\label{eq:3dexchange}
\left( \hd G_{\rm
  d}\hd \right)^3\propto \alpha_{s,k}^6\,. 
\end{align}
Using the fact that $\hd G_{\rm d}\hd$ scales as $\alpha_s^2$ in the UV
finally leads to an overall $\alpha_{s,k}$-scaling of $\alpha_s^6$. 
We conclude that in the UV the
diquark diagrams are suppressed by at least one additional power of
$\alpha_{s,k}$ compared to the quark-gluon ring diagram in \Fig{fig:6qto2q2dq}
that scales as $\alpha_{s,k}^3$.

\begin{figure}[b]
 \includegraphics[width=0.7\columnwidth]{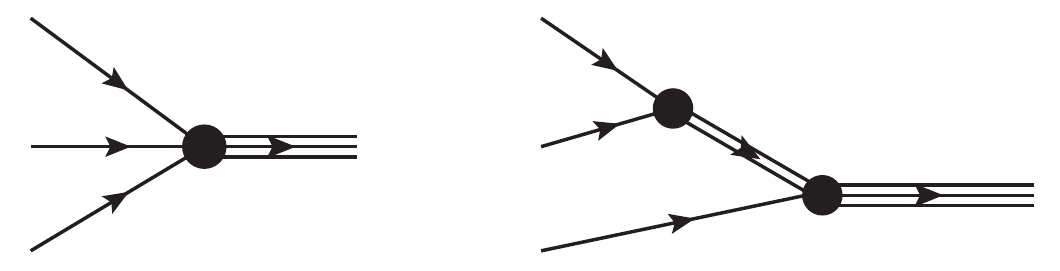}
 \caption{Competing baryon creation processes: direct process
   involving three quarks (left) and 2-step process involving an
   intermediate diquark. \hspace*{\fill}}
 \label{fig:baryoncreation}
\end{figure}

While the diquark diagrams are suppressed in the UV, we expect them to
be relevant in the IR due to phase space arguments, see
\Fig{fig:baryoncreation}.  A related analysis will be published
elsewhere. The baryon channels of the six-quark interaction, as well
as the two-quark--two-diquark interaction are defined by the tensor
structure as well as the baryon momentum routing.

At this point it is instructive to consider three quark interpolation operators for baryons and their chiral transformation properties.  The most obvious choice involving a scalar diquark structure is
\begin{align}
\label{eq:baryonintq3}
(q^T  \epsilon^a \tau^2 C \gamma^5 q) q^a\,,
\end{align}
or correspondingly
\begin{align}
\label{eq:baryonintdq}
\sqrt{2}d^a q^a\,,
\end{align}
which transforms as a quark under axial chiral $SU(2)_A$ transformations as 
the diquark operator is a chiral scalar \cite{Nagata:2007di,Chen:2012vs}. 
Note that we completely disregard the transformation properties with respect to axial
$U(1)_A$ transformations that linearly mixes interpolating operators with
scalar and pseudoscalar diquark structure, see App.~\ref{app:transformations}. We now consider six-quark 
interaction terms that are generated in the effective action corresponding
to a classical Lagrangian of the form,
\begin{align}\label{eq:Lq6}
  \mathcal L_{(6q)}=\lambdab \Zq^3(\bar q \epsilon^a \tau^2 C \gamma^5 \bar q^T) (q^T
  \epsilon^b \tau^2 C \gamma^5 q) (\bar q^a \mathcal{T} q^b)\,,
\end{align}
with a so far undetermined operator $\mathcal{T}$ acting in Dirac and
flavour space.  The simplest and most widely used choice
$\mathcal{T}=\mathbbm{1}$ is obviously not invariant under $SU(2)_A$
transformations. One way of enforcing an $SU(2)_A$-invariant
interaction is to consider momentum-dependent interactions with
$\mathcal{T}=p^\mu \gamma^\mu$ for an appropriate momentum variable
$p$.

\begin{figure}[t]
 \includegraphics[width=\columnwidth]{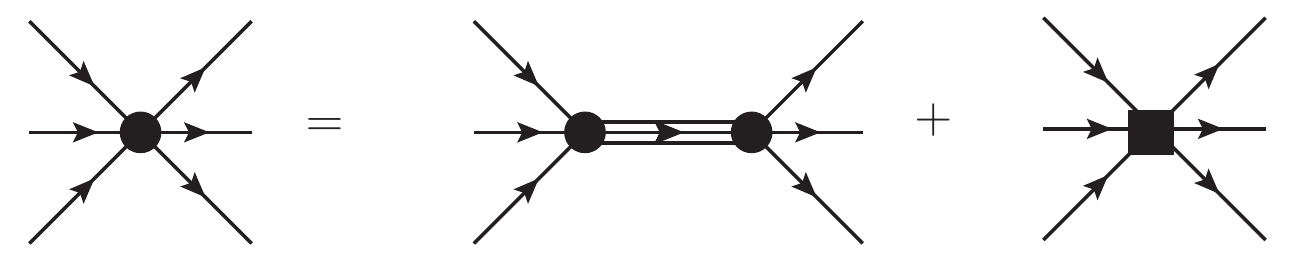}
 \caption{Rewriting the six-quark interaction as an effective baryon
   exchange and a residual six-quark interaction (square). \hspace*{\fill}}
 \label{fig:rebos_definition_6q}
\end{figure}

The interaction in \Eq{eq:Lq6} is now treated with dynamical hadronisation 
analogously to the mesons and diquarks. In
terms of the Bethe-Salpether language used to heuristically describe
the hadronisation of mesons and diquarks we rewrite the baryon
channels in terms of a one-baryon exchange, see
\Fig{fig:rebos_definition_6q}.

\begin{figure}[t]
	\includegraphics[width=\columnwidth]{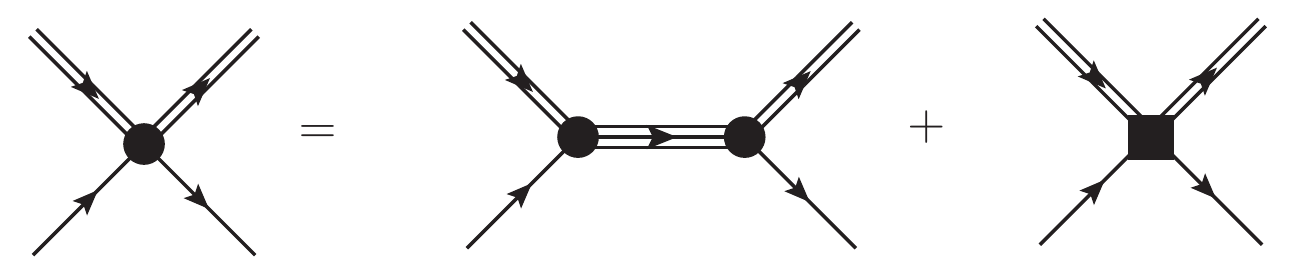}
	\caption{Rewriting the two-quark--two-diquark interaction as an
		effective baryon exchange and a residual two-quark--two-diquark
		interaction (square). \hspace*{\fill}}
	\label{fig:rebos_definition_2q2dq}
\end{figure}

To that end we have to define the baryon propagator. In the chiral
limit at $\sigma=0$ the propagator $G_{\rm b}$ of the exchange
effective field $b$ has to sustain the chiral structure of the
interaction. Accordingly we write
\begin{align} \label{eq:Gb}
G_{\rm b}(p) = \0{1}{\Zb(p)} \0{1}{i \slashed p}\,, 
\end{align}
in the UV{}. The quark-gluon triangle, the
diquark-quark-gluon box and the diquark-quark box diagrams in the
baryon channel generate a two-quark--two-diquark interaction in the
effective action. The classical Lagrangian corresponding to \Eq{eq:Lq6} on
the quark-diquark level reads
\begin{align}\label{eq:Lqdb}
  \mathcal L_{(2q2d)}= \lambdabd\, \Zq
  \Zd\,d^\ast_a d^{\phantom{\ast}}_b (\bar q_a \mathcal{T}q _b) \,.
\end{align}
Analogous to the six-quark interaction we can rewrite the
two-quark--two-diquark interaction as an effective baryon exchange and
a residual interaction, see \Fig{fig:rebos_definition_2q2dq}. It
is important to keep in mind that, although six-quark interaction and
the two-diquark share the same effective quark content, they have to
be considered for the sake of the baryonisation as two independent
interactions.

In summary, the baryonic effective action that gives rise to the
one-baryon exchange diagrams in \Fig{fig:rebos_definition_6q} and
\ref{fig:rebos_definition_2q2dq} on the
equation of motion of the baryon reads
\begin{align}\nonumber 
 &\Gamma^{\rm (b)}_k = \int d^4x \,\biggl\{ \Zb\bar{b} \left(
    \slashed{\partial}-\mub\gamma^0\right) b\\[2ex]
    &\nonumber+2\lambdabd\, \Zq
  \Zd\,d^\ast_a d^{\phantom{\ast}}_b \left(\bar q_a\mathcal{T} q_b\right) \\[2ex] \nonumber &+\lambdab\,\Zq^3\,
  \left(\bar q \epsilon^a \tau^2 C \gamma^5 \bar q^T\right) \left(q^T
  \epsilon^b \tau^2 C \gamma^5 q\right) \left(\bar q_a\mathcal{T} q_b\right)\\[2ex]
  &+ \sqrt{2}\hqdb\,\Zq^{1/2} \Zd^{1/2} \Zb^{1/2}\,\left[ d^{\phantom{\ast}}_a \left(\bar{b}\mathcal{T} q_a\right)+ d_a^\ast \left(\bar q_a\mathcal{T} b\right) \right]
  \nonumber \\[2ex] &+ \hqb\,\Zq^3 \Zb^{1/2} \,\left[ \left( q^T
    \epsilon^a\tau^2 C\gamma^5 q\right) \left(\bar{b}\mathcal{T} q_a\right) + {\rm h.c.}\right] \biggr\} \,.
\label{eq:Gamma_b}
\end{align}
As for the meson and diquark effective action we have pulled
out the appropriate factors of the wave-function renormalisations in
order to have RG-invariant couplings. The second and the third lines contain the terms generated from
quark-gluon-meson-diquark interactions in QCD formulated with
dynamical hadronisation of mesons and diquarks with the effective
action $\Gamma^{(\phi\phi_\mu d)}_{\text{\tiny{QCD}},k}$, see \Eq{eq:effQCDphiVd}. The fourth
and the fifth lines provide the Yukawa terms that describe the scattering
of a quark and a diquark into the effective baryon field and that of
three quarks into the effective baryon field.

We also emphasise that no explicit baryon term is present. In the chiral
limit in the UV, no mass term is present for $\sigma=0$. In turn, for
$\sigma\neq 0$ the six-quark exchange diagram leads to non-chiral
structures in the baryon channel that lead to mass terms in the
dispersion \eq{eq:Gb}.  In the effective action \eq{eq:Gamma_b} this
is encoded in the meson-baryon terms in the last line. In the IR, where spontaneous 
chiral symmetry breaking occurs, the
expectation value of the $\sigma$-field takes a non-trivial value and
the baryon curvature mass in the IR reads 
\begin{align}\label{eq:Mb}
  \Mb(\sigma) =\frac{1}{2}\hb \sigma \qquad {\rm and}\qquad
  \Mb=\frac{1}{2}\hb \sigma_{\text{\tiny EoM}}\,,
\end{align}
analogously to the quark mass. Note that in both cases the structure
$M(\sigma)= \frac{1}{2} h\,\sigma$ seems to imply that the mass
function vanishes for $\sigma= 0$ even in the chirally broken
phase. However, it has been shown in \cite{Pawlowski:2014zaa} that in
the full theory the Yukawa coupling $h(\sigma)$ has a non-trivial
$\sigma$-dependence leading to $M(\sigma)\geq
M(\sigma_{\text{\tiny{EoM}}})$ for all $\sigma$. In any case, the
baryon propagator is given by
\begin{align} \label{eq:fullGb}
G_{\rm b}(p) = \0{1}{\Zb(p)} \0{1}{i \slashed p+\Mb(\sigma)} \,.
\end{align}
This structure immediately leads to the question how the chiral baryon
decouples in the UV\@.  Typically this decoupling, as in the case of
mesons and diquarks, is achieved by a rapidly increasing mass. The
decoupling of the baryon is guaranteed by the equivalence of a baryon
exchange with the Yukawa-type interaction with $\hqb, \hqdb$ to the
respective six-quark and two-quark--two-diquark
interactions. Relying on the approximation \eq{eq:Gamma_b} they read very
schematically,
\begin{align}\label{eq:bexchangeqd}
  \Zq^3 \hqb\mathcal{T} \0{1}{\slashed{p}}\mathcal{T} \hqb\,,
  \qquad \qquad \Zq \Zd \hqdb
  \mathcal{T}\0{1}{\slashed{p}}\mathcal{T} \hqdb \,.
\end{align}
Note that $\Zb$ from the propagator is cancelled by the respective
$\Zb$ in the interaction terms. The wave-function renormalisations of
quarks and diquarks in \Eq{eq:bexchangeqd} take care of the natural
scalings of quarks and diquarks. Accordingly, the normalised,
RG-invariant Yukawa couplings $\hqb, \hqdb$ have to vanish in the
UV{}. In summary, the mass-decoupling for meson and diquark turns
into a vertex-decoupling for the baryon. The link to QCD in the
fundamental degrees of freedom is now given by 
\begin{align}\label{eq:effQCDphiVdb}
  \Gamma_{\text{\tiny{QCD}},k}[\Phifund]=\Gamma_{\text{\tiny{QCD}},k}^{(\phi\phi_\mu
    d)}[\Phifund,\phi_{\text{\tiny{EoM}}},\phi_{\mu,\text{\tiny{EoM}}},d_{\text{\tiny{EoM}}},b=0]\,, 
\end{align}
as the fermionic baryon field $b$ has to vanish on the equations of
motion, $b_{\text{\tiny{EoM}}}=0$. 

\begin{figure}[b]
	\includegraphics[width=\columnwidth]{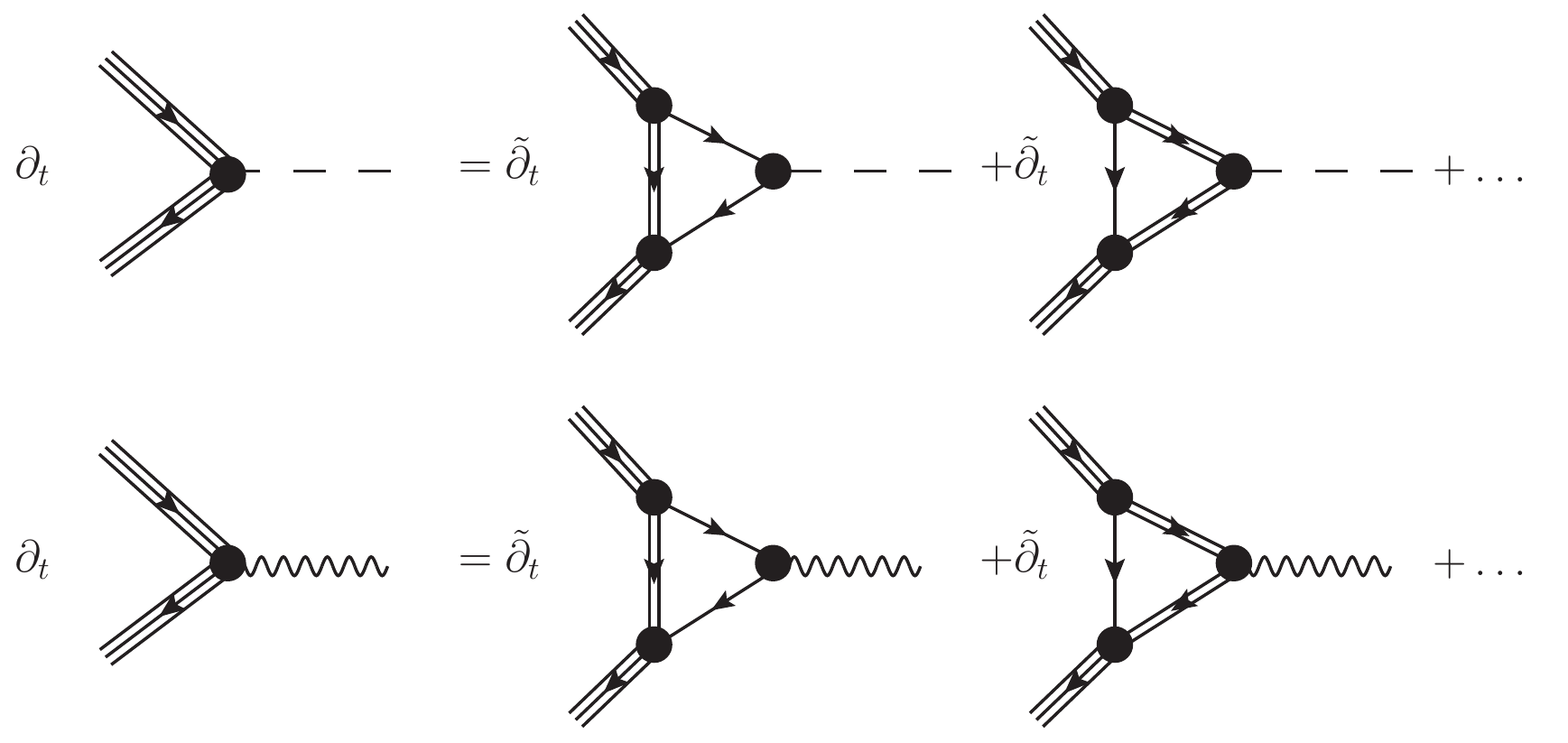}
	\caption{Creation of effective baryon-meson (first line: scalar/pseudoscalar; second line: vector) interaction via a
		quark-diquark loop diagram. \hspace*{\fill}}
	\label{fig:baryonmesonyukawa}
\end{figure}
The Ansatz for the effective action in \Eq{eq:Gamma_b} are all sixth
order in effective quark degrees of freedom. There are of course
higher order terms that are generated during the flow, such as eight-point
interactions in effective quark degrees of freedom, or in terms of
baryonic interactions, baryon-meson and baryon-quark interaction
terms. This is reflected by an Ansatz for the effective action of the
form,
\begin{align}
  & \Gamma^{\rm (bm)}_k = \int d^4x \biggl[\frac{\hb}{2} \Zphi^{1/2}
  \Zb\,\bar{b} \, (\sigma + i\gamma^5
  \vec\pi\cdot\vec\tau) b\nonumber\\[2ex]
  &+ \frac{\hbo}{2} \Zo^{1/2} \Zb\,\bar{b} \, \gamma_\mu\omega_\mu
  b-\lambdabo \Zb \Zq (\bar b\gamma_\mu b)\, (\bar q\gamma_\mu
  q)\biggl]\,,
  \label{eq:baryonmeson}
\end{align}
where we have dropped terms coupling the baryon to the vector mesons
$V_\mu$. The Ansatz \eq{eq:baryonmeson} takes into account the
emission and absorption of mesons, $\phi, \phi_\mu$ from/into a
baryon, see \Fig{fig:baryonmesonyukawa}. Note that the
$\omega_0$-contribution can be absorbed in a shift of the chemical
potential: $\mu\to \mu - \ho \Zo\omega_0$. For later reference
\Eq{eq:baryonmeson} also includes a quark-baryon scattering term. Even
in an approximation scheme with momentum-independent dressing
functions, it is not guaranteed that the complete dynamical
hadronisation of the vector channel $\lambdao$ also leads 
to a complete dynamical hadronisation of the two-quark-two-baryon
interaction in the eighth order in effective quark degrees of
freedom. The omission of this interaction introduces an artificial
dependence on the choice of the basis
in the sector of four-quark interactions.

\subsection{Dynamical hadronisation for baryons}\label{sec:dynhadb}

So far we have discussed the qualitative features of the hadronisation
for baryons. In the present section we put forward the dynamical
hadronisation procedure for baryons. The $k$-dependent change of the
baryon field that allows to remove the effective baryon-exchange
contribution from the six-quark as well as two-quark--two-diquark
channels is given by
\begin{align}\nonumber
  & \partial_t \begin{pmatrix}b_k\\\bar b_k\end{pmatrix}= 
\dot\dynA_{qqq}\,\0{\Zq^{3/2}}{\Zb^{1/2}}\begin{pmatrix}(q^T
    \epsilon^a\tau^2 C\gamma^5 q)q^a\\\bar q^a(\bar q
    \epsilon^a\tau^2 C \gamma^5 \bar q^T )\end{pmatrix}\\[2ex] 
&\qquad +\dot \dynA_{qd}\, \0{\Zq^{1/2} \Zd^{1/2}}{\Zb^{1/2}}\sqrt{2}\begin{pmatrix}d_k^a
  q^a\\d_k^a{}^* \bar q^a\end{pmatrix} +\dot \dynB_b \begin{pmatrix}b_k\\ \bar
  b_k\end{pmatrix}\,,
\label{eq:EFH}\end{align}
where we build on the identification in \Eq{eq:baryonintq3} and \eq{eq:baryonintdq}.
Note that there is no mixing between $\dot{\dynA}_{qqq}$ and
$\dot{\dynA}_{qd}$ on the level of these equation as the former
relates to six-quark and the latter to two-quark--two-diquark flows.
As for mesons and diquarks we restrict ourselves to the vanishing
momentum part for the sake of simplicity. By projecting the full flows
for the six-point and two-quark--two-diquark correlation functions
on the baryon tensor structures,
see \Fig{fig:6qflow} and \ref{fig:2q2dqflow},
we arrive at
\begin{align} \nonumber
  \partial_t \lambdab-3 \eta_{\rm q} \lambdab+2 \hqb\dot{\dynA}_{qqq} =&\,\0{1}{\Zq^3} {\rm
    Flow}^{(6)}_{\rm q}\,,\\[2ex]
  \partial_t \lambdabd- (\eta_{\rm q}+\eta_{\rm d}) \lambdabd+2
  \hqdb\dot{\dynA}_{qd} =&\,\0{1}{\Zq \Zd} {\rm Flow}^{(4)}_{\rm qd}\,,
\label{eq:dotLambdab+d}\end{align}
where ${\rm Flow}^{(6)}_{\rm q}$ and ${\rm Flow}^{(4)}_{\rm q d}$ stand for the
corresponding diagrammatic contributions coming from the right hand side of
\Eq{eq:flow}. For large cutoff scales the six-quark and two-quark--two-diquark
couplings vanish, and in particular we have
$\lim_{k\to \infty}\lambdab \to 0$ and
$\lim_{k\to \infty} \lambdabd\to 0$. If we adjust the dynamical
hadronisation functions $\dot \dynA_{qqq/qd}$ such that both flows vanish, i.e.,
$\partial_t \lambdab\equiv \partial_t\lambdabd\equiv 0$, these couplings
vanish identically also for vanishing cutoff scale $k=0$. This is
obtained by imposing
\begin{align}\nonumber 
  \dot{\dynA}_{qqq} =&\, \frac{1}{2 \hqb}\0{1}{\Zq^3} {\rm
    Flow}^{(6)}_{\rm q}\,,\\[2ex] 
 \dot{\dynA}_{qd} =&\, \frac{1}{2 \hqdb}\0{1}{\Zq \Zd}  {\rm
  Flow}^{(4)}_{\rm qd}\,,
\label{eq:rebosb}
\end{align}
where the right hand sides are RG-invariant by construction. This leads us to  
\begin{align} 
\lambdab\equiv 0\,,\qquad \qquad \lambdabd\equiv 0\,.
\end{align} 
In conclusion we have shown how to eliminate momentum channels of
specific tensor structures of the six-quark interaction with the
choice of the dynamical hadronisation functions $\dot
\dynA_{qqq/qd}$. Furthermore, the scaling function $\dot \dynB_b$ can
be used to reparametrise the baryon propagator. However, as can be
seen from the equations, it has dropped out completely in the
normalised, RG-invariant representation as it should.

Finally, we discuss how the baryonic mass scale arises
diagrammatically from \Fig{fig:6qflow}: the baryonic mass scale
is that of the pole mass. For that purpose we have to Wick-rotate the
Euclidean imaginary time frequencies $\omega_{\textrm{Eucl}}$ to
Minkowski real time frequencies $\omega_{\textrm{Min}}$,
\begin{align}\label{eq:IT-RT}
  \omega_\textrm{Eucl}=\imag \,\omega_\textrm{Min}\,.
\end{align}
Due to the relation between the baryonic propagator and the
quark-gluon ring as well as the quark-diquark box diagrams, a
resonance in the latter corresponds to a pole in the former or a
resonance in the quark-diquark-baryon coupling $\hqdb$ and the
quark-baryon coupling $\lambda_{qqqb}$. This relation 
also extends to the flow of the diagrams. 

Let us now discuss the structure in detail. To begin with, we first
consider a one loop approximation with momentum and cutoff independent
dressings of vertices and propagators. Then resonances in the diagrams
occur at the frequency $\omega_{\text{\tiny Min}} = 3 m_{{\rm q},\text{\tiny con}}$ with a
constituent mass $m_{{\rm q},\text{\tiny con}}=m_{\rm q}/\Zq$ as well
as $\omega_{\text{\tiny Min}} = m_{{\rm q},\text{\tiny con}} + m_{{\rm
    d},\text{\tiny con}}$ with
$m_{{\rm d},\text{\tiny con}}=m_{\rm d}/\Zd$.

Within the next, already non-trivial approximation we consider
cutoff-dependent vertices, wave-function renormalisations and mass
functions. Then the pole mass, or more precisely the first cut or pole
in the complex plane, is obtained from integrating the flow. Seemingly
it is then given by the relations above for the constant dressing at
vanishing cutoff scale. However, the flows of the dressings are not
necessarily monotonous. This feature is indicative for a
non-monotonous momentum-dependence of the baryonic
diagrams. Accordingly the cut or pole may be at a lower frequency as
that obtained with the dressings at vanishing cutoff. This entails 
that the physical baryon mass $m_{\rm b}/\Zb$ has an upper bound by
$3 m_{{\rm q},\text{\tiny con}}$ as well as
$m_{q,\text{\tiny con}} + m_{{\rm d},\text{\tiny con}}$.  These bounds
are close to each other due to the small flow effects in the diquark
sector. 

Furthermore, the large mass scales in the diagrams support a
derivative expansion in the baryonic diagrams.  For that reason the
above Minkowski argument transports straightforwardly to Euclidean
space and the flows of $m_{\rm b}$ and $\Zb$ reflect the above
properties.  In summary we deduce that the baryon mass, up to a small
binding energy derived from the flow, has to agree with
$3 m_{{\rm q},\text{\tiny con}}$ apart from small differences between
$m_{{\rm d},\text{\tiny con}}$ and $2 m_{{\rm q},\text{\tiny
    con}}$. These considerations are also important for the discussion
of the onset of the baryonic density at the liquid-gas transition.

\subsection{Towards basic properties of baryonic matter}

The set-up with dynamical hadronisation enables us to encompass in a
unified approach to both vacuum QCD and the regime with the
nuclear liquid-gas transition. In the present section we discuss this
set-up, and in particular its foundation in the Silver Blaze property
of QCD~\cite{Cohen:2003kd}.
The latter property, roughly speaking, entails that at
vanishing temperature below the onset of baryon density at an onset
chemical potential $\mu_{q,\textrm{onset}}$ no observable depends on
the chemical potential. Here, $\mu_{q,\textrm{onset}}$ equals the
excitation energy of the lowest lying state in QCD\@. More formally
speaking, it signals the distance of the lowest lying pole or cut in
the complex plane for correlation functions that carry baryon charge. 

This approach aims at a first principle access to the baryonic onset
density and the binding energy of nuclear matter from dynamically
hadronised vector mesons, the results of which will be discussed
elsewhere. We begin with a short review of the Silver Blaze property
in QCD in the context of the fRG approach
\cite{Khan:2015puu, Fu:2015naa, Fu:2016tey}. In particular in
\cite{Khan:2015puu} it has been shown from the fRG that at vanishing
temperature and $\mu_q \leq \mu_{q,\textrm{onset}}$ all $q,\bar q$
correlation functions in QCD satisfy the following Silver Blaze
property,
\begin{align}\nonumber 
  &  \Gamma^{(n)}_{q_1\cdots q_n\bar q_{n+1}
    \cdots \bar q_{2n}}(\omega_1,...,\omega_{2n};\mu_q) \\[1ex]
  = & \,\Gamma^{(n)}_{q_1\cdots q_n\bar q_{n+1}
      \cdots \bar q_{2n}}(\omega_1+ i \mu_q  ,...,\omega_{2n} - i\mu_q;
     0 ) \,.
\label{eq:SilverBlazen}\end{align}
In \Eq{eq:SilverBlazen} we have restricted ourselves to
quark--anti-quark correlation functions, the extension to general QCD
correlation functions is straightforward. Moreover, we suppressed the
dependence on spatial momenta, which are
irrelevant for our present discussions.
\Eq{eq:SilverBlazen} simply entails the fact that below
onset the frequency integration $q_0$ in the flow can be shifted with
$q_0+i\mu_q \to q_0$ in order to absorb the chemical potential. With
\Eq{eq:SilverBlazen} the $\mu_q$-independence of observables follows
trivially. For example, the chiral condensate $\Delta_q$ satisfies at $T=0$, 
\begin{align}\nonumber 
  \Delta_q(\mu_q)\propto &\,\int_\R \frac{d q_0 }{2\pi} \tr\,
                           G_{\bar q q}(q_0;\mu_q)\\[2ex]
  =& 
     \, \int_\R \frac{d q_0 }{2\pi} \tr\, G_{\bar q q}(q_0+i\mu_q; 0 )
     = \Delta_q(0)\,.
\label{eq:SilverBlazeDelta}\end{align}
In the last step in \Eq{eq:SilverBlazeDelta} we used that the loop
integration can be shifted with $q_0+ i\mu_q \to q_0$ for
$\mu_q< \mu_{q,\textrm{onset}}$. Note also that in
\Eq{eq:SilverBlazeDelta} we concentrated on the Silver Blaze property
and hence the frequency integration while dropping the spatial
momentum integrals which require renormalisation.  For
practical purposes it is far more convenient to consider the flow of
the condensate, for more details see e.g.\ \cite{Fu:2019hdw}.

The seemingly trivial property \eq{eq:SilverBlazeDelta} has indeed far
reaching consequences. In the formalism developed here we have already
explained how vector mesons emerge and $\omega_0$ particularly has the
same quantum number as the density.  Because $\omega_0$ and the
density share the same quantum number, $\omega_0$ should acquire a
non-zero background at finite density.  Assuming a simultaneous
complete hadronisation on the level of four-quark as well as
eight-quark interactions, we can most efficiently carry out the
resummation with respect to the background by introducing shifted
chemical potentials via
\begin{align}\label{eq:muBo}
  \bar{\mu}_{\rm q/d/b}= \mu_{\rm q/d/b} -\012
  h_{\omega \rm q/d/b} \;\bar \omega\,,
\end{align}
with the normalised RG-invariant field
$\bar\omega = \Zo^{1/2} \omega_0$ and where we for notational
simplicity denoted the $\omega d^* d-$coupling in the effective
potential by $h_{\omega \rm d}$. Indeed, the interactions between the
$\omega_0$-meson and the baryon and diquark originate in triangle
diagrams with a quark-$\omega_0$ Yukawa interaction. Note that the
shift in the chemical potential is momentum-dependent; strictly
speaking, \Eq{eq:muBo} reads
\begin{align}\label{eq:muBofull}
  \bar{\mu}_{\rm q/d/b}(q,p-q)= \mu_{\rm q/d/b}\,\delta(p) -\012
  h_{\omega \rm q/d/b}(q,p-q) \bar \omega(p),
\end{align}
where $p$ is the $\omega_0$ momentum, $q$ is the quark momenta and
$p-q$ is that of the anti-quark. For example, \Eq{eq:muBofull} leads to
$\int_{r,s}\bar q(r-s) \gamma_0 \bar{\mu}_{\rm q}(s,r) q(s)$ in the
effective action action.

For a first investigation we restrict ourselves to a constant
$h_{\omega \rm q/d/b}\,\bar \omega$ and a field-independent
$\omega_0$-mass $m_\omega^2$. Then, \Eq{eq:muBofull} reduces to
\Eq{eq:muBo}, which simply constitutes a shift in the chemical
potential. Then, the whole $\omega_0$-dependence beyond the kinetic
term is that in $\bar{\mu}_{\rm q/d/b}$. This allows us to invoke the
Silver Blaze property for correlation functions, \eq{eq:SilverBlazen},
which now reads
\begin{align}\nonumber 
  \Gamma^{(n)}_{q_1\cdots q_n\bar q_{n+1}
    \cdots \bar q_{2n}}(p_1,...,p_{2n};\mu_q, \omega_0)& \\[1ex]
  &\hspace{-3cm}=\Gamma^{(n)}_{q_1\cdots q_n\bar q_{n+1}
      \cdots \bar q_{2n}}(\bar p_1  ,...,\bar p_{2n};
     0,0 ) \,, 
\label{eq:SilverBlazen-omega}\end{align}
where we have defined $\bar p= (\bar p_0,\vec p)$ with
\begin{align}\label{eq:p0shifted}
  \bar p_0\equiv p_0+i\, \alpha_{\rm q/d/b}\left( \muq-\012
  \hqo\,
    \bar\omega\right)\,.
\end{align} 
In \Eq{eq:p0shifted}, $\alpha_{\Phi_i}$ for $\Phi_i=\rm q,d,b$ is
defined to match the baryon number of $\Phi_i$ given by
$\alpha_{\Phi_i}/3$. This leads us to the surprising property that the
part of the effective potential $V_\omega(\bar \omega)$ of the
$\omega_0$-meson only consists out of the mass term induced by
dynamical hadronisation, 
\begin{align} \label{eq:OmegaPot}
  V_\omega(\bar \omega)= \frac12 m_\omega^2 \bar \omega^2\,.
\end{align}
Despite the approximation of constant
$h_{\omega \rm q/d/b}\,\bar \omega$ and a field-independent
$\omega_0$-mass this is a very strong constraint on the dynamics of
the vector-channel. Now we consider the general case as follows.

Firstly, we emphasise that the present analysis does not entail the
absence of momentum-dependent interactions of the $\omega$-meson with
momentum $p$.  These interactions are present, but are suppressed in
the low energy limit for $p\to 0$.

Secondly, the $\omega$-mass depends on the scalar--pseudoscalar mesons
$\phi=(\sigma,\vec \pi)$ via the field-dependent quark masses
$m_q(\phi)$. Consequently, even for constant fields \eq{eq:OmegaPot}
is an approximation, but interactions are only triggered by
scalar-pseudoscalar loops. In the ultraviolet the $\phi$-dynamics is
strongly suppressed, while in the infrared the $\omega_0-\phi$
interaction is suppressed by powers of the constituent quark mass.  In
summary this amounts to \Eq{eq:OmegaPot} being a very good
approximation of the full potential. 

We now apply the analysis above to the full system of flow equations.
These equations are built upon the propagators, and in particular on
diquark and baryon propagators. With Silver Blaze,
\eq{eq:SilverBlazen-omega}, all propagators are functions of the
shifted frequency \eq{eq:p0shifted} below the baryonic onset, to wit
\begin{align}
\label{eq:propSB}
G_{\Phi_i}(p_0,\vec p;\mu_{\Phi_i}, \bar \omega)=
G_{\Phi_i}(\bar p_0,\vec p;0,0)\, .
\end{align} 
Hence the Yukawa couplings follow from that
of the quark to $\omega_0$ as
\begin{align}\label{eq:Yuko}
  \hdo= 2 \hqo\,,\qquad \hbo=3 \hqo\,,
\end{align}
and with \Eq{eq:muBo} we obtain
\begin{align}
\mudo=2 \muqo\,,\qquad  \mubo=3 \muqo\,,
\end{align}
below the baryonic onset. As already discussed above, strictly
speaking, this relation only holds for constant shifts of the chemical
potential, but we shall apply this to the full Yukawa
couplings. However, even for momentum-dependent Yukawa couplings the
relations \eq{eq:p0shifted} and \eq{eq:Yuko} will remain valid up to
the point where a pole prevents a shift of the integration contour. As
we will see, the main region of interest is at small momenta, where
this is not expected to be an issue.

Now we come back to the effective potential $V_\omega$. We have
derived from the Silver Blaze property, that, in a good approximation,
the potential only includes the mass term of the $\omega_0$ meson, see
\Eq{eq:OmegaPot}. Evidently, the potential should be symmetric in
$\omega_0^2 +\vec \omega^2$ due to $O(4)$-symmetry, also not allowing
for interacting $\vec \omega$-terms.  To see this, we extend our
discussion concerning $\omega_0$ to the spatial components
$\vec \omega$. Indeed, the same Silver Blaze type argument we have
used for $\omega_0$ also applies to the $\vec \omega$-mesons, and we
illustrate this at the example of the coupling to the quark: the
$\vec\omega$ effectively shift the spatial momentum $\vec q$ to
$\vec q - \tfrac{i}{2}\hqo \vec \omega$, and hence below the threshold
of the quark the correlation functions are simply
$\vec \omega$-independent functions of the shifted momentum
argument. Including also the frequency relation \eq{eq:p0shifted} we
are led to the observation that in a general
$\bar \omega_\mu$-background below the onset the propagators will only
depend on
\begin{align}\label{eq:pmushifted}
\bar p_\mu:=   p_\mu+i\, \alpha_{\rm q/d/b}\left( \muq \delta_{0\mu}- \012 \hqo\,
    \bar{\omega}_\mu\right)\,, 
\end{align} 
generalising \Eq{eq:p0shifted} and hence \Eq{eq:propSB}. We emphasise
that this entails that the effective potential below onset does not
receive $\omega$-dependent contributions from charged particle, i.e.,
quark, diquark or baryon loops. This is an essential difference
compared to the $\sigma$-dependent contributions to the
potential. Note however, that a Taylor expansion in $\omega_0$ or
$\mu$ about $\omega=0$ and $\bar\mu$ below onset provides
non-vanishing coefficients due to the non-analyticity of the Silver
Blaze property: for example, expanding in $\mu$ or $\omega_0$ relates
to an expansion in $A_0$ at vanishing momentum, the Taylor
coefficients of the power $2n$ are identical up to $(-)^n$, the odd
ones vanish.

In summary, the grand potential,
\begin{align} \label{eq:Omega}
  \Omega[\Phi;\bar\mu)= \frac{T}{\cal V}\Gamma[\Phi]\,,
\end{align}
with the spatial volume $\cal V$ at vanishing pion field,
$\vec \pi=0$, has the following unique decomposition,
\begin{align}\label{eq:Omegaob}
  \Omega(\sigma,\bar{\omega},\bar\mu) =
  \Omega_\phi(\sigma)+\Omega_{\bar \omega}(\bar{\omega}) +
  \Delta \Omega_{\bar \mu}(\sigma,\bar\omega, \bar\mu)\,,
\end{align}
where $\bar\mu$ stands for momentum-dependent $\muqo,\mudo,\mubo$ in
\Eq{eq:muBofull} and the last term in \Eq{eq:Omegaob} stands for the density-contributions, 
\begin{align}\label{eq:DeltaOmega}
  \Delta \Omega_{\bar\mu}(\sigma,\bar\omega, \bar\mu) =
  \Omega(\sigma,\bar\omega, \bar\mu)-
  \Omega(\sigma,\bar\omega, 0)\,.
\end{align}
In turn, the first two terms are simply the grand potential
at vanishing chemical potential and density, 
\begin{align} \label{eq:Omegaso} \Omega(\sigma,\bar{\omega},0) =
  \Omega_\phi(\sigma)+\Omega_{\bar \omega}(\bar{\omega}) \,.
\end{align}
The first term originates from the quark-gluon and scalar-pseudoscalar
loops in the flow equation. The second part is generated directly by
the dynamical hadronisation of the vector meson channel. As we have
already discussed before, for constant fields and Yukawa couplings and
field-independent quark mass terms, the effective potential of the
$\omega$-meson is only given by the mass term below onset chemical
potential, see \Eq{eq:OmegaPot}. In this regime, both the
momentum-dependent interactions as well as that triggered by the
exchange of $\sigma$'s and $\vec \pi$'s can be considered to be small
and we conclude
\begin{align}\label{eq:Omegao}
  \Omega_{\bar \omega}(\bar{\omega})\approx \012 m^2_{\bar
    \omega}\,\bar\omega^2\,, 
\end{align}
where $ m_{\bar \omega}$ is the curvature mass of the
$\omega$-meson. This entails that the full $\omega$-dependence on the
action is carried by a free action $\Gamma_{\bar\omega}[\bar\omega]$
with a general dispersion,
\begin{align}\label{eq:Gao}
  \Gamma_{\bar\omega}[\bar\omega]=\012 \int_p Z_\omega(p^2)\,\omega (p)\,
\left[ p^2 + m^2_\omega(p^2)\right]\, \omega(-p)\,. 
\end{align} 
The dispersion is fully determined by the vector four-quark coupling
$\lambdao(p)$ with 
\begin{align} \label{eq:lambdao} 
\lambdao(0,p)=\012 \hqo(0,p)  G_{\omega}(p) \hqo(0,-p)\,,
\end{align}
with $\lambdao(0,p)=\lambda(0,p,0,-p)$ and 
\begin{align}\label{eq:Gomega} 
  G_{\omega}(p) = \0{1}{Z_\omega(p^2)}\0{1}{p^2 + m^2_\omega(p^2)}\,.
\end{align} 
Here we obtained \Eq{eq:lambdao} by evaluating the analogue of
\Eq{eq:pion4} in the vector channel,
\begin{align}\nonumber 
\lambdao(q,-q+p,-q) & \\[1ex]  
& \hspace{-1.5cm} =\012 \hqo(q,p-q)  G_{\omega}(p) \hqo(-q,q-p)\,,
\label{eq:lambdaofull} 
\end{align}
for vanishing quark momentum $q=0$. Note, however, that an immediate
consequence of the above Silver Blaze discussion is, that also the
four-fermi couplings depend on the frequencies $\bar
p_0$. Accordingly, \Eq{eq:lambdaofull} has the dependence
\begin{align}\nonumber 
  \lambdao(\bar q,-\bar q+p,-\bar q)& \\[1ex]  
  & \hspace{-1.5cm} =\012 \hqo(\bar q,p-\bar q)
  G_{\omega}(p) \hqo(-\bar q,\bar q-p)\,.
\label{eq:lambdaofullcor} 
\end{align}
In conclusion the results of this section allow for a simple and
quantitative access to the physics of the liquid-gas transition in QCD
within functional first principles approaches: this access rests on
the triviality of the $\bar \omega$-potential enforced by the Silver
Blaze property, and the access to the frequency dependence of the four-quark
interaction in the $\omega$ channel. Accordingly it only requires the
quantitative computation of vacuum correlation functions of QCD as
done in \cite{Mitter:2014wpa,Cyrol:2017ewj}.

\section{Quark-hadron mixing}\label{sec:QuarkHadron}

As soon as the mass threshold of diquarks is exceeded, they
form a condensate. Inevitably, this causes a mixing effect of the
energy dispersion relations of quarks and baryons. Also, a
diquark condensate would induce a further mixing effect between mesons
and diquarks.  We can see such a mixing pattern very transparently in
the flow equation for the full effective potential $U_k(\rhos,\rhod)$. Its flow at low energies 
is generated from the sum of the quark-loop, the baryon-loop, and the meson-diquark-loop, 
\begin{align}\label{eq:Uk-qdb}
 \partial_t U_k
 = \partial_t U^{\rm (q)}_k + \partial_t U^{\rm (b)}_k
  + \partial_t U^{\rm (m\text{-}d)}_k \,.
\end{align}
For the sake of the present argument we restrict ourselves to the flow equation of the effective
potential in the local potential approximation (LPA). We present all loop contributions in Appendix~\ref{app:flow} utilising the three-dimensional analogue of the LPA-optimised regulator~\cite{Litim:2000ci}. This should facilitate own computation of interested readers. 

Moreover, the phase diagram of the related quark-meson-diquark model (without nucleons) at finite temperature and density has been worked out in \cite{Khan:2015etz} on the basis of the LPA-flow equation for $ U_k$ in \eq{eq:Uk-qdb}, and can be compared to the respective computation in \cite{Khan:2015etz, Khan:2015puu}: the results for the phase structure in the three-color quark-diquark meson model resemble that of the two-colour quark-diquark-meson model, though it shows a much smaller 'BEC-BCS--regime (in terms of $\mu_B$) at low temperatures. This should be attributed to the absense of the Pauli-Gursey symmetry in three-colour, physical, QCD. The latter symmetry leads to $m_\pi=m_d$ in the two-color case and the baryonic onset starts at much smaller baryon (diquark)  chemical potential. 

Interestingly, also the three-color phase diagram shows the phenomenon of precondensation at large densities already found and discussed in the two-colour quark-diquark model in \cite{Khan:2015puu}. Whether such a precondensation phase is also present in the quark-meson-diquark-nucleon model with \eq{eq:Uk-qdb}, is an exciting and phenomenologically relevant open question and will be discussed elsewhere. 

\begin{figure}
\centering
	\subfloat[$q,\bar q, b,\bar b$-masses at $\muq=0\GeV$.]{\includegraphics[width=.95\columnwidth]{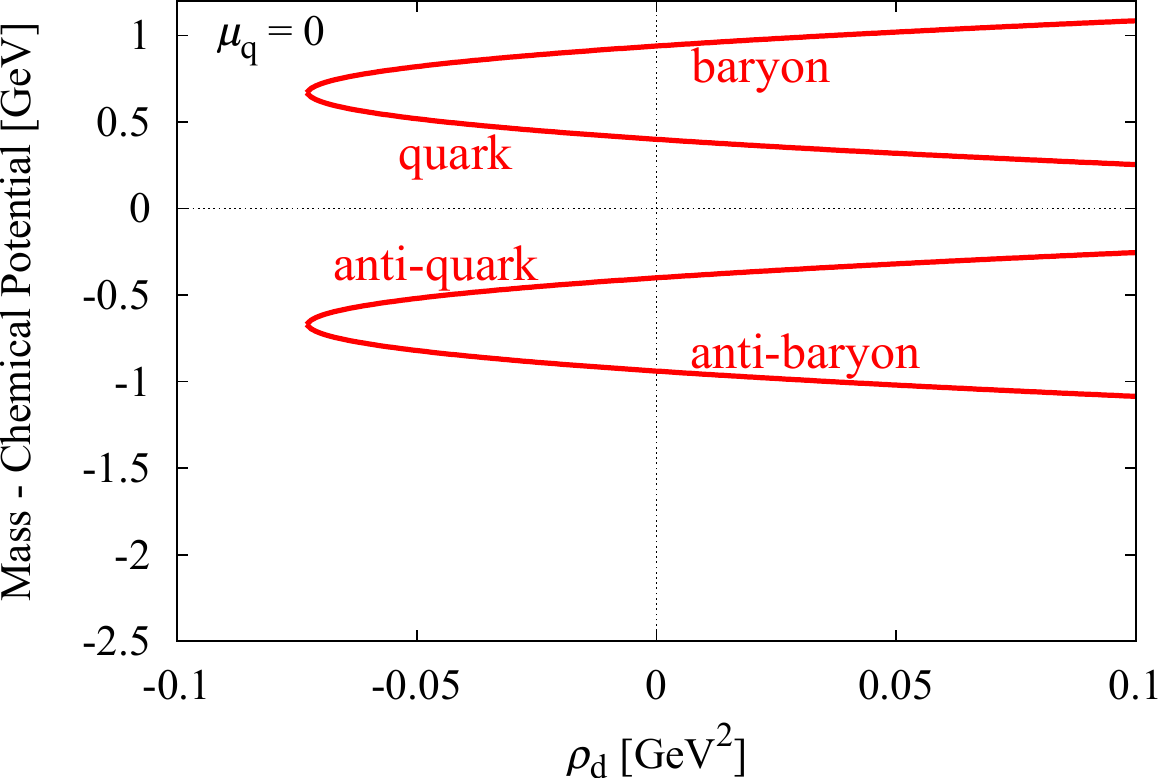}\label{fig:mu000}}
	
	\subfloat[$q,\bar q, b,\bar b$-masses at the baryon threshold $\muq=M_{\rm b}/3=0.313\GeV$.]{\includegraphics[width=.95\columnwidth]{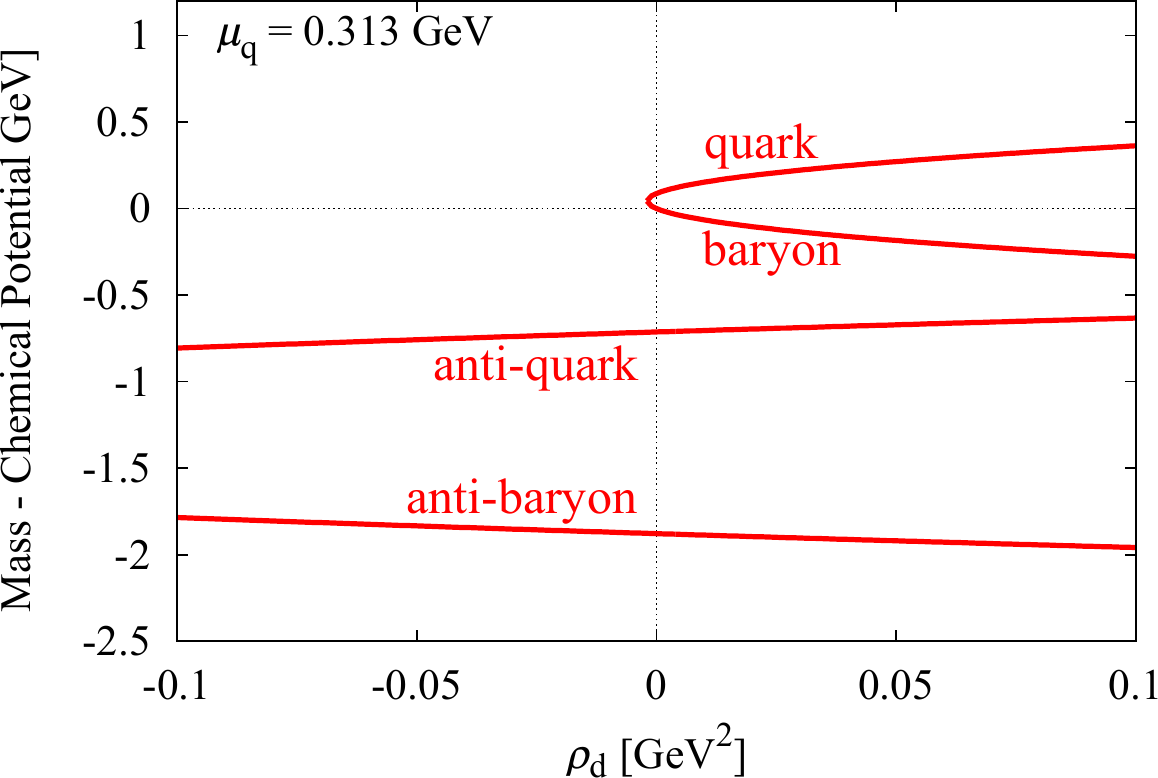}\label{fig:mu313}}

	\caption{$a,\bar q$ and $b,\bar b$ masses relative to the Fermi surface. For simplicity, $h_{qdb}$ is taken to be unity. \hspace*{\fill} }
	\label{fig:mu}
\end{figure}

One part of the contribution of the quark-loop is characterised by
familiar energy dispersion relations, 
\begin{align}\label{eq:familiarE}
 \omega = \sqrt{(\energyq \pm \muq)^2 + 2\hd^2\rhod} \;,\quad
 \energyq \equiv \sqrt{k^2 + \Mq^2} \,.
\end{align}
In a phase with a vanishing diquark condensate, $\rhod=0$, \eq{eq:familiarE} reduces to the standard expressions with dynamical masses. However, there is another more intricate
contribution that arises from the mixing of quarks and baryons in the
presence of a diquark condensate, see \Eq{eq:flow_U_quark} and 
\eq{eq:flow_U_baryon}.  For $\rhod=0$, there is no mixing between quarks and baryons (except for
higher-order fluctuations). Then we find two distinct dispersion
relations,
\begin{align}
 \omega = \pm\sqrt{k^2+\Mq^2} - \muq\;,\quad
 \omega = \pm\sqrt{k^2+M_{\rm b}^2} - \mub\,,
\end{align}
for quarks, anti-quarks, baryons, and anti-baryons, respectively.  The mixing 
of these energy dispersion relations for $\rhod\neq0$ provides relevant information about the 
phenomenology in this phase. The mixing can be deduced from \Eq{eq:omega}, and is depicted in \Fig{fig:mu}. 
There we show the quark and bayron masses at functions of the momentum scale $k$.  At $\rhod=0$ and $\muq=0$, as shown in \Fig{fig:mu000}, this indicates $\pm 0.939\GeV$ for baryons and $\pm 0.4\GeV$ for quarks, which is our input.

We notice from \Fig{fig:mu000}, that the quark and the baryon masses smoothly meet in a region analytically continued to $\rhod<0$.  This link between quarks and baryons is absent in the physical $\rhod>0$ region. It hints at a duality of quarks and baryons, in particular when the diquarks condensate. This also suggests that the relation between the quark and baryon mass branches dictates the binding energy of baryons. \Fig{fig:mu313} shows the masses (relative to the quark/baryon chemical potential) at the baryon mass threshold, $\muq=M_{\rm b}/3$. We see that the baryon branch touches zero, and hence $M_{\rm b}-\mub=0$, at $\rhod=0$.  The quark branch also approaches zero, but it remains finite because $\Mq-\muq=0.087\GeV>0$ there.

\begin{figure}
	\centering
	\subfloat[$\sigma, \tilde d,d$-masses below the diquark
	threshold at $\muq=0.2\GeV$]{\includegraphics[width=.95\columnwidth]{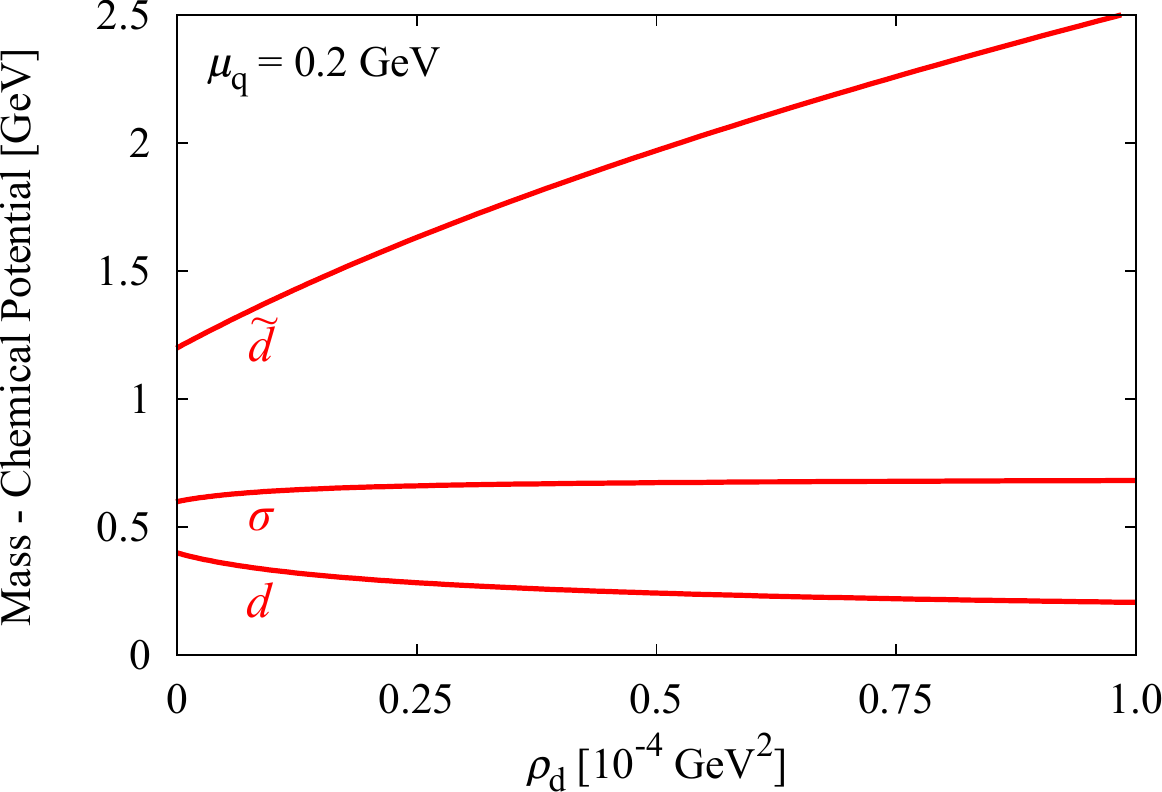}\label{fig:dmu200}}
	
	\subfloat[$\sigma, \tilde d,d$-masses above the diquark
	threshold at $\muq=0.2\GeV$]{\includegraphics[width=.95\columnwidth]{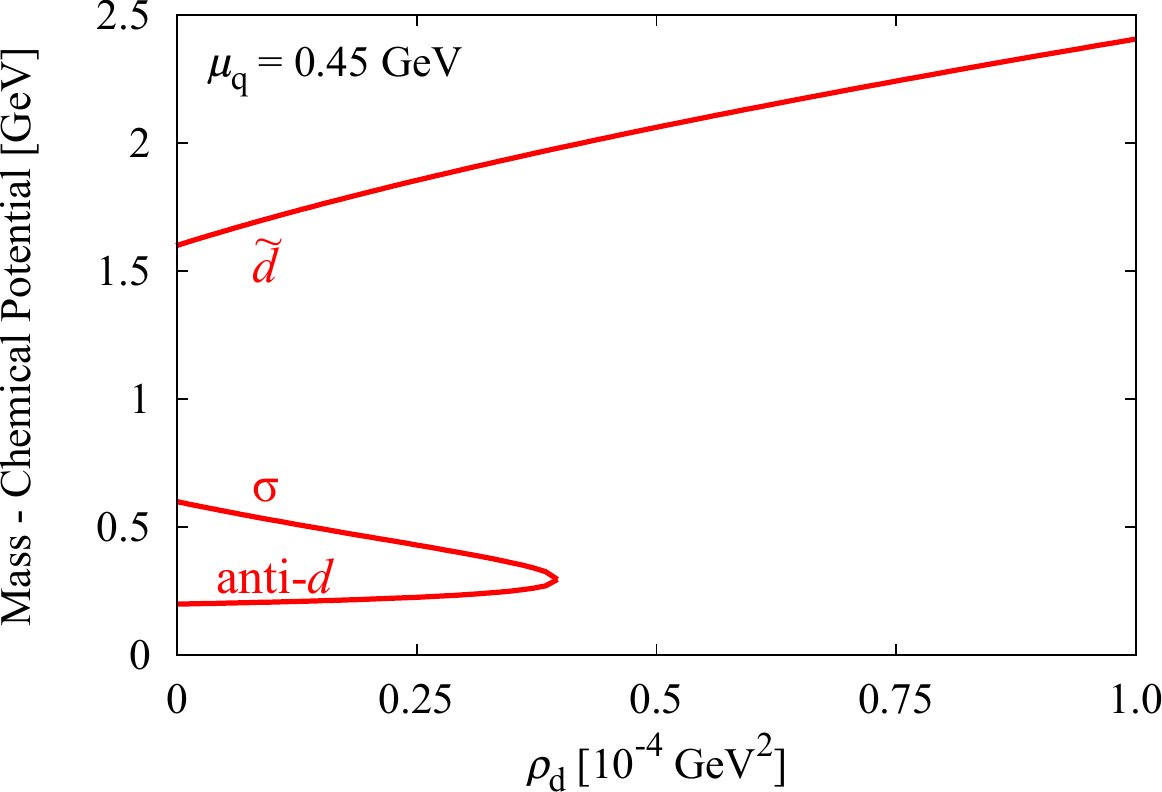}\label{fig:dmu450}}
	
	\caption{Masses of $\sigma$-meson and $\tilde d, d$-diquarks relative to the Fermi surface.\hspace*{\fill} }
	\label{fig:dmu}
\end{figure}
Next we discuss the meson- and the diquark-loop contributions. The respective 'physical' bosonic degrees of freedom are the ($\pi,\sigma$)-mesons and the diquarks. However, be aware of the fact that the diquarks can
have different dispersions just like the ($\pi,\sigma$) mesons, once $\rhod$ becomes nonzero.  The $\sigma$ and pion masses are given by
\begin{align}
 M_\pi^2 = \frac{\partial U_k}{\partial\rhos}\;,\qquad
 M_\sigma^2 = \frac{\partial U_k}{\partial\rhos}
  + 2\rhos\frac{\partial^2 U_k}{\partial \rhos^2} \,.
\label{eq:mass_meson}
\end{align}
Similarly, the two diquark masses read,
\begin{align}
 M_d^2 = \frac{\partial U_k}{\partial\rhod}\,,\qquad
 M_{\tilde d}^2 = \frac{\partial U_k}{\partial\rhod}
   + 2\rhod\frac{\partial^2 U_k}{\partial \rhod^2} \,,
\end{align}
and $M_d^2<M_{\tilde d}^2$.

For $\rhos\neq 0$ and $\rhod\neq 0$, there are also off-diagonal
components in the mass matrix, that induce a mixing effect between $\sigma$ and
$\tilde{d}$ (heavier diquark), see \Eq{eq:bosonicmixing} for an explicit
expression. Then, the physical masses of $\sigma$, $\tilde{d}$, and $d$ are defined by the eigenvalues 
of the mass matrix. 

Their mixing is quite non-trivial. In \Fig{fig:dmu} we have depicted an example of the behaviour of these masses as functions of the diquark condensate $\rhod$. For the results shown in \Fig{fig:dmu} we used the following lowest order Ansatz for the potential, 
\begin{align}\label{eq:U-Ansatz}
 U_k = -m^2 \Bigl(\rhos+\frac{\rhod}{\gamma}\Bigr)
   + g \Bigl(\rhos+\frac{\rhod}{\gamma}\Bigr)^2 - h \sigma \,.
\end{align}
\Eq{eq:U-Ansatz}, while being rather qualitative, should capture the generic properties well. 
We have four parameters for $U_k$, $\gamma$, $g$, $\rhos$, $h$, and we fix these parameters with physical observables: $\sigma=f_\pi=93\MeV$, $M_\pi=135\MeV$, $M_\sigma=600\MeV$, $M_{\rm d}=800\MeV$ at $\rhod=0$. The latter identifications of parameters with observables are the standard ones for  LPA-approximations to LEFTs, the QCD-adjustments for quantitatively reliable approximations in the spirit of QCD-assisted LEFTS can be found in \cite{Mitter:2014wpa, Braun:2014ata, Rennecke:2015eba, Cyrol:2017ewj, Fu:2019hdw}. 

Then, the parameters in our model are straightforwardly determined from the relations, $h=2f_\pi M_\pi^2$,
$\gamma=M_\pi^2/M_d^2$, $g=(M_\sigma^2-M_\pi^2)/(8f_\pi^2)$, and $m^2=M_\sigma^2/4-5M_\pi^2/4+(M_\sigma^2-M_\pi^2)\rhod/(4f_\pi^2\gamma)$. For an analysis of the mixing pattern at finite $\rhod$, we simply study the $\sigma$ and $\tilde{d}, d$ masses as a function of the size of the diquark condensate $\rhod$, while keeping all other parameters fixed.

The three lines in \Fig{fig:dmu200} represent $M_d$, $M_\sigma$,
and $M_{\tilde d}$ from bottom to top.  The $\sigma$ mass is
not directly affected by the chemical potential, so $M_\sigma$ at
$\rhod=0$ stays at $0.6\GeV$.  The diquark dispersions are shifted
by $2\muq=0.4\GeV$. Accordingly, we get $M_d-2\muq=0.4\GeV$ at
$\rhod=0$.  The other diquark is found to be quite massive and
hence decouples from the dynamics.

With increasing $\muq$, we have smaller $M_d-2\muq$ and eventually a
diquark condensate should develop when $M_d-2\muq<0$.  Then, the
diquark branch labelled with $d$ in \Fig{fig:dmu200} moves below
zero.  At the same time, the anti-diquark branch moves up and it is
connected to the $\sigma$ branch, as shown in \Fig{fig:dmu450} 
for $\muq=0.45\GeV$ which is greater than the diquark threshold. 

In summary this leaves us with a very appealing, simple and coherent picture of 
baryonic properties at vanishing finite density. Its embedding in the first principles 
fRG-approach to QCD should allow for a determination of hadron masses and the 
physics of the liquid-gas transition at vanishing and small temperatures.

\section{Conclusions}
In the present work we have discussed dynamical hadronisation in QCD within the functional renormalisation group approach at the explicit example of $N_f=2$ flavour QCD\@. To date, dynamical hadronisation has mostly been applied to the light scalar-pseudoscalar $\sigma, \vec \pi$ mesons. The latter are introduced as effective fields for the resonant scalar-pseudoscalar channel of the four-quark correlation function. This allows us to monitor the dynamical emergence of these low energy degrees of freedom within QCD. 

Here we have, for the first time, set up the dynamical hadronisation approach also for an effective diquark field $d$ for the scalar diquark channel. While the latter is not an asymptotic state it can be understood as a collective offshell degree of freedom that may become resonant at large density in the colour-superconducting phase, see e.g.\ \cite{Braun:2019aow} for an fRG study.

The final new ingredient is the extension of the dynamical hadronisation approach to baryons, i.e.\ the nucleons. The respective
effective field takes care of both, the baryon channel in the six-quark interaction as well as the baryon channel in the
two-quark--two-diquark interaction. 

The approach of emergent hadrons and diquarks put forward in the present work has natural applications to the hadron resonance spectrum, the nuclear liquid-gas transition as well as the physics of nuclear and quark matter at large density.

\acknowledgments

We thank N.~Khan for discussions and collaboration in an early stage of this work. We thank J.~Braun, W.-j.~Fu, F.~Ihssen, F.~Rennecke, F.~Sattler, B.-J.~Schaefer, N.~Wink and C.~Zelle for discussions and a collaboration on related subjects. J.~M.~P.\ thanks the other members of the fQCD-collaboration, \cite{fQCD}, for discussions and work on related subjects. 
K.~F.\ is supported by JSPS KAKENHI Grant No.\ 18H01211 and 19K21874. K.~F.\ is grateful for a warm hospitality at Institut f\"ur
Theoretische Physik, Universit\"at Heidelberg, where he stayed as a visiting professor of the ExtreMe Matter Institute EMMI at GSI and where a part of this work was completed.  The work is supported by EMMI, the BMBF grant 05P18VHFCA, and is part of and supported by the DFG Collaborative Research Centre "SFB 1225 (ISOQUANT)" as well as by Deutsche Forschungsgemeinschaft (DFG) under Germany’s Excellence Strategy EXC-2181/1 - 390900948 (the Heidelberg Excellence Cluster STRUCTURES).

\appendix

\section{Four-quark interactions}

\label{app:4f}
We rely on the basis for $U(2)_L\times U(2)_R$ symmetric four-quark
interactions already used in \cite{Mitter:2014wpa,Cyrol:2017ewj}, see
also \cite{Jaeckel:2002rm}, and \cite{Braun:2011pp} for a review,
\begingroup
\allowdisplaybreaks
\begin{align}
  \mathcal{L}_{(\bar q q)^2}^{(S-P)_+^{\phantom{adj}}}&=(\bar q T^0
  q)^2\!-
  \!(\bar q \gamma^5 T^f q)^2\!-\!(\bar q \gamma^5 T^0  q)^2\!+\!(\bar q T^f q)^2\,,\nonumber\\[2ex]
  \mathcal{L}_{(\bar q
    q)^2}^{(V-A)_{\phantom{-}}^{\phantom{adj}}}&=(\bar q \gamma^\mu
  T^0 q)^2\!+\!(\bar q \gamma^\mu\gamma^5 T^0 q)^2\,,\nonumber\\[2ex]
  \mathcal{L}_{(\bar q
    q)^2}^{(V+A)_{\phantom{-}}^{\phantom{adj}}}&=(\bar q
  \gamma^\mu T^0 q)^2\!-\!(\bar q \gamma^\mu\gamma^5 T^0 q)^2\,,\nonumber\\[2ex]
  \mathcal{L}_{(\bar q q)^2}^{(V-A)_{\phantom{-}}^{\text{adj}}}&=(\bar
  q \gamma^\mu T^0 T^a q)^2\!+\!(\bar q \gamma^\mu\gamma^5 T^0 T^a
  q)^2\,,\nonumber \\[2ex]
  \mathcal{L}_{(\bar q q)^2}^{(S+P)_-^{\phantom{adj}}}&=(\bar q T^0
  q)^2\!-
  \!(\bar q \gamma^5 T^f q)^2\!+\!(\bar q \gamma^5 T^0  q)^2\!-\!(\bar q T^f q)^2\!,\nonumber\\[2ex]
  \mathcal{L}_{(\bar q q)^2}^{(S+P)_-^{\text{adj}}}&=(\bar q T^0 T^a q)^2\!-\!(\bar q \gamma^5 T^f T^a q)^2\!\nonumber\\[1ex]
  &\hspace{1.5cm}+\!(\bar q \gamma^5 T^0 T^a q)^2\!-\!(\bar q T^f T^a q)^2\,,\nonumber \\[2ex]
  \mathcal{L}_{(\bar q q)^2}^{(S+P)_+^{\phantom{adj}}}&=(\bar q T^0
  q)^2\!+\!(\bar q
  \gamma^5 T^f q)^2\!+\!(\bar q \gamma^5 T^0  q)^2\!+\!(\bar q T^f q)^2\!,\nonumber\\[2ex]
  \mathcal{L}_{(\bar q q)^2}^{(S+P)_+^\text{adj}}&=(\bar q T^0 T^a
  q)^2\!+\!(\bar q
  \gamma^5 T^f T^a q)^2\nonumber\\[1ex]
  &\hspace{1.5cm}+\!(\bar q \gamma^5 T^0 T^a q)^2\!+\!(\bar q T^f T^a q)^2\,,\nonumber\\[2ex]
  \mathcal{L}_{(\bar q q)^2}^{(S-P)_-^{\phantom{adj}}}&=(\bar q T^0
  q)^2\!+\!(\bar q
  \gamma^5 T^f q)^2\!-\!(\bar q \gamma^5 T^0  q)^2\!-\!(\bar q T^f q)^2\!,\nonumber\\[2ex]
  \mathcal{L}_{(\bar q q)^2}^{(S-P)_-^{\text{adj}}}&=(\bar q T^0 T^a
  q)^2\!+
  \!(\bar q \gamma^5 T^f T^a q)^2\nonumber\\[1ex]
  &\hspace{1.5cm}-\!(\bar q \gamma^5 T^0 T^a q)^2\!-\!(\bar q T^f T^a q)^2\,.
    \label{eq:fourfermi_sym}
\end{align}
  \endgroup
We denote the generators of flavour $U(1)_f$ and $SU(N)_f$ by
$T^0=\mathbbm{1}/\sqrt{2N_f}$ and $T^{f}$ whereas $T^a$ are the
generators of colour $SU(3)_c$. In addition we define combined
channels like the scalar-pseudoscalar and the (axial)vector channel via
\begin{align}
  \mathcal{L}_{(\bar q q)^2}^{(\phi)}&=\mathcal{L}_{(\bar q
    q)^2}^{(S-P)_+} + \mathcal{L}_{(\bar q q)^2}^{(S+P)_-}=2 (\bar q
  T^0 q)^2-2(\bar q
  \gamma^5 T^f q)^2\!,\nonumber\\[2ex]
  \mathcal{L}_{(\bar q q)^2}^{(\eta')}&=\mathcal{L}_{(\bar q
    q)^2}^{(S-P)_+} - \mathcal{L}_{(\bar q q)^2}^{(S+P)_-}=2 (\bar q
  T^f q)^2-2(\bar q \gamma^5 T^0 q)^2\!,\nonumber\\[2ex]
  \mathcal{L}_{(\bar q q)^2}^{(V)}&=\mathcal{L}_{(\bar q
    q)^2}^{(V-A)} + \mathcal{L}_{(\bar q q)^2}^{(V+A)}=2 (\bar q
  T^0 \gamma^\mu q)^2\,,\nonumber\\[2ex]
  \mathcal{L}_{(\bar q q)^2}^{(AV)}&=\mathcal{L}_{(\bar q
    q)^2}^{(V-A)} - \mathcal{L}_{(\bar q q)^2}^{(V+A)}=2 (\bar q
  T^0 \gamma^\mu\gamma^5 q)^2 \,.
  \label{eq:sigmapiT} \end{align}

\section{Transformation properties of diquark and baryon interpolation operators}
\label{app:transformations}

We review the transformation properties of diquark and baryon interpolation operators \cite{Nagata:2007di,Dmitrasinovic:2011yf}. Here we restrict ourselves 
to (pseudo-)scalar diquark interpolating operators, i.e., to 
\begin{align}
D_1^a=q^T \epsilon^a\tau^2 C\gamma^5 q\,,\qquad \quad 
D_2^a=q^T \epsilon^a\tau^2 C q\,,
\end{align}
and to $I=\frac{1}{2}$ baryon interpolation operators, where for local interactions it is
sufficient due to Fierz identities to consider the two operators \cite{Nagata:2007di},
\begin{align}
B_1=(q^T  \epsilon^a \tau^2 C \gamma^5 q) q^a\,,\quad
B_2=(q^T  \epsilon^a \tau^2 C q) \gamma^5 q^a\,.
\end{align}
Under $U(1)_A$ transformations of the form,
\begin{align}
q\to \exp(\imag \gamma^5 \alpha) q\,,
\end{align}
the above interpolation fields transform as 
\begin{align}
\delta^5_\alpha D_1^a&=2 \imag \alpha D_2\,,\nonumber\\[1ex]
\delta^5_\alpha D_2^a&=2 \imag \alpha D_1\,,\nonumber\\[2ex]
\delta^5_\alpha B_1&=\imag \alpha \gamma^5 (B_1+2 B_2)\,,\nonumber\\[1ex]
\delta^5_\alpha B_2&=\imag \alpha \gamma^5 (B_2+2 B_1)\,. 
\end{align}
Under infinitesimal $SU(N_f)_A$ transformations 
corresponding to 
\begin{align}
q\to \exp(\imag \gamma^5 \vec \alpha\vec \tau ) q \,,
\end{align}
we have 
\begin{align}
\delta^5_{\vec{\alpha}} D_i&=0\,,\nonumber\\[1ex]
\delta^5_{\vec{\alpha}} B_i&=\imag \gamma^5 \vec \alpha \vec \tau B_i\,, 
\end{align}
which implies that the diquarks are chiral scalars and the quarks 
transform in the fundamental representation. The linear combinations 
$D_n\equiv D_1+D_2$, $D_m\equiv D_1-D_2$, $B_n=B_1+B_2$, $B_m=B_1-B_2$
transform under irreducible representations of $U(1)_A$: 
\begin{align}
\delta^5_\alpha D^a_n&=2 \imag \alpha D^a_n\,,\nonumber\\[1ex]
\delta^5_\alpha D^a_m&=-2\imag \alpha D^a_m\,,\nonumber\\[2ex]
\delta^5_\alpha B_n&=3 \imag \alpha \gamma^5 B_n\,,\nonumber\\[1ex]
\delta^5_\alpha B_m&=- \imag \alpha \gamma^5 B_m\,, 
\end{align}
where the subscripts refer to naive/mirror assignment. Note that
interpolation operators of the form $B_1$ is the standard choice for
lattice investigations of the proton ground state whereas $B_n$ and
$B_m$ are frequently used in sum rule analyses
\cite{Leinweber:1994nm}. Considering 6-quark interactions constructed
from $D_1$ and $B_1$ and even disregarding $U(1)_A$ for the moment,
the obvious choice $\bar B_1 B_1$ and correspondingly
$D^*{}^a D{}^b(q^a q^b)$ obviously break chiral symmetry. This can be
fixed by considering momentum-dependent interactions of the type
$\bar B_1 \slashed p B_1$.

\section{LPA flow equations}
\label{app:flow}

We provide the details of the flow equation \eq{eq:Uk-qdb} of the effective potential in the quark-meson-diquark-nucleon model. Its quark-meson-diquark reduction has been discussed in detail in \cite{Khan:2015etz}. There, also the phase diagram of this model has been computed on the basis of \eq{eq:Uk-qdb} without nucleons: it resembles that of the two-colour quark-diquark model, and in particular shows the phenomenon of precondensation at large densities already found and discussed in the two-colour quark-diquark model in \cite{Khan:2015puu}. 

\begin{widetext}
\subsection{Fermionic contributions}

The inverse fermionic propagator for a combined fermionic field
$\Psi=(q_1,\tau_2 C \bar q_2^T,q_3,b)$, where explicit indices denote
colour indices, in a background where $d_1=d_2=0$ takes the following
block-diagonal form,
\begin{align}
\Gamma^{(2)}_{\bar \Psi \Psi}=\begin{pmatrix}M_{q_1 q_2}&0\\0& M_{q_3 b}\end{pmatrix}\,,
\end{align}
where

\begin{align}\nonumber 
  M_{q_1 q_2}&=\begin{pmatrix}i \slashed{p}-\muq \gamma^0+\frac{\hs}{2}(\sigma+i
    \gamma^5 \vec\tau\vec \pi) & -\sqrt{2}\hd  \gamma^5d_3^* \\ \sqrt{2}\hd
    \gamma^5 d_3 &\imag\slashed{p}+\muq \gamma^0+\frac{\hs}{2}(\sigma-
    i \gamma^5 \vec\tau\vec \pi)\end{pmatrix}\,,\\[2ex]
  M_{q_3 b}&=\begin{pmatrix}i \slashed{p}-\mu \gamma^0+\hs(\sigma
    +i \gamma^5 \vec\tau\vec \pi)& \sqrt{2}\hqdb d_3^* \\
    \sqrt{2}\hqdb d_3 &\imag\slashed p- \mub \gamma^0+
    \frac{\hb}{2} (\sigma +i \gamma^5 \vec\tau\vec \pi) \end{pmatrix}.
\end{align}
The quark-loop yields in the local potential approximation (LPA) with
the optimised cutoff,
\begin{align}
 \partial_t U^{\rm (q)}_k=-\frac{8 N_f k^5 T}{6\pi^2}\sum_{p_0}
    \frac{2\rhod \hd^2 + k^2 - \muq^2 + p_0^2 +
    \Mq^2}{(p_0^2+\xi_+^2)(p_0^2+\xi_-^2)}-\frac{4 N_f k^5 T}{6\pi^2} \sum_{p_0} \frac{k^2 +
    \Mb^2 - (\mub - i p_0)^2 + 2 \hqdb^2 \rhod}{p_0^4-i
    b_3 p_0^3-b_2 p_0^2 +i b_1 p_0+b_0}\,,
\label{eq:potqbeforemat}
\end{align}
whereas the baryon-loop contributes with 
\begin{align} 
  \partial_t U^{\rm (b)}_k =-\frac{4 N_f k^5 T}{6\pi^2} \sum_{p_0}
  \frac{k^2 + \Mq^2 - (\muq - i p_0)^2 + 2 \hqdb^2
    \rhod}{p_0^4-i b_3 p_0^3-b_2 p_0^2 +i b_1 p_0+b_0}\,.
\label{eq:potbbeforemat}
\end{align}
Evaluating the Matsubara sum in \eq{eq:potqbeforemat} leads to  
\begin{align}
 \partial_t U^{\rm (q)}_k =& -\frac{4 N_f k^5}{12\pi^2}
  \Biggl[ \frac{1}{\xi_-}\biggl(1-\frac{\muq}{\energyq}
  \biggr)\Bigl(1-2 n_f(\xi_-)\Bigr) + \frac{1}{\xi_+}
  \biggl(1+\frac{\muq}{\energyq}\biggr)
  \Bigl(1-2 n_f(\xi_+)\Bigr) \notag\\[1ex]
  &+\sum_{i=1}^4
  \frac{1-2n_f(\xi_i)}{(\xi_i-\xi_{i+1})(\xi_i-\xi_{i+2})
  (\xi_i-\xi_{i+3})} \Bigl\{ k^2+M_{\rm b}^2-(\mub-\xi_i)^2+ 2\hqdb^2\rhod \Bigr\}
  \Biggr]\;,
\label{eq:flow_U_quark}
\end{align}
\end{widetext} 
where the energies are defined as
\begin{align}
 \energyq = \sqrt{k^2 + \Mq^2} \;,\qquad
 \xi_\pm = \sqrt{(\energyq \pm \muq)^2 + 2\hd^2\rhod}
\end{align}
and
\begin{align}
n_f(x)=\frac{1}{e^{x/T}+1}\;,\quad \Mqb^2=\frac{1}{2} \hsb^2 \rhos\;.
\end{align}
The $\xi_i$'s in \eq{eq:flow_U_quark} represent the four roots of the equation
\begin{align}
 \xi^4 + b_3\xi^3 +b_2\xi^2 +b_1\xi + b_0 = 0 \;,
\label{eq:omega}
\end{align}
where we set in addition $\xi_i=\xi_{i-4}$ for $i=5,6,7$. The
coefficients are given by
\begin{align}
  b_0 &= (k^2+\Mq^2-\muq^2)(k^2+M_{\rm b}^2-\mub^2) \nonumber\\[0.7ex]
  &\quad + 4\hqdb^2(k^2-\muq\mub-\Mq \Mb) \rhod +
  4\hqdb^4\rhod^2 \;,\nonumber\\[1ex]
  b_1 &= 2\mub(k^2+\Mq^2-\muq^2) +2\muq(k^2+
  M_{\rm b}^2-\mub^2) \nonumber\\[0.7ex]
  &\qquad\qquad -4\hqdb^2\rhod(\muq
  +\mub) \;,\nonumber\\[1ex]
  b_2 &= -(k^2+\Mq^2-\muq^2)-(k^2
  +M_{\rm b}^2-\mub^2) \nonumber\\[0.7ex]
  &\qquad\qquad -4\rhod\hqdb^2 + 4\muq\mub \;,\nonumber\\[1ex]
  b_3 &= -2(\muq+\mub) \;.
\end{align}
Similarly, by carrying out the Matsubara sum in \eq{eq:potbbeforemat}
one obtains,
\begin{align}
  & \partial_t U^{\rm (b)}_k = -\frac{2 N_f k^5}{12\pi^2}\sum_{i=1}^4
  \frac{1-2n_f(\xi_i)}{(\xi_i-\xi_{i+1})(\xi_i-\xi_{i+2})
    (\xi_i-\xi_{i+3})} \notag\\[1ex]
  &\times \Bigl\{ k^2+M_{\rm q}^2-(\muq-\xi_i)^2 + 2\hqdb^2\rhod
  \Bigr\} \;.
\label{eq:flow_U_baryon}
\end{align}
To confirm that both flow equations carry the correct number of
physical degrees of freedom it is instructive to consider the limit of
a vanishing diquark condensate $\rhod=0$, where we find
\begin{align}\label{eq:qb-flowd=0}
  \partial_t U_k^{({\rm q})}&=-\frac{2k^5 \nu_q}{12\pi^2}\frac{1}{\energyq}
  \left[ 1-n_f(\energyq+\muq)-n_f(\energyq-\muq)\right]\,,
  \notag\\[2ex]
\partial_t U_k^{({\rm b})}&=-\frac{2k^5 \nu_b}{12\pi^2}\frac{1}{\energyb}
\left[ 1-n_f(\energyb+\mub)-n_f(\energyb-\mub)\right]\,,
\end{align}
with $\nu_q=2N_c N_f$ and $\nu_b=2N_f$ for $N_f=2$,
$N_c=3$ and $\energyb=\sqrt{k^2+\Mb^2}$ and clearly identify
quark and baryon contributions.

\subsection{Bosonic contributions}
The inverse bosonic propagator for $\Phi=(\vec
\pi,\,\sigma,\,\text{Re}\,d^3,\, \text{Im}\, d^3,\,\text{Re}\,d^1,\,\text{Im}\,d^1,\,\text{Re}\,d^2,\,\text{Im}\,d^2)$ in a
background where $d^a=\delta^{a3}\sqrt{2\rhod}$ takes the
block-diagonal form,
\begin{align}
\Gamma^{(2)}_B=\begin{pmatrix}(p^2+U_{,\rhos})\mathbbm{1}_{3\times3}&&&\\&M_{\sigma \tilde d d}&&\\&&M_{dd}&\\&&&M_{dd}\end{pmatrix}\,,
\end{align}
where we use the shorthand notation $U_{,\rho_i}\equiv \frac{\partial U_k}{\partial \rho_i}$ and $U_{,\rho_i\rho_j}\equiv \frac{\partial^2 U_k}{\partial \rho_i\partial \rho_j}$.
The non-trivial submatrices are given by
\begin{widetext}
\begin{align}
M_{\sigma \tilde d d}&=\begin{pmatrix}
p^2+U_{,\rhos}+2 U_{,\rhos\rhos}\rhos&U_{,\rhos\rhod}\sqrt{\rhos \rhod}&0\\
 U_{,\rhos\rhod}\sqrt{\rhos\rhod}&p^2+U_{,\rhod}+2U_{,\rhod\rhod}\rhod-\mud^2& -2 p_0\mud\nonumber\\
0&2 p_0\mud&p^2+U_{,\rhod}- \mud^2
\end{pmatrix},\nonumber\\[2ex]
M_{dd}&=\begin{pmatrix}p^2+U_{,\rhod}-\mud^2& -2 p_0\mud\\
2 p_0\mud&p^2+U_{,\rhod}- \mud^2
\end{pmatrix}\,.
\label{eq:bosonicmixing}
\end{align}
\end{widetext}
In the LPA for the flat regulator we can now evaluate the bosonic contributions
to the flow of the effective potential that read,
\begin{align}
 \partial_t U_k^{(\rm m\mbox{-}d)} =&\, \frac{k^5}{12\pi^2}\Biggl[
  \frac{3}{\varepsilon_\pi}\Bigl(1+2 n_b(\varepsilon_\pi)\Bigr) \notag\\[1ex]
 &\hspace{.8cm} +\frac{4}{\varepsilon_{\rm d}} \Bigl(1+n_b(\varepsilon_{\rm d}+\mud)+n_b(\varepsilon_{\rm d}-\mud)\Bigr)\notag\\[1ex]
 &\hspace{-2cm}+ \sum_{i=1}^3 \frac{\alpha_4\omega_i^4+\alpha_2\omega_i^2
  +\alpha_0}{\beta_6(\omega_{i+1}^2\!-\!\omega_i^2)(\omega_{i+2}^2
  \!-\!\omega_i^2)} \frac{1}{\omega_i}\Bigl(1+2n_b(\omega_i)\Bigr)
  \Biggr]\, ,
\label{eq:flow_m}
\end{align}
where
\begin{align}
\varepsilon_\pi\equiv\sqrt{k^2+M_\pi^2}\,,\quad \varepsilon_{\rm d}\equiv\sqrt{k^2+Md^2}\,,
\end{align}
and
\begin{align}
n_b(x)=\frac{1}{e^{x/T}-1}\,,\quad M_\pi^2=U_{,\rhos},\quad \Md^2=U_{,\rhod}\,.
\end{align}
Here we defined $\omega_i^2$ as the three roots of the cubic equation in $p_0^2=-\omega^2$,
\begin{align}
\beta_6 p_0^6+\beta_4 p_0^4+\beta_2 p_0^2+\beta_0=0\,,
\end{align}
which in turn relates to the $(3\times 3)$-submatrix $M_{\sigma d d}$ via
\begin{align}
\label{eq:alphabeta}
\text{Tr} M_{\sigma \tilde d d}^{-1}|_{\vec p^2\to k^2} =\frac{\alpha_4 p_0^4+\alpha_2 p_0^2+\alpha_0}{\beta_6 p_0^6+\beta_4 p_0^4+\beta_2 p_0^2+\beta_0}\,.
\end{align}
As in the case of the $\xi_i$'s we identify $\omega_i=\omega_{i-3}$ for $i=4,5$. At this point we refrain from giving explicit expressions for the coefficients $\alpha_i$ and $\beta_i$, which can be easily calculated from \eq{eq:alphabeta} and are given explicitly for the closely related case of two-colour QCD in \cite{Strodthoff:2011tz}. Again it is instructive to consider the case of a vanishing diquark condensate $\rhod=0$, where the bosonic flow reads
\begin{align}
\label{eq:bosonicrhod0}
\partial_t U_k^{(\rm m\mbox{-}d)} =&\, \frac{k^5}{12\pi^2}\Biggl[
  \frac{3}{\varepsilon_\pi}\Bigl(1+2 n_b(\varepsilon_\pi)\Bigr) +\frac{3}{\varepsilon_\sigma}\Bigl(1+2 n_b(\varepsilon_\sigma)\Bigr) \notag\\[1ex]
 &\hspace{-.7cm} +\frac{6}{\varepsilon_{\rm d}} \Bigl(1+n_b(\varepsilon_{\rm d}+\mud)+n_b(\varepsilon_{\rm d}-\mud)\Bigr)\Biggr]\,,
\end{align}
where 
\begin{align}
\varepsilon_\sigma=\sqrt{k^2+U_{,\rhos}+2 U_{,\rhos\rhos}\rhos}\,.
\end{align}
 In \eq{eq:bosonicrhod0} one easily identifies
contributions from three pions, one sigma meson and three complex diquark degrees of freedom.

\bibliographystyle{apsrev4-1}
\bibliography{diquark}
\end{document}